\documentclass{article}[12pt]
\pdfoutput=1
\usepackage[totalwidth=6.0in, totalheight=9.0in]{geometry}
\usepackage{amsmath,amssymb}
\usepackage{graphicx}
\usepackage[toc,page,titletoc]{appendix}

\usepackage{cite}
\usepackage[pdftex]{hyperref}
\hypersetup{
	linktoc=all,
	breaklinks=true,
	colorlinks,
	citecolor=black,
	filecolor=black,
	linkcolor=black,
	urlcolor=black
}
\def\eq{\begin{equation}}
\def\en{\end{equation}}
\def\eqa{\begin{eqnarray}}
\def\ena{\end{eqnarray}}
\def\aeq#1{\begin{align}#1\end{align}}  
\def\ateq#1#2{\begin{alignat}{#1}#2\end{alignat}}  

\def\Reals{\mathbb{R}}
\def\Integers{\mathbb{Z}}
\def\Complexes{\mathbb{C}}
\def\partialby#1{\frac{\partial\hfill}{\partial#1}}
\def\tr{\mathrm{tr}}
\def\rI{\rho_{+}}
\def\rIbar{\rho_{-}}

\def\xI{x_{+}}
\def\xIbar{x_{-}}

\def\fI{f_{+}}
\def\fIbar{f_{-}}

\def\DI{D_{+}}
\def\DIbar{D_{-}}
\def\AI{A_{+}}
\def\AIbar{A_{-}}
\def\YM{\mathit{YM}}

\def\gtwist{g_{\mathit{tw}}}
\def\zb{\mathbf{z}}
\def\eb{\mathbf{e}}
\def\yb{\mathbf{y}}
\def\wb{\mathbf{w}}
\def\tw{\mathit{tw}}
\def\slow{\mathit{slow}}
\def\Dslow{D^{\slow}}
\def\Dpm{D_{\pm}}
\def\Round{R^{2}}
\begin{document} 
\begin{titlepage}
\def\thepage{}
\vspace*{0.5in}

\begin{center}
{\Large \bf A loop of $SU(2)$ gauge fields\\[2ex] 
stable under the Yang-Mills flow}

\vskip .75in
{\large Daniel Friedan}

\vskip 0.2in
Department of Physics and Astronomy\\
Rutgers, The State University of New Jersey\\
Piscataway, New Jersey 08854-8019 USA\\[0.5ex]
and\\[0.5ex]
Natural Science Institute\\
The University of Iceland\\
Reykjavik, Iceland\\[1ex]
friedan@physics.rutgers.edu\\[1.5ex]

\vskip .5in

\begin{abstract}
\normalsize
The gradient flow of the Yang-Mills action
acts pointwise on closed loops of gauge fields.
We construct
a topologically nontrivial loop of $SU(2)$ gauge fields on $S^{4}$
that is locally stable under the flow.
The stable loop is written explicitly as a path
between two gauge fields equivalent under
a topologically nontrivial $SU(2)$ gauge transformation.
Local stability is demonstrated by calculating the flow
equations to leading order in perturbations of the loop.
The stable loop might play a role in physics as a
classical winding mode of the {lambda model}, a 2-d quantum field
theory that was proposed as a mechanism for generating spacetime
quantum field theory.
We also present evidence
for 2-manifolds of $SU(3)$ and $SU(2)$ gauge fields
that are stable under the Yang-Mills flow.
These might provide 2-d instanton corrections in the
lambda model.
\vskip2ex

For Isidore M.\ Singer in celebration of his eighty-fifth birthday.
\end{abstract}
\end{center}
\end{titlepage}

\newpage
\tableofcontents

\newpage
\section{Introduction}

We are interested in the long time behavior of the Yang-Mills
flow acting on topologically nontrivial loops and 2-spheres of 
$SU(2)$ and $SU(3)$ gauge fields
on $S^{4}$.  
Singer \cite{Singer1978} noted that the homotopy groups $\pi_{n}(\mathcal{A}/\mathcal{G})$ of the space
$\mathcal{A}/\mathcal{G}$ of gauge fields on $S^{4}$ modulo gauge equivalence 
are given by the homotopy groups of the gauge group $G$,
\eq
\pi_{n}(\mathcal{A}/\mathcal{G})=\pi_{n-1}\mathcal{G}=\pi_{n+3}G
\,.
\en
There are
nontrivial loops of $SU(2)$ gauge fields becausen
$\pi_{4}SU(2)=\Integers_{2}$ \cite{Hopf1931,Freudenthal1937}.
For $SU(3)$,
there are no nontrivial loops because
$\pi_{4}SU(3)=0$ \cite{Eckmann1942thesis}.
There are topologically nontrivial 2-spheres of gauge fields for both $SU(2)$ and 
$SU(3)$ because
$\pi_{5}SU(3)=\Integers$ \cite{Eckmann1942thesis} and
$\pi_{5}SU(2)=\Integers_{2}$ \cite{Pontryagin1950,Whitehead1950}.

Our motivation is a hypothetical effect in a speculative theory of
physics.  The lambda model \cite{Friedan2003} is a 2-dimensional
nonlinear model whose target space is the manifold of spacetime
fields.  The short distance fluctuations in a 2-d nonlinear model
generate a measure on its target manifold, called the \emph{a priori}
measure.  In the lambda model, the \emph{a priori}
measure is a measure on the manifold of
spacetime fields: a quantum field theory.  The \emph{a priori} measure
of the lambda model is generated by a diffusion process in the loop
space of the target manifold, driven by the gradient flow of the
classical spacetime action.  We are pursuing the possibility that the
quantum field theory generated by the lambda model will be different
from the canonically quantized field theory because of nonperturbative
2-dimensional effects.  The dominant nonpertubative effects at weak
coupling will be due to winding modes, which are associated with
topologically nontrivial loops in the target manifold, and instantons,
which are associated with topologically nontrivial 2-spheres in the
target manifold.  Winding modes in the lambda model might give rise to
non-canonical physical states in the spacetime quantum field theory.
Instantons in the lambda model might produce non-canonical
interactions.

We are motivated by these possibilities to investigate the
concentration points of the gradient flow of the Yang-Mills action as
it acts on loops and on 2-spheres of $SU(2)$ and $SU(3)$ gauge fields
on $\Reals^{4}$.  We replace $\Reals^{4}$ by its conformal
compactification $S^{4}$, studying gauge fields in topologically 
trivial bundles over $S^{4}$.
As it turns out, 
our results will be applicable to gauge fields on
$\Reals^{4}$ because they will
concern fixed points of the Yang-Mills flow,
i.e., critical points of the Yang-Mills action,
which is conformally invariant.

We find that the Yang-Mills flow concentrates on a nontrivial loop of
singular $SU(2)$ gauge fields made out of a zero-size
instanton and a zero-size anti-instanton.  We find evidence that the
Yang-Mills flow concentrates on a nontrivial 2-sphere of singular $SU(3)$ gauge
fields also made from a zero-size instanton and
a zero-size anti-instanton.  We find evidence that the flow
concentrates on a nontrivial 2-sphere of $SU(2)$ gauge fields made
from configurations of two zero-size instantons and two zero-size
anti-instantons.  These singular gauge fields live in the boundary of
the manifold of gauge fields.

The natural metric on the manifold of gauge fields degenerates at the
boundary, so the stable loop of gauge fields has zero length and the presumptive
stable 2-spheres have zero area.  This keeps alive the hope that they
might have observable effects at low energy in the quantum field
theory.  
Loops or 2-spheres of nonsingular gauge fields,
with nonzero length or area, would make
contributions in the lambda model only visible at
extreme small distance in spacetime.

Let ${\mathcal{A}}$ be the space of connections (gauge fields)
in the trivial $SU(2)$ principle bundle over $S^{4}$.
A connection is described by its corresponding covariant derivative
$D=d+A$,
where $A$ is an $su(2)$-valued 1-form on $S^{4}$.
The curvature 2-form is
\eq
F= D^{2} = dA + A^{2}
\,.
\en
The group
of gauge transformations, $\mathcal{G}$,
is the group of maps $\phi: S^{4}\rightarrow SU(2)$
acting on connections by
\eq
d+A \mapsto \phi(d+A)\phi^{-1}
\qquad
A \mapsto  \phi d(\phi^{-1}) +\phi A \phi^{-1}
\,.
\en
${\mathcal{A}}/\mathcal{G}$ is
the space of gauge equivalence classes of connections.

The Yang-Mills (Y-M) action is
\eq
S_{\YM}(A) = \frac1{8\pi^{2}}\int_{S^{4}} \tr(-F *F)
\en
where ${*}$ is the Hodge operator, which takes $k$-forms to $(4-k)$-forms
and satisfies ${*}^{2} = (-1)^{k}$.
The action $S_{\YM}$ is normalized so that the 
BPST instanton \cite{Belavin1975} has action $1$.
The Yang-Mills flow on the space of connections 
is the gradient flow
of the Y-M action \cite{Donaldson1985,Struwe1994,Schlatter1996,Schlatter1997,Schlatter1998},
\eq
\label{eq:gradient}
\frac{d A\hfil}{d t} = -\nabla S_{\YM}
=*\, D * F
= *\, \left ( d  * F + [A,\, *F]\right )
\,.
\en
The sign is such that $S_{\YM}$ decreases along the flow.
The gradient is taken with respect to the $L_{2}$ metric on 
variations $\delta A$ of $A$,
\eq
(ds)^{2}_{\mathcal{A}}
= \frac1{4\pi^{2}}\int_{S^{4}}  \tr(-\delta A *\delta A)
\,.
\label{eq:metric}
\en
The Y-M flow is gauge invariant (commutes with gauge transformations), so it acts
on the gauge equivalence classes ${\mathcal{A}}/\mathcal{G}$.

The Y-M flow acts pointwise on
parametrized loops in $\mathcal{A}/\mathcal{G}$,
acting simultaneously on each connection along the loop.
This action on parametrized loops is invariant under reparametrizations of the loop,
so the Y-M flow acts on the unparametrized loops
$\mathit{Maps}(S^{1}\rightarrow 
\mathcal{A}/\mathcal{G})/\mathit{Diff}(S^{1})$.
We are interested in the long time behavior of the Y-M flow acting on 
the unparametrized loops in $\mathcal{A}/\mathcal{G}$.
We expect that each connected component of the loop space
contains a stable loop that is the generic attractor for the Y-M flow.
There is an obvious stable attractor
among the topologically trivial loops:
the constant loop at the flat connection.
All nearby connections are driven to the flat connection,
so all nearby loops are driven to the constant loop.

The connected components of the loop space are the elements of the 
fundamental group
$\pi_{1}({\mathcal{A}}/\mathcal{G})$.  As Singer \cite{Singer1978} 
pointed out,
the long exact sequence of homotopy groups implies
$\pi_{n}({\mathcal{A}}/\mathcal{G})=\pi_{n-1}\mathcal{G}$,
since $\mathcal{A}$ is a contractible space.
In particular, 
\eq
\pi_{1}({\mathcal{A}}/\mathcal{G})=\pi_{0}\mathcal{G} = 
\pi_{4}SU(2) = \Integers_{2}
\,.
\en
The loop space of $\mathcal{A}/\mathcal{G}$ thus has
two connected components: the trivial 
(contractible) loops and the 
nontrivial (non-contractible) loops.
The nontrivial loops in ${\mathcal{A}}/\mathcal{G}$ lift to paths in $\mathcal{A}$ whose endpoints 
are gauge equivalent under a nontrivial gauge transformation,
i.e., one that belongs to the nontrivial connected component of $\mathcal{G}$.

Heuristically, we expect a stable nontrivial loop to be associated
with an index 1 fixed point --- a fixed point whose unstable manifold 
is one-dimensional.
The unstable manifold will consist of two
outgoing branches.  We expect each of the two branches to flow to a flat
connection, the two flat connections being gauge equivalent under a
nontrivial gauge transformation.  The unstable manifold will thus form a
nontrivial loop in $\mathcal{A}/\mathcal{G}$.  This loop will be
locally stable because any nearby loop will intersect the codimension
1 stable manifold of the fixed point.

Here, we use elementary methods to find a locally stable attractor
among the nontrivial loops.  We start out completely ignorant of the
long time fate of a generic nontrivial loop of connections under the
Y-M flow.  In hope of relieving our ignorance, we pick a particular
nontrivial loop of connections,
derived from the homogeneous space $SU(3)/SU(2)=S^{5}$,
then try by numerical calculation to
discover its long time behavior under the Y-M flow.  The numerical
results suggest the existence of an index 1 fixed point lying within
the space of singular connections that consist of a zero-size
instanton at one point and a zero-size anti-instanton at a second
point and are flat everywhere else.  A nontrivial loop of such
singular connections is written explicitly.  The loop is parametrized 
by the angle
$\sigma$ that measures the relative rotation between the instanton and
the anti-instanton.  The Y-M flow is calculated asymptotically near
these \emph{twisted pairs}.  The stable loop of twisted pairs is found by examining the
flow lines.

Sibner, Sibner and Uhlenbeck \cite{Sibner1989} study a 
related
problem.  They consider the submanifold 
$(\mathcal{A}/\mathcal{G})_{\mathit{inv}}\subset
\mathcal{A}/\mathcal{G}$ consisting of the $SU(2)$ connections on 
$S^{4}$ invariant under a certain $U(1)$ symmetry group.
The submanifold $(\mathcal{A}/\mathcal{G})_{\mathit{inv}}$ separates into a series of connected components, 
indexed by $m\ge 1$.
Each connected component has nontrivial $\pi_{1}$.
For each $m$, they write a nontrivial loop
consisting of a zero-size $m$-instanton at one pole in $S^{4}$ glued to a zero-size 
$m$-anti-instanton at the other pole.
For $m=1$, their loop is exactly the loop of twisted pairs considered 
here.
They point to \cite{Donaldson1986} for references on the 
nontriviality of such loops.
They apply a min-max procedure:
minimizing the maximum value of $S_{\YM}$ along the loop, over
all nontrivial loops
in $(\mathcal{A}/\mathcal{G})_{\mathit{inv}}$ that belong to
the same homotopy class.
For $m\ge 2$, they are able to make
a small perturbation of the loop
of singular connections to obtain a loop of nonsingular connections
that has $S_{\YM}<2m$ everywhere on the loop.
They then
prove that the min-max connection
provides a non-singular critical point of the Y-M action
that is neither self-dual nor anti-self-dual ---
the first examples of such in 4 dimensions.
Their min-max connections should have index 1 within the submanifold 
$(\mathcal{A}/\mathcal{G})_{\mathit{inv}}$
and should correspond to globally stable loops under the Y-M flow
acting on $(\mathcal{A}/\mathcal{G})_{\mathit{inv}}$.
Here, we treat
a much more elementary
question: the local stability of the loop of twisted pairs
(their $m=1$ loop)
within the full 
$\mathcal{A}/\mathcal{G}$.

We present a summary of our results on the stable loop of 
$SU(2)$ gauge fields, then some preliminaries on notation
and basic formulas, then the computer calculation, then the explicit 
loop of twisted pairs and its nontriviality,
then the calculation of the flow asymptotically nearby and the 
demonstration of local stability.
We present evidence of
stable 2-manifolds for the gauge groups 
$SU(3)$ and $SU(2)$.
For $SU(3)$ we expect this to be
a stable 2-sphere.
For $SU(2)$ we expect either a 2-torus or 2-sphere.
At the end, we raise some
mathematical questions and make some very preliminary remarks about
possible effects in the lambda model.

\section{Summary of the result}

\subsection{BPST instantons}

The BPST instanton \cite{Belavin1975} is the self-dual gauge field,
${*}F =F$, in the $SU(2)$ bundle of Pontryagin index ${+}1$ over $S^{4}$.
The anti-instanton is the anti-self-dual gauge field, ${*}F =-F$, in the
bundle of Pontryagin index $-1$.
Explicit formulas are given
in section~\ref{sect:instanton} below.
The instantons are parametrized by
a point in $S^{4}$ --- the \emph{location} of the instanton ---
and by a nonnegative real number --- the \emph{size} of the instanton 
---
and by an element in $SU(2)/\{\pm\mathbf{1}\}$ --- the orientation of the 
instanton.
Strictly speaking, all the orientations of an isolated instanton are 
gauge equivalent.
The orientation becomes significant
when instantons are combined, the relative 
orientations being gauge invariant.

In the limit where the size of the instanton goes to $0$, the 
instanton becomes a singular connection whose action density
$({8\pi^{2}})^{-1}\tr(-F *F)$ is a Dirac delta-function concentrated at the 
location of the instanton,
while $F=0$ everywhere else.

\subsection{Twisted pairs}
A \emph{twisted pair} is
a singular gauge field in the trivial bundle consisting
of a zero-size instanton at one point in $S^{4}$ and a zero-size
anti-instanton at a second point.  The zero size limit is taken with the
ratio of the sizes held fixed.
A twisted pair is everywhere either self-dual or anti-self-dual or flat,
so each twisted pair is a fixed point of the flow.
The twisted pairs are parametrized
by the location $x_{+}\in S^{4}$ of the instanton,
by the location $x_{-}\in S^{4}$ of the anti-instanton,
by the ratio $\rho_{+}/\rho_{-}$ of the size of the instanton to the size of 
the anti-instanton, and by the relative orientation or
\emph{twist},  $\gtwist\in SU(2)/\{\pm1\}$.
Of the two orientations,
the instanton's and the anti-instanton's,
one is eliminated by
a gauge transformation,
leaving only the relative orientation
to parametrize the twisted pairs.
We establish by
an explicit calculation that
a loop of twisted pairs is nontrivial in $\mathcal{A}/\mathcal{G}$
if $\gtwist$ traverses a nontrivial
loop in the space $SU(2)/\{\pm1\}$ of relative orientations.

\subsection{Conformal symmetry}

The Hodge ${*}$-operator acting on 2-forms is conformally invariant in
four dimensions, so the conformal symmetry group of
$S^{4}$, which is $SO(1,5)$, acts on the space of critical points of the Yang-Mills
action, in particular on the space of twisted pairs.  Using conformal
transformations, we can move the zero-size instanton to the south pole
in $S^{4}$ and the zero-size anti-instanton to the north pole.  We can
make the sizes of the instanton and the anti-instanton equal.  The
remaining subgroup of the conformal group is $SO(4)$, which acts on
the twist $\gtwist\in SU(2)/\{\pm\mathbf{1}\}$ by conjugations.  So we
can diagonalize $\gtwist$.  The twisted pair is invariant under the
remaining $U(2)$ subgroup of $SO(4)$.  The conformal equivalence
classes of twisted pairs form a one parameter family labelled by the
conjugacy classes of $SU(2)/\{\pm\mathbf{1}\}$.  Each twisted pair has
a $U(2)$ symmetry.

The metric on $\mathcal{A}$ is not conformally invariant, so the Y-M
flow is not conformally invariant away from the fixed points.  Near
the fixed points, the conformal group acts merely by rescaling
parameters, so the qualitative behavior of the flow in the
neighborhood of the fixed points \emph{is} conformally invariant.  It
is enough to study the Y-M flow near a slice of the conformal 
equivalence classes, consisting of a
representative in each conformal equivalence class of twisted pairs.

\subsection{The Y-M flow near the twisted pairs}

Instantons and anti-instantons are individually stable under the Y-M
flow, so the Y-M flow very near the twisted pairs reduces to a flow in
a \emph{slow manifold} parametrized by an asymptotically small 
instanton and an asymptotically
small anti-instanton.

We represent the conformal equivalence classes by
puting the instanton at the south pole and 
the anti-instanton at the north pole,
by making their sizes equal, $\rho_{+}=\rho_{-}=\rho\approx 0$,
and by diagonalizing the twist,
\eq
\gtwist =
\begin{pmatrix}
e^{ \frac12 i \sigma}& 0 \\
0 & e^{-\frac12 i \sigma}
\end{pmatrix}
\,,
\qquad \sigma\in [0, 2\pi]
\,.
\en
The slow manifold is represented by a two dimensional space
of connections parametrized by $\rho\approx 0$ and by $\sigma$.
The twisted pairs are at $\rho=0$.

\begin{figure}[ht]
\centering
\vskip2ex
\includegraphics[width=3.00in]{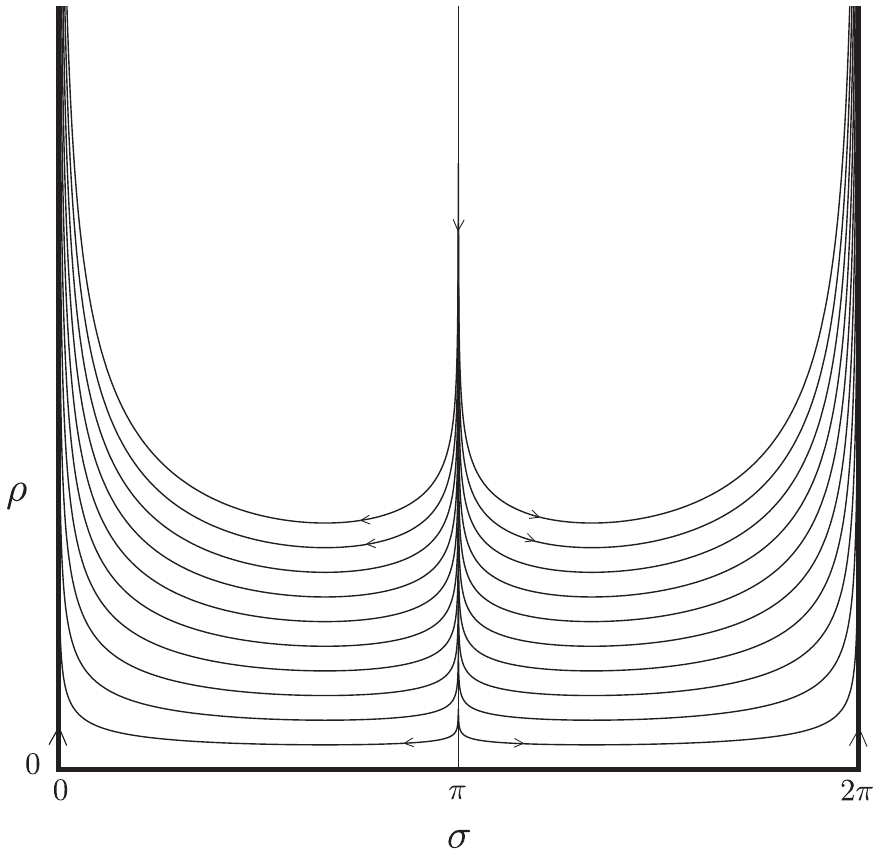}
\caption{The flow on the slow manifold.
The twisted pairs are represented by the horizontal axis.
The vertical axes at $\sigma=0$ and $\sigma=2\pi$ are identified 
under a nontrivial gauge transformation.}
\label{fig:flow_lines_1}
\end{figure}

We calculate the Y-M flow equations to leading order in $\rho$,
\eq
\frac{d\rho}{dt} =  \rho^{3}
(1+2\cos\sigma) +O(\rho^{5})\,,\qquad
\frac{d\sigma}{dt} = - 8\rho^{2}\sin\sigma+O(\rho^{4})\,.
\en
The flow lines follow the curves
\eq
\rho^{8} (1-\cos\sigma) \sin \sigma = C
\,.
\en
as pictured in Figure~\ref{fig:flow_lines_1}.
The twisted pairs lie on the horizontal axis, $\rho=0$.
The vertical axes at $\sigma=0$ and at $\sigma=2\pi$ are
identified by a nontrivial gauge transformation
$\phi:S^{4}\rightarrow SU(2)$.
The \emph{maximally} twisted pairs are those with $\tr(\gtwist) =0$,
represented in Figure~\ref{fig:flow_lines_1} by the point $\sigma=\pi$ on 
the horizontal axis.
The attracting (stable) manifold of the maximally twisted pairs
is represented in Figure~\ref{fig:flow_lines_1} by 
the vertical line $\sigma=\pi$.
It has codimension 1 in $\mathcal{A}/\mathcal{G}$.
If a loop of connections intersects it,
the Y-M flow drives
the intersection point to a maximally twisted pair.
An infinitesimal neighborhood within the 
loop around the intersection point
is driven to the unstable manifold of that maximally twisted pair,
which is represented in Figure~\ref{fig:flow_lines_1} by the three 
axes: the horizontal axis $\rho=0$ and the vertical axes $\sigma = 0$ and $\sigma=2\pi$.
The unstable manifold of the maximally twisted pair is one 
dimensional.
One outgoing branch consists of the segment of the 
horizontal axis going from $\sigma = \pi$ to 
$\sigma=0$, followed by the outgoing trajectory along the vertical 
axis at $\sigma=0$.
The other branch consists of the segment of the horizontal axis going from $\sigma = \pi$ to 
$\sigma=2 \pi$, followed by the outgoing trajectory along the vertical 
axis at $\sigma=2 \pi$.
The unstable manifold of the maximally twisted pair has to be
constructed asymptotically in the limit $\rho\rightarrow 0$.
In the limiting unstable manifold,
the first segment of each branch --- on the 
horizontal axis in Figure~\ref{fig:flow_lines_1} ---
is in fact a line of fixed points.
Effectively, the maximally twisted pairs are fixed points of index 1.

The stable loops are indexed by the maximally twisted pairs.
The stable loop passing through a general maximally twisted pair 
$\tr (\gtwist) =0$
is obtained from the stable loop in Figure~\ref{fig:flow_lines_1} by 
the inverting the conjugation that diagonalized $\gtwist$.
The segment of the stable loop lying within the twisted pairs
consists of the shortest geodesic loop in $SU(2)/\{\pm\mathbf{1}\}$ 
that starts and ends at $\pm1$ and that passes through 
$\pm\gtwist$.
This segment of fixed points is preceeded and followed by the 
outgoing trajectory leaving from the twisted pair at $\sigma=0,2\pi$,
with twist $\pm1$,  
the \emph{untwisted} pair.

\begin{figure}[ht]
\centering
\includegraphics[width=3.00in]{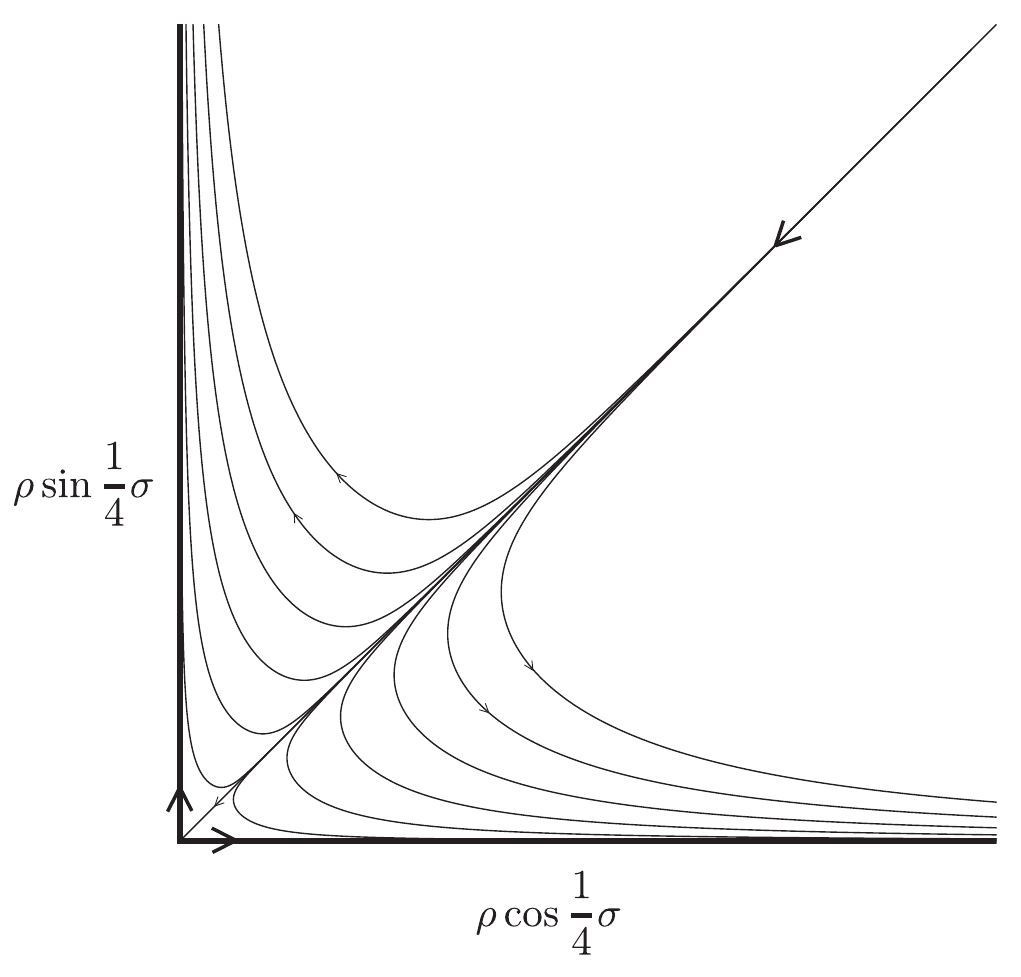}
\caption{Conical geometry of the slow manifold.  
The twisted pairs are represented by the vertex.
The two axes are identified under a nontrivial gauge transformation.}
\label{fig:flow_lines_2}
\end{figure}

The twisted pairs look more literally like fixed points of index 1
when pictured in the riemannian geometry of the space of gauge fields.
To leading order in $\rho$,
the metric on the slow manifold in ${\mathcal{A}/\mathcal{G}}$ is
\eq
(ds^{2})^{\slow}_{\mathcal{A}/\mathcal{G}}
= 
16 (d\rho)^{2}
+
\rho^{2}  (d\sigma )^{2} 
\,.
\en
Geometrically,
the space of connections is a cone,
as pictured in Figure~\ref{fig:flow_lines_2}.
The loop of twisted pairs --- the horizontal axis in 
Figure~\ref{fig:flow_lines_1} ---
collapses to the vertex of the cone.
The vertex of the cone looks like an index 1 fixed point lying on the 
boundary of ${\mathcal{A}/\mathcal{G}}$.

The outgoing trajectories at $\sigma = 0$ and $\sigma=2\pi$, i.e., at
$\gtwist=+1$ and $\gtwist=-1$, are gauge equivalent.  The orientations
of the small instanton and the small anti-instanton are lined up
within $SU(2)$.  Under the Y-M flow, the instanton and anti-instanton
grow larger, presumably merging together and annihilating, flowing
eventually to the flat connection.  It remains to be proved that this
does in fact happen in general, that the outgoing trajectory from the
untwisted pair, $\gtwist = \pm 1$, ends at the flat connection, and
not at some other fixed point with $S_{\YM}>0$.  Here, we prove this
only for the special case where the instanton and anti-instanton of
the untwisted pair are located at opposite poles in $S^{4}$.  Then the
entire outgoing trajectory is $SO(4)$-invariant, and we can show that
there are no $SO(4)$-invariant fixed points besides the flat
connection and the untwisted pair itself.

\subsubsection{Global stability}
We have only established the local stability of the loop of twisted
pairs.  We can argue for global stability based on a theorem of Taubes
\cite{Taubes1983} which states that, for connections in the trivial
$SU(2)$ bundle over $S^{4}$, the hessian of $S_{\YM}$ must have
at least 2 negative eigenvalues at any smooth solution of the
Yang-Mills equation (fixed point of the Y-M flow).  It follows that there are no smooth fixed points
with unstable manifolds of dimension 0 or 1.  
Any index 1 fixed point must be completely singular, so it must be a 
twisted pair or must have $S_{\YM}\ge 4$.
Therefore any loop of
gauge fields with $S_{\YM}<4$ must flow to the stable loop of twisted
pairs.  
This argument does not work for stable 2-spheres, since Taubes'
theorem allows smooth fixed points of Morse index 2.

In retrospect, Taubes' theorem and/or the paper of Uhlenbeck, Sibner,
and Sibner could have made our numerical explorations unnecessary,
leading directly to consideration of the loop of twisted pairs.

\section{Preliminaries}

We will be doing elementary, explicit calculations with
the $U(2)$-invariant connections over $S^{4}$.
In this section, we establish notation and collect some basic formulas.
More detail is given in Appendix~\ref{app:InvConnections}.

\subsection{Parametrization of \texorpdfstring{$S^{3}$}{S3}}
We realize $S^{3}$ as the unit sphere in $\Complexes^{2}$,
parametrized by the unit vectors
\eq
\zb=
\begin{pmatrix}
z_{1}\\
z_{2}
\end{pmatrix}
\,,
\qquad
\zb^{\dagger}\zb = \bar z_{1}z_{1}+ \bar z_{2} z_{2} =1
\,.
\en
We write the complementary projection matrices
\eq
P(\zb) = \zb \zb^{\dagger}\,, \qquad Q(\zb) =1 - P(\zb)
\,.
\en
The volume form on $S^{3}$ is
\eq
\mathrm{dvol}_{S^{3}} = - \frac12 (\zb^{\dagger}d\zb) 
(d\zb^{\dagger}d\zb)\,,
\qquad
\int_{S^{3}} \mathrm{dvol}_{S^{3}}= 2\pi^{2}
\,.
\en

\subsection{Parametrization of \texorpdfstring{$S^{4}$}{S4}}
We realize $S^{4}$ as the unit sphere in 
$\Reals\oplus\Complexes^{2}$, parametrized by the unit vectors
\eq
\vec y = (y_{0},\yb) \qquad y_{0}^{2}+\yb^{\dagger}\yb = 1
\,.
\en
In polar coordinates,
\eq
\vec y = (\cos \theta, \zb \sin \theta)\,,\qquad \zb\in S^{3}
\,.
\en
Most often, we use coordinates $(x,\zb)$ where
\eq
x = \ln \tan \left ( \frac{\theta}2 \right )\,,\qquad -\infty\le x \le 
\infty
\,.
\en
The north pole of $S^{4}$ is at $\theta=0$, $x=-\infty$.
The south pole is at $\theta=\pi$, $x=\infty$.

\subsection{Round metric on \texorpdfstring{$S^{4}$}{S4}}
The round metric
on $S^{4}$ is
\eq
(ds)_{S^{4}}^{2}= 
\Round(x) \left [ (dx)^{2} + d\zb^{\dagger}d\zb \right ]
\,,\qquad
\Round(x) = (\cosh x)^{-2}\,.
\label{eq:roundmetric}
\en

\subsection{Action of \texorpdfstring{$U(2)$ on $S^{3}$ and 
$S^{4}$}{U(2) on S3 and S4}}
$U\in U(2)$ acts on $S^{3}\subset \Complexes^{2}$ by
$\zb\mapsto U\zb$
and acts on $S^{4}\subset \Reals\oplus\Complexes^{2}$ by
$(y_{0},\yb)\mapsto (y_{0},U\yb)$
or
$
(x,\zb) \mapsto (x, U\zb)
$.

\subsection{\texorpdfstring{$S^{3}$ identified with $SU(2)$}{S3 
identified with SU(2)}}
$S^{3}$ is identified with $SU(2)$ by
\eq
g(\zb)  \eb  = \zb
\,,
\qquad
\eb =
\begin{pmatrix}
1\\
0
\end{pmatrix}
\,,
\qquad
g(\zb) = \begin{pmatrix}
z_{1} & -\bar z_{2} \\
z_{2} & \bar z_{1}
\end{pmatrix}
\,.
\en
The action of $U(2)$ on $S^{3}$ becomes
\eq
g(U\zb) = U g(\zb) 
\begin{pmatrix}
1 & 0 \\
0 & (\det U)^{-1}
\end{pmatrix}
\,.
\en
The rotation group $SO(4)$ is identified with $SU(2)\times 
SU(2)/\Integers_{2}$ via
\eq
g(O (\zb)) = g_{L}g(\zb) g_{R}^{-1}
\,,
\qquad O\in SO(4)
\,, \qquad (g_{L},g_{R})\in SU(2)\times SU(2)
\,.
\en

\subsection{\texorpdfstring{$U(2)$-invariant $su(2)$-valued 1-forms 
on $S^{3}$}{U(2)-invariant su(2)-valued 1-forms on S3}}

The general $U(2)$-invariant $su(2)$-valued 1-form on $S^{3}$ is
\eq
f \eta - \bar f \eta^{\dagger} + f_{3}\eta_{3}\,,
\qquad f_{3}\in\Reals\,, f\in\Complexes
\en
where
\begin{alignat}{2}
\eta &= -PdP
   &\quad&= -( \zb^{\dagger}d \zb)\zb  \zb^{\dagger}  -\zb d \zb^{\dagger} \\
\eta^{\dagger} &= -dP P &&= (\zb^{\dagger} d\zb)  \zb \zb^{\dagger}-d \zb \zb^{\dagger}
\\
\eta_{3} &= (\zb^{\dagger}d \zb)(P-Q) 
        &&= (\zb^{\dagger}d \zb) (2\zb\zb^{\dagger}-1)
\end{alignat}
is a natural basis that diagonalizes the $U(1)$ generated by $i(P-Q)$,
\eq
[P-Q,\, \eta] = 2 \eta
\,,
\qquad
[P-Q,\, \eta^{\dagger}] = -2 \eta^{\dagger}
\,,
\qquad
[P-Q,\, \eta_{3}] = 0
\,.
\en

\subsection{The Maurer-Cartan form \texorpdfstring{$\omega$ on 
$SU(2)$}{omega on SU(2)}}
The Maurer-Cartan form $\omega$ on $SU(2)$ is 
\eq
\omega  =g d(g^{-1}) = 
-\eta+\eta^{\dagger} -\eta_{3} 
\,,
\en
satisfying
\eq
d\omega +\omega^{2} = 0
\,,\qquad
-\frac16 \omega  ^{3}  = \mathrm{dvol}_{S^{3}}\,1
\,.
\en

\subsection{\texorpdfstring{$U(2)$-invariant connections on 
$S^{4}$}{U(2)-invariant connections on S4}}

A connection in the trivial $SU(2)$ bundle over $S^{4}$
is described by its covariant derivative
\eq
D = d +A(x,\zb)
\en
where $A(x,\zb)$ is an $su(2)$-valued 1-form on $S^{4}$.
Regularity at the poles requires
\eq
A(\pm \infty,\zb)=0
\,.
\en
Invariance under $U(2)$ is the condition
\eq
A(x,U\zb) = U A(x,\zb) U^{-1} \qquad U\in U(2)
\,.
\en
Define
\eq
d_{\omega} =   g d g^{-1} = d + \omega
\en
which satisfies
\eq
d_{\omega}^{2}=0
\,,\qquad
[d_{\omega},\, P-Q]= 0
\,.
\label{eq:U1covariant}
\en
The $U(1)$ generated by $i(P-Q)$ thus leaves $d_{\omega}$ invariant.
It is convenient to write
the $U(2)$-invariant connections in the $U(1)$-covariant form
\eq
D = d+A = d_{\omega} +\Delta A
\en
\eq
\Delta A =  
A_{0}(x) dx \,i(P-Q)
+f(x) \eta -\bar f(x) \eta^{\dagger}+ f_{3}(x) \eta_{3}\
\en
with
\eq
A_{0}(x)\,,f_{3}(x)\in\Reals\,,\; f(x)\in \Complexes
\,.
\en
The $U(2)$-invariant gauge transformations act by
\eq
D\mapsto e^{i\varphi (x)(P-Q)}De^{-i\varphi (x)(P-Q)}
=d_{\omega} - (\partial_{x}\varphi  dx)  \,i (P-Q) + e^{i\varphi (x)(P-Q)}\Delta Ae^{-i\varphi (x)(P-Q)}
\en
\eq
e^{i\varphi(x)(P-Q)}\,\Delta A \,e^{-i\varphi(x)(P-Q)}
=
A_{0}(x) dx \,i(P-Q)
+
e^{2i\varphi(x)}f(x)\eta -  e^{-2i\varphi(x)}\bar f(x)\eta^{\dagger}+f_{3}(x) \eta_{3}
\,.
\en
Connections with $A_{0}(x)=0$
are said to be in $A_{0}=0$ gauge.
Any connection can be brought to $A_{0}=0$ gauge
by the gauge transformation with $\partial_{x}\varphi = A_{0}(x)$,
perhaps at the cost of introducing singularities at the poles $x=\pm\infty$.

\subsection{The Yang-Mills action}

The curvature 2-form is
\eq
F = D^{2} = dA +A^{2}
\,.
\en
The Yang-Mills action is
\eq
S_{YM} = \frac1{2 \pi^{2}} \int_{S^{4}} \frac14 \tr(-F{*}F)
\en
where $*$ is the Hodge operator
taking $k$-forms to $(4-k)$-forms
and satisfying ${*}^{2} = (-1)^{k}$.

For $U(2)$-invariant connections, the integrand
is an invariant volume form
\eq
\frac14 \tr(-F{*}F) =   dx \;L_{\YM}(x) \,  \mathrm{dvol}_{S^{3}}
\en
and the Y-M action is
\eq
S_{YM} = \int_{-\infty}^{\infty} dx \,L_{\YM}(x)
\,.
\en

\subsection{Hodge duality}
The (anti-)self-dual curvature is
\eq
F_{\pm} = \frac12 \left (F \pm {*}F \right )
\en
The Y-M action splits into contributions of the two 
chiralities,
\eq
S_{\YM} =  S_{+}+S_{-}
\qquad
S_{\pm} = \frac1{2 \pi^{2}} \int \frac14 \tr(-F_{\pm}{*}F_{\pm})
\,.
\en
The integer instanton number is
\eq
S_{\mathit{top}} = S_{+}-S_{-}
\,.
\en
The instanton number vanishes for connections in the trivial bundle over $S^{4}$.

\subsection{Hodge duality for \texorpdfstring{$U(2)$}{U(2)}-invariant connections}
For $U(2)$-invariant connections,
\eq
L_{\YM}(x) = L_{+}(x) + L_{-}(x)
\en
\eq
\frac14 \tr(-F_{\pm}{*}F_{\pm}) = dx  \, L_{\pm}(x) \; \mathrm{dvol}_{S^{3}}
\en
\eq
S_{\pm} = \int_{-\infty}^{\infty} dx \,L_{\pm}(x)
\en
\eq
S_{\YM} = \int_{-\infty}^{\infty} dx \, \left [ 
L_{+}(x)+L_{-}(x)\right ]\,,
\qquad
S_{\mathit{top}}  = \int_{-\infty}^{\infty} dx \, \left [ 
L_{+}(x)-L_{-}(x)\right ]
\,.
\en
For a $U(2)$-invariant connection in $A_{0}=0$ gauge,
\eq
D= d_{\omega}+\Delta A\,,
\qquad \Delta A = f(x) \eta - \bar f(x)\eta^{\dagger}+ f_{3}(x) \eta_{3}
\,,
\en
\eq
L_{\pm} =
\frac14
\left [\partial_{x}f_{3}\pm 2\left (f_{3} -|f|^{2}\right )\right ]^{2} 
+\frac12 \left |\partial_{x}f\pm 2\left (1 -f_{3}\right )f)\right |^{2}
\label{eq:Lpm}
\en
\eq
L_{\YM} =
\frac12 (\partial_{x}f_{3})^{2}
+ \left |\partial_{x}f\right |^{2}
+ 2\left (f_{3} -|f|^{2}\right )^{2} 
+ 4 \left(1 -f_{3}\right )^{2} \left | f\right |^{2}
\label{eq:LYM}
\en
\eq
L_{+}-L_{-} = \partial_{x}
\left (
f_{3}^{2} + 2\left | f \right |^{2} - 2f_{3}\left | f \right |^{2}
\right )
\,,\qquad
S_{\mathit{top}} =  
\left . \left (f_{3}^{2} + 2\left | f \right |^{2} - 2f_{3}\left | f \right |^{2}\right ) 
\right |_{x=-\infty}^{x=\infty}
\,.
\label{eq:Stop}
\en

\subsection{Connections 
\texorpdfstring{$d_{\omega}-f(x)\omega$}{d sub omega -f(x)}}

The $U(2)$-invariant connections
that are invariant under the full
$O(4)=SU(2)\times SU(2)/\{\pm 1\}$ are of the form
\eq
D = d_{\omega}-f(x)\omega  =d_{\omega} + f(x) (\eta-\eta^{\dagger}+\eta_{3})
\,.
\en
Substituting in equations~\ref{eq:Lpm} and \ref{eq:Stop},
\eq
L_{\pm}(x) =  \frac34 \left [ \partial_{x}f \pm 2f(1-f)\right]^{2}
\,.
\label{eq:fomegaLpm}
\en
\eq
S_{\mathit{top}} =  \left . \left (3f^{2}-2f^{3}\right ) \right |_{x=-\infty}^{x=\infty}
\,.
\en

\subsection{The basic instanton}
\label{sect:instanton}
The BPST instanton \cite{Belavin1975} is the self-dual connection,
$F=F_{+}$, of instanton number $1$.
For us,
the basic instanton is the self-dual $U(2)$-invariant connection of the form
\eq
\DI = d_{\omega} - \fI(x) \omega  
\,.
\en
The self-duality equation $F_{-}=0$ becomes the ordinary differential equation
\eq
\partial_{x} \fI =  2 \fI(1-\fI)
\,.
\label{eq:selfduality}
\en
The general solution is
\eq
\fI(x)
=\frac1{1+e^{-2(x-\xI)}}
=\frac1{1+\rI^{-2}e^{-2x}}
\,.
\label{eq:basicinstanton}
\en
The parameter $\rI = e^{-\xI}$ is the \emph{size} of the instanton.
The Y-M action density is
\eq
L_{\YM}(x) = L_{+}(x) 
 = \frac3{4 \cosh^{4}(x-\xI)}
\en
and the Y-M action is
\eq
S_{\YM}= S_{+}=S_{\mathit{top}} =1
\,.
\en
The basic instanton is regular at the south pole ($x=\infty$),
because $\DI\rightarrow d$ there.
Near the north pole,
$\DI \rightarrow d_{\omega}=g d g^{-1} $,
so the instanton lives in the nontrivial bundle
formed from trivial bundles
over the two hemispheres, patched together at the equator using the index $+1$ map
$\zb\mapsto g(\zb)$
from $S^{3}$ to $SU(2)$.
When the instanton size goes to zero,
when $\xI\rightarrow \infty$,
the action density becomes a delta-function concentrated at the south 
pole, at $x=\infty$.
We say that the basic instanton is \emph{located} at the south pole.

\subsection{The basic anti-instanton}
The anti-instanton is the anti-self-dual connection
of instanton number $-1$.
Our basic anti-instanton is
\eq
\DIbar = d_{\omega} - \fIbar(x) \omega  
\en
\eq
\partial_{x} \fIbar = - 2 \fIbar(1-\fIbar)
\en
\eq
\fIbar(x)
=\frac1{1+e^{2(x-\xIbar)}}
=\frac1{1+\rIbar^{-2}e^{2x}}
\,.
\label{eq:basicantiinstanton}
\en
The parameter $\rIbar = e^{\xIbar}$ is the size of the 
anti-instanton.
The basic instanton and anti-instanton are related
by the orientation reversing map $x\mapsto -x$,
$\rI \leftrightarrow \rIbar$.
The Y-M action is
\eq
L_{\YM}(x) = L_{-}(x) 
 = \frac3{4 \cosh^{4}(x-\xIbar)}
\,, \qquad
S_{\YM}= S_{-}= - S_{\mathit{top}} =1
\,.
\en
The basic anti-instanton is regular at the south pole ($x=-\infty$),
because $\DI\rightarrow d$.
Near the north pole,
$\DI \rightarrow g d g^{-1} $,
so the anti-instanton lives in the nontrivial bundle formed from
trivial bundles on the hemispheres patched together at the equator
using the index $-1$ map $\zb\mapsto g(\zb)^{-1}$.
When the anti-instanton size goes to zero,
when $\xIbar\rightarrow - \infty$,
the action density becomes a delta-function concentrated at the north
pole, at $x=-\infty$.
The basic anti-instanton is \emph{located} at the north pole.

\subsection{Twisted (anti-)instantons}

The $U(2)$-invariant twisted instanton
with twist angle $\sigma_{+}\in[0,2\pi]$ is
\eq
e^{ i \alpha_{+}(x) (P-Q)}\DI e^{-i \alpha_{+}(x)  (P-Q)}
\en
where $\alpha(x)$ satisfies
\eq
\alpha_{+}(-\infty)=\frac12\sigma_{+}
\,,\qquad
\alpha_{+}(\infty)=0
\,.
\en
and the $U(2)$-invariant twisted (anti-)instanton 
with twist angle $\sigma_{-}$
is
\eq
e^{ i \alpha_{-}(x) (P-Q)}\DIbar e^{-i \alpha_{-}(x)  (P-Q)}
\en
\eq
\alpha_{-}(-\infty)=0\,,\quad
\alpha_{-}(\infty)=\frac12\sigma_{-}
\,.
\en
These are of course merely gauge transforms of the basic (anti-)instanton.
The twist angle $\sigma_{\pm}$ is gauge invariant only if we 
restrict our notion of gauge equivalence to the group of pointed 
gauge transformations, that act as the identity at a base-point in 
$S^{4}$,
here the gauge transformations $e^{i\varphi(x)(P-Q)}$ with 
$\varphi(\pm \infty)=0$.

The (anti-)instanton twisted by a general element $g_{\tw}\in 
SU(2)/\{\pm1\}$ is
\eq
D_{\pm}(g_{\tw}) = \phi(g_{\tw}) D_{\pm} \phi(g_{\tw})^{-1}
\en
where
\eq
\phi(g_{\tw})(x,\zb) \rightarrow g(\zb) g_{\tw}g(\zb)^{-1}\,, \qquad x\rightarrow 
\pm \infty\,.
\en
The $U(2)$-invariant twisted (anti-)instanton corresponds to
\eq
g_{\tw} = 
\begin{pmatrix}
e^{\frac12 i\sigma_{\pm}} & 0\\
0 & e^{-\frac12 i\sigma_{\pm}}
\end{pmatrix}
\,, \qquad
g(\zb) g_{\tw}g(\zb)^{-1} = e^{i\frac12 \sigma_{\pm}(P-Q)}
\en
Rotations $O=(g_{L},g_{R})$ in $SO(4)=SU(2)\times SU(2)/\Integers_{2}$
transform the twisted (anti-)instanton by
\eq
D_{\pm}(g_{\tw})(x,O\zb)
=g_{L} 
D_{\pm}(g_{R}^{-1}g_{\tw}g_{R})(x,\zb)
g_{L}^{-1}
\,.
\en
The $g_{R}$ act by conjugation 
on the twist $g_{\tw}$,
so every twisted instanton can be taken to a $U(2)$-invariant one by 
a rotation in $SO(4)$.
The $g_{L}$ are symmetries,
as are the $g_{R}$ that commute with $g_{tw}$.

\subsection{Nontrivial \texorpdfstring{$U(2)$-invariant maps 
$\phi_{h}:S^{4}\rightarrow SU(2)$}{U(2)-invariant maps phi sub 
h:S4->SU(2)}}
\label{sect:nontrivialmap}

The Hopf fibration \cite{Hopf1931} is the map $h:S^{3}\rightarrow 
S^{2}\subset \Reals \oplus \Complexes$,
\eq
h(\zb)=(|z_{1}|^{2}-|z_{2}|^{2},2 \bar z_{1} z_{2})
\,.
\en
The nontrivial element in $\pi_{4}SU(2)=\Integers_{2}$ is represented 
by the suspension, $S h : S^{4}\rightarrow S^{3}=SU(2)$, of the Hopf fibration
\cite{Freudenthal1937}.
In particular,
the $U(2)$-invariant maps $\phi_{h}:S^{4}\mapsto SU(2)$
of the form
\eq
\phi_{h} (x,\zb) = e^{ i\varphi_{h}(x)(P-Q)}
=
g(\zb)
\begin{pmatrix}
e^{ i \varphi_{h}(x)}& 0 \\
0 & e^{-i\varphi_{h}(x)}
\end{pmatrix}
g(\zb)^{-1}
\en
with
\eq
\varphi_{h}(-\infty)=\pi\,,\qquad\varphi_{h}(\infty)=0
\en
represent
the nontrivial element in $\pi_{4}SU(2)=\Integers_{2}$
\cite{Puettmann2003}.
Explictly,
\eq
\phi_{h} (x,\zb)
=
\cos   \varphi_{h}(x)
\begin{pmatrix}
1&0 \\
0&1
\end{pmatrix}
+ i \sin \varphi_{h}(x)
\begin{pmatrix}
|z_{1}|^{2}-|z_{2}|^{2}& 2z_{1} \bar z_{2} \\
2\bar z_{1} z_{2} & - |z_{1}|^{2}+|z_{2}|^{2}
\end{pmatrix}
\,.
\en

\section{Computer calculation}

We start with a numerical calculation, looking for a clue to the 
long term behavior of the Y-M flow on the nontrivial loops.
We pick a particular nontrivial loop and
try to discover what it flows to.
The calculation is sketched here.
Details are given in Appendix \ref{app:DetailsComputer}.

\subsection{Rationale}
We use the homogeneous space $SU(3)/SU(2)=S^{5}$
to construct a nontrivial loop of connections
on $S^{4}$.
The
$SU(2)$ bundles over $S^{5}$
are classified topologically by $\pi_{4}SU(2)$,
since they are made by gluing two trivial bundles along the equator in 
$S^{5}$ by a map from the equator, $S^{4}$, to $SU(2)$.
The bundle $SU(2)\rightarrow SU(3)\rightarrow S^{5}$ represents the nontrivial
element in $\pi_{4}SU(2)$ \cite{Eckmann1942thesis}.

There is a canonical invariant connection $D_{\mathit{inv}}$ in
$SU(2)\rightarrow SU(3)\rightarrow S^{5}$.  We pull back
$D_{\mathit{inv}}$ along a certain map $[-1,1]\times S^{4}\rightarrow
S^{5}$
to obtain a one parameter family $D(s)$ of connections over $S^{4}$.
The map is chosen so that the endpoint connections $D(\pm 1)$ are 
both flat, so $s\mapsto D(s)$ forms a closed loop in $\mathcal{A}/\mathcal{G}$.
The nontriviality of the loop is verified explicitly
in Appendix \ref{app:DetailsComputer}.
The map $[-1,1]\times S^{4}\rightarrow S^{5}$ preserves
a $U(2)$ subgroup of the symmetries of $D_{\mathit{inv}}$,
so each $D(s)$ is a $U(2)$-invariant connection over $S^{4}$.

We want to see what happens to this particular nontrivial loop 
under the Y-M flow.
The Y-M flow preserves symmetry, so loop will remain
within the $U(2)$-invariant connections on $S^{4}$.
Two additional discrete symmetries of $D_{\mathit{inv}}$ are likewise 
preserved by our construction,
one taking each $D(s)$ to itself, the other taking $D(s)$ to $D(-s)$.
The connection $D(0)$ at the midpoint of the loop thus
has an extra discrete symmetry.
Again, the Y-M flow preserves these discrete symmetries.

In order to simplify the computational problem, we assume a
plausible-seeming scenario.  We assume that the midpoint $D(0)$ of the
initial loop will flow to a fixed point in $\mathcal{A}/\mathcal{G}$
of Morse index 1, while the rest of the loop will flow to the one
dimensional unstable manifold of the fixed point.  The discrete
symmetry that takes $D(s)$ to $D(-s)$ will exchange the two outgoing
branches of the unstable manifold.  Now we do not need to run the Y-M
flow on the entire loop, but only on the single connection $D(0)$.  We
simplify still further by assuming that $D(0)$ will flow to a
connection that minimizes $S_{\YM}$ among all the $U(2)$-invariant
conections with the same two discrete symmetries as $D(0)$.  Assuming
this scenario, there is no need to run the Y-M flow at all.  We need
only minimize $S_{\YM}$ on this class of invariant connections, which
is quite easy to do numerically.  There is a fairly extensive
literature on minimizing $S_{YM}$ over connections with specific
prescribed
symmetries~\cite{Urakawa1988,Sibner1989,Bor1990,Sadun1990,Sadun1991,
Sadun1992,Sadun92sta,Bor1992,Parker1992a,Parker1992,Sadun1994,Schlatter1998},
but seemingly not the $U(2)\times \Integers_{2}^{2}$ symmetry of
interest here.  The closest seems to be \cite{Parker1992a}, which
studies $U(2)$ invariant connections on non-round $S^{4}$ and finds a
solution of the Yang-Mills equation which degenerates, in the round
limit, to a zero-size instanton/anti-instanton pair.

The purpose of the numerical calculation is only heuristic.  The
simplifying assumptions are justified by the clue that emerges from 
the computation.
It could have turned out otherwise.  In particular, it could have 
turned out that
an initial loop with such special symmetries
would not detect generic properties of the Y-M flow acting on loops.

\subsection{Numerical results}
\begin{figure}[t]
\label{fig:Lplusminus}
\centering
\includegraphics{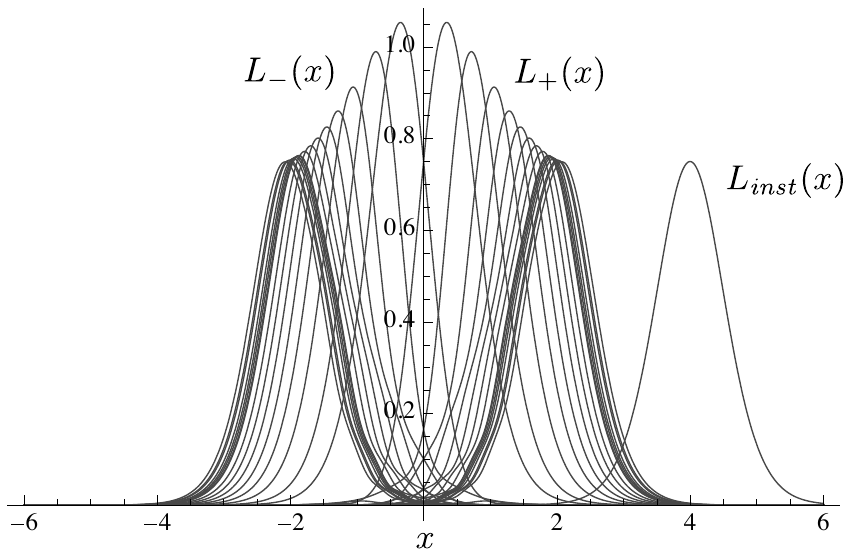}
\caption{Plots of $L_{\pm}(x)$ for the connections numerically 
minimizing $S_{\YM}$ for $N\le 30$.
The curves move away from 
the origin as $N$ increases.  For comparison, the rightmost curve is $L_{+}(x)$ for the instanton
of size $\rI= e^{-x_{+}}$, $x_{+}=4$.}
\end{figure}

The midpoint connection $D(0)=d_{\omega}+\Delta A(0)$ of the initial loop 
is calculated in Appendix \ref{app:DetailsComputer},
\eq
\Delta A(0) =  \cos \theta \,(\eta -\eta^{\dagger})
+ \left ( 1-\frac12 \sin^{2}\theta \right ) \eta_{3}
\,.
\en
It is convenient to use the polar angle $\theta$ here, 
rather than
$x = \ln \tan \frac12 \theta$
which we use elsewhere.
The two discrete symmetries of $\Delta A(0)$ are derived in Appendix 
\ref{app:DetailsComputer}.
The general $U(2)$-invariant connection with these two additional discrete symmetries
has the form
\eq
\Delta A =  f(\theta) (\eta- \eta^{\dagger}) +f_{3}(\theta) \eta_{3}
\en
with
\eq
f=\bar f \qquad f(\pi-\theta) = - f(\theta) \qquad f_{3}(\pi-\theta) = f_{3}(\theta)
\,.
\en
Regularity at the poles requires the boundary conditions
\eq
f =  f_{3} = 1 \;\text{ at } \theta =0
\,.
\en
We change variables again, to
\eq
t=\cos \theta 
\en
The Y-M action is given by equation~\ref{eq:LYM},
\eq
S_{\YM} = \int_{-1}^{1}dt\,(1-t^{2})^{-1} L_{\YM}\,
\en
\eq
L_{\YM}= \frac12 (1-t^{2})^{2}( \partial_{t}f_{3})^{2} + 2(f_{3}-f^{2})^{2}
+ (1-t^{2})^{2} (\partial_{t}f)^{2} + 4 (1-f_{3})^{2}f^{2}
\qquad
\,.
\en
The initial connection $D(0)$ has
\eq
L_{YM}= \frac32 (1-t^{2})(1-t^{4})\,,
\qquad
S_{\YM} = 2.4\,.
\en
To minimize $S_{\YM}$ numerically,
we use a finite mode approximation \cite{Sadun92sta}.
We write $f_{3}$ and $f$ as
polynomials in $t$
obeying the symmetry and boundary conditions,
\eq
f_{3}= 1 + \sum_{n=1}^{N/2} (t^{2n}-1) a_{2n}
\qquad
f = t + \sum_{n=1}^{N/2} (t^{2n+1}-t) a_{2n-1}
\en
where $N$ is an even number.
The $N$ real variables $a_{k}$ parametrize an 
affine subspace of $\mathcal{A}$ of dimension $N$.
We are approximating $\mathcal{A}$ by an increasing family of finite 
dimensional affine subspaces.
On each subspace,
$S_{\YM}$ evaluates to a quartic 
polynomial in the $a_{k}$,
which is minimized numerically using
mathematical software
such as Sage\cite{sage}.
Typical results are shown in 
Table~\ref{table:SYM}.
\begin{table}[t]
\centering
\begin{tabular}{clrlrl}
$N$ & $\min(S_{\mathit{YM}})$ \\[0.5ex]
2 &   2.15627 &12 &   2.00723  &  22 &   2.00286 \\
4 &   2.06011 &14 &   2.00504  & 24 &   2.00251\\
6 &   2.03019 &16 &   2.00368 & 26 &   2.00202\\
8 &   2.01735 &18 &   2.00346 &28 &   2.00186\\
10 &   2.01086 & 20 &   2.00313 &30 &   2.00147
\end{tabular}
\caption{Results of numerical minimization of $S_{\YM}$
on affine subspaces of $\mathcal{A}$ of dimension $N$.}
\label{table:SYM}
\end{table}
The numerical results suggest that there is a global minimum
with $S_{\YM}=2$.
The possibility of an integer global minimum
motivates examining the self-dual and anti-self-dual action densities
$L_{\pm}(x)$ of the approximate minima obtained from the computer 
calculations.
Figure~\ref{fig:Lplusminus} plots the evolution
of $L_{\pm}(x)$ as $N$ increases.
It looks like the global minimum is
a connection that consists of a zero-size instanton at the south pole
and a zero-size anti-instanton at the north pole,
and is otherwise flat.
Closer inspection suggests that the minimum is attained at the 
connection given by
\ateq{2}{
f_{3}&=f = \fIbar = \frac1{1+e^{2(x-\xIbar)}}\,, &\qquad& x<0 \\
f_{3}&=- f = \fI = \frac1{1+e^{-2(x-\xI)}}\,, &\qquad& x>0
}
in the limit $\xI\rightarrow \infty$, $\xIbar\rightarrow -\infty$.
This is the zero-size basic anti-instanton at the north pole combined 
with a twisted zero-size instanton at the south pole,
twisted by $\pi$.

\section{Twisted pairs}

Motivated by the numerical calculation, we investigate the long time
behavior of the Y-M flow near the singular
connections that consist
of a zero-size instanton and a zero-size anti-instanton
patched together on a 3-sphere separating their locations.
We are calling such connections \emph{twisted pairs}.
The general twisted pair is parametrized by the locations of the 
instanton and anti-instanton and by their relative twist $g_{\tw}$.
The $U(2)$-invariant twisted pair has the instanton at the south pole 
and the anti-instanton at the north pole and has diagonal $g_{\tw}$,
so is parametrized by the twist angle $\sigma\in[0,2\pi]$.
We write the $U(2)$-invariant twisted pair explicitly in the next 
section.
The general twisted pair is obtained by making a conformal 
transformation of $S^{4}$.

\subsection{The \texorpdfstring{$U(2)$}{U(2)}-invariant twisted pairs}
A $U(2)$-invariant twisted pair
combines a $U(2)$-invariant twisted instanton of small size $\rI$  at the south 
pole with a $U(2)$-invariant twisted
anti-instanton of the same size $\rIbar$ at the north pole,
in the limit $\rho_{\pm}\rightarrow 0$,
\eq
D_{\tw}(\alpha_{+},\alpha_{-}) =\lim_{\rho_{\pm}\rightarrow 0} \left \{
\begin{matrix}
e^{  i \alpha_{+}(x) (P-Q)} \DIbar e^{- i \alpha_{+}(x)  (P-Q)} \quad 
& x>0 \hfill\\[1ex]
e^{i \alpha_{-}(x) (P-Q)}\DI e^{ - i \alpha_{-}(x) (P-Q)} \quad &x<0
\,.
\end{matrix}
\right .
\en
The functions $\alpha_{\pm}(x)$ should vanish fast enough at the poles to ensure
that the connection is regular there,
\eq
\alpha_{\pm}(x) = O(e^{\mp 2x})\,, \qquad x \rightarrow \pm\infty
\,.
\en
The connection $D_{\tw}(\alpha_{+},\alpha_{-})$ and its curvature are
discontinuous at the equator as long as $\rho_{\pm}>0$, but the
discontinuities disappear in the zero-size limit.

The relative twist is
\eq
\sigma = 2\alpha_{+}(0)-2\alpha_{-}(0)
\,.
\en
The $U(2)$-invariant gauge transformations $e^{i\varphi(x)(P-Q)}$ act by $\alpha_{\pm}(x)\mapsto 
\alpha_{\pm}(x)-\varphi(x)$, so the relative twist $\sigma$ is gauge 
invariant.  Any two $U(2)$-invariant twisted pairs with the same 
twist $\sigma$ are gauge equivalent.

\subsection{The nontrivial loop of twisted pairs}

The $U(2)$-invariant twisted pairs with
twist $\sigma\in [0,2\pi]$ form a nontrivial closed loop in 
$\mathcal{A}/\mathcal{G}$
(for references on the nontriviality of such loops,
reference \cite{Sibner1989} refers to reference \cite{Donaldson1986}).
To see this explicitly,
let $D_{\tw}(2\pi)= D_{\tw}(\alpha_{+},\alpha_{-})$ be any twisted pair with 
$\sigma=2\alpha_{+}(0)-2\alpha_{-}(0)=2\pi$,
and let $D_{\tw}(0)= D_{\tw}(0,0)$, which has
twist $0$.
Then
\eq
D_{\tw}(2\pi)= \phi_{h} D_{\tw}(0) \phi_{h}^{-1}
\en
where $\phi_{h}$ is one of the nontrivial maps $S^{4}\rightarrow 
SU(2)$ described in section~\ref{sect:nontrivialmap} above,
\eq
\phi_{h}(x,\zb) =e^{i\varphi_{h}(x)(P-Q)}\,,\qquad
\varphi_{h}(x) =
\left \{
\begin{matrix}
\alpha_{+}(x)\hfill \quad 
& x>0\hfill \\[1ex]
\pi+\alpha_{-}(x) \quad &x<0\,.
\end{matrix}
\right .
\en

\section{The slow manifold}

A twisted pair is everywhere self-dual or anti-self-dual or flat,
so the twisted pairs are all fixed points under the Y-M flow.
Instantons in isolation are stable under the Y-M flow, as are
anti-instantons.  All perturbations of the instanton transverse to the
space of instantons are driven rapidly to zero under the Y-M flow.
Therefore, the Y-M flow,
acting on a small neighborhood of the twisted pairs,
rapidly compresses the neighborhood down to a space of approximate fixed
points, the \emph{slow manifold}, which is parametrized by an asymptotically
small instanton and an asymptotically small anti-instanton.
The long
time behavior of the Y-M flow near the twisted pairs is
determined by the flow on the slow manifold,
which can be represented as a flow on the parameter space of the 
instanton-anti-instanton pair.
To find the long time behavior,
it will be enough to calculate
the asymptotic expansion of the flow equation on the
slow manifold to leading order in the sizes of the instanton and
anti-instanton, at least if the leading order flow is robust against
small perturbations.

The slow manifold is parametrized by the location, size and twist
of the instanton and by the location, size and twist of the
anti-instanton.  A gauge transformation eliminates one of the twists,
leaving the relative twist.  The slow manifold is thus parametrized by
the two locations, the two sizes and the relative twist $g_{\tw}$.

The Y-M action $S_{\YM}$ is invariant under the 15 parameter conformal group
$SO(1,5)$, so $S_{\YM}$ on the slow manifold,
as a function of the parameters of the instanton-anti-instanton pair,
is invariant under $SO(1,5)$.
Therefore it suffices to calculate the
generator of the Y-M flow, which is the gradient of $S_{\YM}$, on a
representative slice through the orbits of the conformal
group.

We move the instanton to the south pole in $S^{4}$ using
a conformal transformation,
and the anti-instanton to the north pole
using another.
The remaining subgroup of $SO(1,5)$ consists of the rotation group 
$SO(4)$ and the translations in $x$ (which are the dilations of
$\Reals^{4}$ in the stereographic projection).
A rotation in $SO(4)$ diagonalizes the relative twist 
$g_{\tw}$.
We now have a $U(2)$-invariant instanton and a $U(2)$-invariant 
anti-instanton.
A translation $x\mapsto x+a$ takes $\rI$ to $e^{-a}\rI$ and 
$\rIbar$ to $e^{a}\rIbar$, so we can use a translation in $x$
to set $\rI=\rIbar=\rho$.
We now have a representative slice of the slow manifold parametrized 
by
the $U(2)$-invariant instanton-anti-instanton pairs of equal size $\rho$
and relative twist $\sigma$.

\section{The Y-M flow equation on the slow manifold}

To calculate the asymptotic expansion of the Y-M flow equation on 
the slow manifold,
we start with a larger than necessary slice of the slow 
manifold: all the $U(2)$-invariant 
instanton-anti-instanton pairs,
parametrized by their asymptotically small sizes $\rho_{\pm}$
and their twist angles $\sigma_{\pm}\in 
[0,2\pi]$.
This slice of the slow manifold is
a family of $U(2)$-invariant connections
\eq
\Dslow
=
\left \{
\begin{matrix}
e^{i \alpha_{+}(x) (P-Q)} \,(\DI+\delta \AI)\, 
e^{- i \alpha_{+}(x) (P-Q)}
\quad &x>0\hfill
\\[1ex]
e^{ i \alpha_{-}(x) (P-Q)}
\,( \DIbar+\delta \AIbar )\, e^{- i \alpha_{-}(x) (P-Q)} \hfill  
&x<0\,,
\end{matrix}
\right .
\en
\eq
\alpha_{\pm}(0)=\frac12\sigma_{\pm}
\,.
\en
The $\delta A_{\pm}$ are asymptotically small perturbations of 
the instanton and anti-instanton, to be determined
by the condition that
the family of connections $\Dslow$,
parametrized by $\rho_{\pm}$ and $\sigma_{\pm}$,
is preserved under the Y-M flow.
There must be velocity vector fields
\eq
\frac{d}{dt} \rho_{\pm} = \dot \rho_{\pm}\,, \qquad 
\frac{d}{dt} \alpha_{\pm}(x) = \dot \alpha_{\pm}(x)
\,,\qquad 
\frac{d}{dt} \sigma_{\pm} =\dot\sigma_{\pm}= 2\dot \alpha_{\pm}(0)
\en
such that the Y-M flow equation is satisfied on the slow manifold
\eq
\frac{d}{dt}\Dslow = {*} \Dslow {*} F^{\slow}
\label{eq:slowflow}
\en
where $F^{\slow}$ is the curvature of $\Dslow$.
We solve for the velocities in two steps:
\begin{enumerate}
\item First we solve equation~\ref{eq:slowflow},
the Y-M flow equation,
separately in each open hemisphere.
The general solution in each hemisphere
depends on several undetermined parameters,
including $\dot \rho_{\pm}$ and $\dot \sigma_{\pm}$.
\item
Then we require
$\Dslow$ and $F^{\slow}$ to be continuous at the 
equator, $x=0$, so that the flow equation holds there as well.
At this stage, to simplify the calculation, we specialize to the 
subfamily
where there is an $x\rightarrow -x$ symmetry, where
\eq
\rI=\rIbar=\rho
\,,\qquad
\alpha_{+}(x) = -\alpha_{-}(-x) = \alpha(x)
\,,\qquad
\sigma_{+}=-\sigma_{-}=\frac12\sigma
\,.
\en
The symmetric twisted pairs still represent
every orbit of the conformal group.

The continuity conditions at $x=0$ fix all parameters in the 
separate solutions on the two hemispheres, thereby determining
the velocity vectors $\dot \rho$, $\dot \alpha(x)$, and $\dot \sigma$.
\end{enumerate}
All calculations are to leading order 
in $\rho$.  With more work, the 
method would produce the velocity vectors on the slow manifold to all orders in 
$\rho$.

\subsection{The flow equation in each open hemisphere}

In this section, we solve the flow equation, 
equation~\ref{eq:slowflow}, in each hemisphere separately,
to leading order in $\rho_{\pm}$.
The calculation is the same in each hemisphere, so,
for the sake of legibility, we temporarily write 
$\rho$ instead of $\rho_{\pm}$ and $f$ instead of 
$f_{\pm}$.

The lhs of equation~\ref{eq:slowflow} is,
to leading order,
\eq
\frac{d}{dt}\Dslow = 
e^{i \alpha_{\pm} (x) (P-Q)} \,
\left (
\dot \rho \partialby{\rho } \Dpm 
+ \left [\Dpm ,\, -i\dot\alpha_{\pm} (x)(P-Q)\right ]
\right )\, 
e^{- i \alpha_{\pm} (x) (P-Q)}
\en
We only need the rhs of equation~\ref{eq:slowflow} 
expanded
to first order in $\delta A $.
To this order,
the connection $D= \Dpm +\delta A $ has curvature
$F = \Dpm ^{2} + \Dpm \delta A $
where $\Dpm ^{2}$ is the curvature of the 
(anti-)instanton.  So
\eq
{*} D {*} F = {*} [\delta A ,\, \Dpm ^{2}] 
+ {*} \Dpm  {*} \Dpm \delta A 
= \mp {*} \Dpm ^{2}\delta A + {*} \Dpm  {*} \Dpm \delta 
A  
=  {*} \Dpm  ({*}\mp1) \Dpm \delta A 
\en
so the rhs of equation~\ref{eq:slowflow} becomes
\eq
{*} \Dslow {*} F^{\slow} = 
e^{ i \alpha_{\pm}(x) (P-Q)}
\left [
{*} \Dpm  ({*}\mp1) \Dpm \delta A 
\right ]
e^{- i \alpha_{\pm}(x) (P-Q)} 
\,.
\en
The flow equation,
equation~\ref{eq:slowflow}, is now
\eq
\dot \rho \partialby{\rho } \Dpm 
+\left [\Dpm ,\,-  i\dot\alpha_{\pm} (x)(P-Q)\right ]
=
{*} \Dpm  ({*}\mp1) \Dpm \delta A 
\,.
\label{eq:flowred}
\en
This takes a particularly simple form if we make a change of 
basis
\eq
\omega =  -\eta+\eta^{\dagger}-\eta_{3}
\,,\qquad
\omega_{1} = \eta-\eta^{\dagger}-2\eta_{3}
\,,\qquad
\omega_{2} = -i(\eta+\eta^{\dagger})
\,.
\en
Recall that
\eq
f=\frac{1}{1+\rho^{-2}e^{\mp 2x}}
\,,\qquad
\frac1{f } -1 =\rho ^{-2}e^{\mp 2x}
\en
so
\eq
\rho \partialby{\rho } \Dpm 
= \rho \partialby{\rho } (d_{\omega}-f \omega) =
-2 f (1-f )\omega
\,.
\en
The $U(2)$-invariant infinitesimal gauge transformations of the (anti-)instanton are
\aeq{
\left [\Dpm ,\, - i \varphi (x)(P-Q)\right ]
&= - \partial_{x}\varphi (x) dx\,  i (P-Q) - \varphi (x) 
\left [-f  \omega,\, i (P-Q)\right ] \\
&=
- \partial_{x}\varphi (x) dx  \,i (P-Q) - 2f \varphi (x)\omega_{2}
\,.
\label{eq:gaugetransformations}
}
Using this formula with $\varphi=\dot\alpha_{\pm}$,
equation~\ref{eq:flowred} becomes
\eq
-2 \rho ^{-1}\dot \rho 
f (1-f )\omega
-
\partial_{x}\dot \alpha_{\pm}  dx\,  i (P-Q) 
- 2 f \dot \alpha_{\pm} \omega_{2}
=
{*} \Dpm  ({*}\mp1) \Dpm \delta A 
\,.
\label{eq:flowequation}
\en
We expand in the new basis
\eq
\delta A  =
\delta A_{0}(x) dx\, i(P-Q)
+ \delta f (x) \omega
+\delta A_{1}(x) \omega_{1}
+\delta A_{2}(x) \omega_{2}
\,.
\en
The laplacian, derived in 
Appendix~\ref{app:laplacianformulas},
is diagonal in this basis,
\aeq{
{*} \Dpm  ({*}\mp1) \Dpm \delta A  &=
- 8 \Round(x)^{-1}f ^{2} \delta \tilde A_{0} 
i dx (P-Q)
-2\Round(x)^{-1}f ^{-1}\partial_{x}
\left (
f ^{2}
\delta \tilde A_{0}
\right )
\omega_{2}
\\
&\qquad{}
+\Round(x)^{-1}f ^{-1}(1-f )^{-1}\partial_{x}
f ^{2}(1-f )^{2}
\partial_{x}f ^{-1}(1-f )^{-1}\delta f\,  \omega
\\
&\qquad{}
+\Round(x)^{-1}f ^{-1}(1-f )^{2}\partial_{x}
f ^{2}(1-f )^{-4}
\partial_{x}f ^{-1}(1-f )^{2}
\delta A_{1}\,\omega_{1}
}
where $\Round(x)=(\cosh x)^{-2}$ is the conformal factor in the round metric on 
$S^{4}$ as written in equation~\ref{eq:roundmetric}
and where
\eq
\delta \tilde  A_{0} = \delta A_{0}
-\frac 12 \partial_{x}(f ^{-1}\delta A_{2})
\,.
\en
Note that $\delta \tilde  A_{0}$ vanishes for 
perturbations of the form $\delta A_{0}=-\partial_{x}\varphi$, $\delta A_{2}= -2 f \varphi$,
which are the infinitesimal gauge 
transformations of the (anti-)instanton
as given in equation~\ref{eq:gaugetransformations}.
So the laplacian annihilates the infinitesimal gauge 
transformations,
as it should.

We take advantage of the infinitesimal gauge transformations to
set $\delta A_{0} =0$,
keeping the perturbation $\delta A$ in $A_{0}=0$
gauge.
Then
\eq
\delta \tilde  A_{0} =
-\frac 12 \partial_{x}(f ^{-1}\delta A_{2})
\,.
\label{eq:deltatildaA}
\en
The flow equation, equation~\ref{eq:flowequation}, is now four ordinary equations
\aeq{
- \partial_{x}\dot \alpha_{\pm} &=
-8 \Round(x)^{-1} f ^{2} \delta \tilde A_{0} 
\label{eq:flow1i}\\
-2 f \dot \alpha_{\pm} &= -2\Round(x)^{-1}f ^{-1}\partial_{x}
\left (
f ^{2}
\delta \tilde A_{0}
\right )
\label{eq:flow2i}\\
-2 \rho ^{-1}\dot \rho 
f (1-f )
&=\Round(x)^{-1}f ^{-1}(1-f )^{-1}\partial_{x}
f ^{2}(1-f )^{2}
\partial_{x}f ^{-1}(1-f )^{-1}\delta f 
\label{eq:flow3i}\\
0&=
\Round(x)^{-1}f ^{-1}(1-f )^{2}\partial_{x}
f ^{2}(1-f )^{-4}
\partial_{x}f ^{-1}(1-f )^{2}
\delta A_{1}
\label{eq:flow4i}
\,.
}
Since we are solving the flow equation only to leading order,
we expand
\eq
\Round(x) = (\cosh x)^{-2}
=
\frac{4 f (1-f )}{\left [
\rho ^{-1}f +\rho (1-f )
\right ]^{2}}
= 4\rho ^{2} f ^{-1}(1-f ) + O(\rho ^{4})
\label{eq:R2approx}
\en
and keep only the leading order term.
This approximation expresses the fact that, in the limit 
$\rho \rightarrow 0$, only the metric at the location of the 
(anti-)instanton enters into the solution of the flow equation.
The four equations~\ref{eq:flow1i}--\ref{eq:flow4i} become 
\aeq{
\partial_{x}\dot \alpha_{\pm} &=
2\rho ^{-2} f^{3} (1-f )^{-1} \delta \tilde A_{0} 
\label{eq:flow1}\\
\dot \alpha_{\pm} &= \frac14 \rho ^{-2} f^{-1} (1-f )^{-1}\partial_{x}
\left (
f ^{2}
\delta \tilde A_{0}
\right )
\label{eq:flow2}\\
\rho ^{-1}\dot \rho 
&=-\frac18 \rho ^{-2} f ^{-1}(1-f )^{-3}\partial_{x}
f ^{2}(1-f )^{2}
\partial_{x}f ^{-1}(1-f )^{-1}\delta f 
\label{eq:flow3}\\
0&=
\partial_{x}
f ^{2}(1-f )^{-4}
\partial_{x}f ^{-1}(1-f )^{2}
\delta A_{1}
\label{eq:flow4}
\,.
}

For $\Dslow$ to be regular at the pole,
the perturbations must vanish at $x=\pm\infty$,
\eq
\delta f = \delta A_{1} =\delta A_{2} =O(e^{\mp 2x}) = O(1-f) \quad
\qquad
x\rightarrow \pm\infty\,,
\; f \rightarrow 1
\,.
\en

\subsubsection{The first two flow equations}
The first of the four flow equations, equation~\ref{eq:flow1}, is 
trivially solved to give
\eq
\delta \tilde A_{0}  =
\frac12 \rho ^{2} f ^{-3}(1-f )  \partial_{x}\dot \alpha_{\pm}
\,.
\label{eq:A0}
\en
Then the second of the four flow equations, equation~\ref{eq:flow2}, becomes an equation on $\dot 
\alpha_{\pm}(x)$,
\eq
8 f (1-f ) \dot \alpha_{\pm} = \partial_{x}\left [
f ^{-1}(1-f ) \partial_{x}\dot \alpha_{\pm}
\right ]
\,.
\en
If we change independent variable from $x$ to $f (x)$,
\eq
df  = \pm 2 f (1-f ) dx
\,,
\en
this becomes
\eq
2 \dot \alpha_{\pm} = \partial_{f } (1-f )^{2} \partial_{f }\dot \alpha_{\pm} 
\en
which has two independent solutions, $1-f $ and 
$(1-f )^{-2}$.  The latter is singular at the pole $x=\pm 
\infty$, so we must have
\eq
\dot \alpha_{\pm} = C_{\alpha{\pm}} (1-f )
\,.
\en
At $x=0$, this is
\eq
\dot \alpha_{\pm}(0)   = \frac{C_{\alpha{\pm}}}{1+\rho ^{2}}
\en
so, to leading order,
\eq
C_{\alpha{\pm}} = \frac12 \dot \sigma_{\pm} 
\,,\qquad
\dot \alpha_{\pm} = \frac12 \dot \sigma_{\pm}  (1-f )
\,.
\label{eq:dotalpha}
\en
Using equations~\ref{eq:dotalpha} and \ref{eq:deltatildaA}
in equation~\ref{eq:A0}, we get
\eq
\partial_{x}(f ^{-1}\delta A_{2})
=
-\frac 14  \rho ^{2}\dot\sigma_{\pm}  \partial_{x}
\left [ f ^{-2}(1-f)^{2} \right ]
\,.
\en
The unique solution that goes to zero at the pole, where $f =1$,
is
\eq
\delta A_{2} = - \frac14 \rho ^{2}\dot \sigma_{\pm}
f ^{-1}(1-f )^{2}
\,.
\en
We now have the general solution of the first two equations,
\eq
\delta A_{2} = -\frac14 \rho ^{2}\dot \sigma_{\pm}
f ^{-1}(1-f )^{2}
\,,\qquad
\dot \alpha_{\pm} = \frac12 \dot \sigma_{\pm} (1-f )
\,.
\en

\subsubsection{The last two flow equations}
The last two of the four flow equations, equations \ref{eq:flow3} 
and~\ref{eq:flow4}, become,
after the change of independent variable from $x$ to $f (x)$,
\aeq{
- 2\rho \dot \rho(1-f )^{2}
&=\partial_{f }
f ^{3}(1-f )^{3}
\partial_{f }f ^{-1}(1-f )^{-1}\delta f 
\\
0&=
\partial_{f }
f ^{3}(1-f )^{-3}
\partial_{f }f ^{-1}(1-f )^{2}
\delta A_{1}
\,.
}
Integrating once, we get
\aeq{
\frac23 \rho\dot \rho f^{-3}
&=
\partial_{f }f ^{-1}(1-f )^{-1}\delta f 
\\
-2 C_{1}\rho^{2}f^{-3}(1-f )^{3}
&=
\partial_{f }f ^{-1}(1-f )^{2}
\delta A_{1}
\,.
}
The integration constant in the first equation is fixed by the 
boundary condition that $\delta f$ should go to zero at $f=1$.
We write the integration constant in the second 
equation as $-2 C_{1}\rho^{2}$ for later convenience.

Integrating again, we get
\aeq{
\delta f 
&=
- \frac13 \rho \dot \rho f^{-1}(1-f )
+C_{f}f(1-f )
\\
\delta A_{1}
&=
C_{1}\rho^{2}
\left [
f^{-1} +2 -6(1-f)^{-1}-6 f(1-f)^{-2} \ln f
\right ]
\,.
}
The integration constant $C_{f}$  can be absorbed into a redefinition 
of $\rho$, so we set $C_{f}=0$.  The new integration constant 
in the second equation is fixed by the boundary condition at $f=1$.

\subsubsection{Summary: the general solution in each hemisphere}
Now we restore the $\pm$ subscripts to $\rho$ and $f$, indicating the 
hemisphere in which they obtain.
The general solution to the flow equation in each hemisphere,
with the gauge fixing condition $\delta A_{0\pm}=0$, is
\aeq{
\delta f_{\pm} 
&=
- \frac13 \rho_{\pm} \dot \rho_{\pm} f_{\pm}^{-1}(1-f_{\pm} )
\\
\delta A_{1{\pm}}
&=
C_{1{\pm}}\rho_{\pm}^{2}
\left [
f_{\pm}^{-1} +2 -6(1-f_{\pm})^{-1}-6 f_{\pm}(1-f_{\pm})^{-2} \ln f_{\pm}
\right ]
\\
\delta A_{2\pm} &= -\frac14 \rho_{\pm}^{2}\dot \sigma_{\pm}
f_{\pm}^{-1}(1-f_{\pm})^{2}
\\
\dot \alpha_{\pm}(x) &= \frac12 \dot \sigma_{\pm} (1-f_{\pm} )
\,.
}
The solution in each hemisphere is parametrized by three quantities,
$\dot r_{\pm}$, $\dot 
\sigma_{\pm}$ and $C_{1\pm}$,
which are to be determined by the continuity equations at the equator.

\subsection{Continuity conditions at the equator}

Now we specialize to the subfamily of connections $\Dslow$ with
\eq
\rho_{+}=\rho_{-}=\rho\,,
\qquad
\sigma_{+} = - \sigma_{-} =  \frac12\sigma\,,
\qquad
\alpha_{+}(x) = -\alpha_{-}(-x) = \alpha(x)
\,.
\en
These connections $\Dslow$ have an $x\mapsto -x$ symmetry that simplifies the calculations.
The general solution to the flow equation in each hemisphere is now
\aeq{
\delta A_{0\pm}&=0\\
\label{eq:deltaf}
\delta f_{\pm} 
&=
- \frac13 \rho \dot \rho f_{\pm}^{-1}(1-f_{\pm} )
\\
\label{eq:deltaA1}
\delta A_{1{\pm}}
&=
C_{1{\pm}}\rho^{2}
\left [
f_{\pm}^{-1} +2 -6(1-f_{\pm})^{-1}-6 f_{\pm}(1-f_{\pm})^{-2} \ln f_{\pm}
\right ]
\\
\delta A_{2\pm} &= \mp \frac18 \rho^{2}\dot \sigma
f_{\pm}^{-1}(1-f_{\pm})^{2}
\label{eq:deltaA2}
\\
\dot \alpha_{\pm} &= \pm \frac14 \dot \sigma (1-f_{\pm} )
\\
\alpha_{\pm}(0) &= \pm \frac14 \sigma
}
We will need, at $x=0$, the values
\eq
\delta f_{\pm}(0)= -\frac13 \rho ^{-1}\dot \rho
\,,
\qquad
\delta A_{1{\pm}}(0)
=
C_{1{\pm}}
\,,
\qquad
\delta A_{2\pm}(0) = \mp \frac18 \dot \sigma
\,,
\label{eq:valuesatzero}
\en
and the first derivatives
\eq
\partial_{x}\delta f_{\pm}(0)= \pm\frac23 \rho ^{-1}\dot \rho
\,,
\qquad
\partial_{x}\delta A_{1{\pm}}(0)
=
\mp 2 C_{1{\pm}}
\,,
\qquad
\partial_{x}\delta A_{2\pm}(0) = \frac14 \dot \sigma
\,.
\label{eq:derivativesatzero}
\en

\subsubsection{Continuity of \texorpdfstring{$\Dslow$}{Dslow}}
The continuity of $\Dslow$ at the equator is the condition,
at $x=0$, 
\eq
e^{i \alpha_{+}(x) (P-Q)} \,(\DI+\delta \AI)\, 
e^{- i \alpha_{+}(x) (P-Q)}
=e^{i \alpha_{-}(x) (P-Q)}
\,( \DIbar+\delta \AIbar )\, e^{-i \alpha_{-}(x) (P-Q)}
\,.
\en
This is equivalent to
\eq
e^{\frac14 i \sigma (P-Q)} \,(-\rho^{2}\omega +\delta \AI)\, 
e^{-\frac14 i \sigma (P-Q)}
=e^{-\frac14 i \sigma (P-Q)}
\,( -\rho^{2}\omega +\delta \AIbar )\, 
e^{\frac14 i \sigma (P-Q)} 
\en
since $D_{\pm}=d_{\omega}-f_{\pm}\omega$ and
$
\partial_{x}\alpha_{+}(0) = \partial_{x}\alpha_{-}(0) 
$
by the symmetry $\alpha_{-}(x) = -\alpha_{+}(-x)$
and
\eq
f_{+}(0)=f_{-}(0) = \rho^{2} +O(\rho^{4})
\,.
\en
At $x=0$,
\aeq{
-\rho^{2}\omega+\delta A_{\pm} &= \left (-\rho^{2} - \frac13 \rho ^{-1}\dot 
\rho\right ) \omega
+ C_{1{\pm}} \omega_{1} \mp \frac18 \dot \sigma \omega_{2}\\
&=\left (\rho^{2}+\frac13 \rho ^{-1}\dot 
\rho+ C_{1{\pm}}\pm  \frac18 i\dot \sigma\right ) \eta
- \left (\rho^{2}+\frac13 \rho ^{-1}\dot 
\rho+ C_{1{\pm}}\mp \frac18 i \dot \sigma \right ) \eta^{\dagger}
\\&\qquad{}
+ \left (\rho^{2}+\frac13 \rho ^{-1}\dot 
\rho -2 C_{1{\pm}}\right) \eta_{3}
}
so the continuity condition becomes the two equations
\aeq{
e^{\frac12 i \sigma}
\left (\rho^{2}+\frac13 \rho ^{-1}\dot 
\rho+ C_{1{+}}+  \frac18 i\dot \sigma\right ) 
&=
e^{-\frac12 i \sigma}
\left (\rho^{2}+\frac13 \rho ^{-1}\dot 
\rho+ C_{1{-}}-  \frac18 i\dot \sigma\right ) 
\\
\rho^{2}+\frac13 \rho ^{-1}\dot 
\rho -2 C_{1{+}}
&=
\rho^{2}+\frac13 \rho ^{-1}\dot 
\rho -2 C_{1{-}}
}
which are equivalent to the two equations
\eq
C_{1+}=C_{1-}
\en
and
\eq
\left (\rho^{2}+\frac13 \rho ^{-1}\dot 
\rho+ C_{1{+}}\right ) 
\sin \frac12 \sigma
=  -\frac18 \dot \sigma \cos \frac12 \sigma
\,.
\label{eq:Dslowcont}
\en

\subsubsection{Continuity of \texorpdfstring{$F^{\slow}$}{Fslow}}

Continuity of $F^{\slow}$ at $x=0$ is
\eq
e^{\frac14 i \sigma (P-Q)} \,(D_{+}^{2}+D_{+}\delta \AI)\, 
e^{-\frac14 i \sigma (P-Q)}
=e^{-\frac14 i \sigma (P-Q)}
\,(D_{-}^{2}+D_{-}\delta \AIbar )\, 
e^{\frac14 i \sigma (P-Q)} 
\en
The self-dual part of this condition is
\eq
e^{\frac14 i \sigma (P-Q)} \,(2D_{+}^{2}+({*}+1)D_{+}\delta \AI)\, 
e^{-\frac14 i \sigma (P-Q)}
=e^{-\frac14 i \sigma (P-Q)}
\,({*}+1)D_{-}\delta \AIbar \, 
e^{\frac14 i \sigma (P-Q)} 
\label{eq:selfdual}
\en
while the anti-self-dual part is
\eq
e^{\frac14 i \sigma (P-Q)} \,({*}-1)D_{+}\delta \AI\, 
e^{-\frac14 i \sigma (P-Q)}
=e^{-\frac14 i \sigma (P-Q)}
\,(-2D_{-}^{2}+({*}-1)D_{-}\delta \AIbar )\, 
e^{\frac14 i \sigma (P-Q)} 
\,.
\label{eq:antiselfdual}
\en
The self-dual and anti-self-dual continuity equations
are equivalent under the $x\mapsto -x$ symmetry.

From Appendix~\ref{app:FFpm},
the (anti-)instanton curvature is
\eq
D_{\pm}^{2} = -2 f_{\pm}(1-f_{\pm}) ({*}\pm 1) dx \omega
\en
which is, at $x=0$,
\eq
D_{\pm}^{2} = -2\rho^{2} ({*}\pm 1) dx \omega 
\,.
\en
From Appendix~\ref{app:laplacianformulas},
\aeq{
\Dpm \delta f_{\pm} \omega &=
\partial_{x}\delta f_{\pm}dx \omega + \lambda \delta f_{\pm}{*}dx \omega
\\
\Dpm \delta  A_{1\pm} \omega_{1} &=
\partial_{x}\delta A_{1\pm} dx \omega_{1} + \lambda_{1} \delta A_{1\pm}{*}dx \omega_{1}
\\
\Dpm \delta A_{2\pm} \omega_{2} &= \partial_{x}\delta 
A_{2\pm} dx \omega_{2}
+ \lambda_{2}\delta A_{2\pm}{*}dx \omega_{2}
}
where, to leading order,
\eq
\lambda=\lambda_{1}=\lambda_{2}=2
\,.
\en
At $x=0$, using the values collected in 
equations~\ref{eq:valuesatzero} and \ref{eq:derivativesatzero},
\aeq{
\Dpm \delta f_{\pm}\omega &=- \frac23 \rho^{-1}\dot \rho  ({*} \mp 1 ) dx \omega
\\
\Dpm \delta  A_{1\pm} \omega_{1} &=2 C_{1\pm} ( {*}\mp 1)dx \omega_{1}
\\
\Dpm \delta A_{2\pm} \omega_{2} &= \mp \frac14 \dot \sigma
({*} \mp 1 ) dx \omega_{2}
}
so
\eq
D_{\pm}\delta A_{\pm} = 
({*} \mp 1 ) dx
\left (
-\frac23 \rho^{-1}\dot \rho \omega +
2 C_{1\pm} \omega_{1}
\mp \frac14 \dot \sigma
\omega_{2}
\right )
\,.
\en
Equation~\ref{eq:selfdual}, the self-dual continuity condition, 
becomes
\eq
e^{\frac14 i \sigma (P-Q)} \,\rho^{2}  \omega \, 
e^{-\frac14 i \sigma (P-Q)}
=e^{-\frac14 i \sigma (P-Q)}
\, \left (
\frac13 \rho^{-1}\dot \rho \omega
-C_{1\pm} \omega_{1}
- \frac18 \dot \sigma
\omega_{2}
\right )
\, 
e^{\frac14 i \sigma (P-Q)} 
\,.
\en
Equation~\ref{eq:antiselfdual}, the anti-self-dual continuity condition, 
becomes the equivalent equation
\eq
e^{\frac14 i \sigma (P-Q)} 
\left (
\frac13 \rho^{-1}\dot \rho \omega 
- C_{1\pm} \omega_{1}
+\frac18 \dot \sigma
\omega_{2}
\right )
e^{-\frac14 i \sigma (P-Q)}
= e^{-\frac14 i \sigma (P-Q)}
\,\rho^{2}  \omega \, 
e^{\frac14 i \sigma (P-Q)} 
\,.
\en
Their solution is
\aeq{
C_{1+}&= \frac13 \rho^{2}(\cos \sigma -1)\\
\rho^{-1}\dot \rho &= \rho^{2}(1+2\cos \sigma)\\
\dot \sigma &= -8 \rho^{2} \sin \sigma
\,.
}
where we have used $C_{1+}=C_{1-}$ which was required for continuity of 
$\Dslow$.
Finally, we check that
the remaining continuity condition on $\Dslow$,
equation~\ref{eq:Dslowcont}, is now also satisfied.

\subsection{Summary: the Y-M flow equation on the slow manifold}
The slow manifold is represented by
the family of $U(2)$-invariant connections
\eq
\Dslow
=
\left \{
\begin{matrix}
e^{i \alpha_{+}(x) (P-Q)} \,(\DI+\delta \AI)\, 
e^{- i \alpha_{+}(x) (P-Q)}
&\qquad x>0
\\[1ex]
e^{ i \alpha_{-}(x) (P-Q)}
\,( \DIbar+\delta \AIbar )\, e^{- i \alpha_{-}(x) (P-Q)} \hfill  &\qquad x<0
\,.
\end{matrix}
\right .
\en
obeying the symmetry condition
\eq
\rho_{+}=\rho_{-}=\rho
\qquad
\alpha_{+}(x) = -\alpha_{-}(-x) = \alpha(x)
\,.
\en
The relative twist of the instanton and anti-instanton is
\eq
\sigma = 4 \alpha(0)
\,.
\en
The slow manifold is parametrized by the instanton size $\rho$, the 
relative twist $\sigma$, and by the gauge function $\alpha(x)$,
$\alpha(0)=\frac14\sigma$.
The gauge transformations
\eq
\Dslow \mapsto e^{i \varphi (x) (P-Q)}
\Dslow  e^{- i \varphi (x) (P-Q)}
\,,
\qquad
\varphi (x) = -\varphi (-x)
\en
act on the slow manifold by
\eq
\alpha(x) \mapsto \alpha(x) + \varphi (x)
\en
so the slow manifold in $\mathcal{A}/\mathcal{G}$ is parametrized by 
$\rho$ and $\sigma$ alone.

The Y-M flow equations on the slow manifold are
\aeq{
\dot \rho &= \rho^{3}(1+2\cos \sigma) + O(\rho^{5}) \\
\dot \sigma &= - 8 \rho^{2} \sin \sigma+ O(\rho^{4}) \\
\dot \alpha &= \frac14 \dot \sigma (1-f_{+})+ O(\rho^{4}) 
\,.
}
The perturbation $\delta A_{\pm}$ of the (anti-)instanton is
\eq
\delta A_{\pm} = \delta f_{\pm}(x) \omega + \delta A_{1\pm}(x)
\omega_{1}+\delta A_{2\pm}(x) \omega_{2}
\en
where
\aeq{
\delta f_{\pm} 
&=
- \frac13 \rho^{4}(1+2\cos \sigma)  f_{\pm}^{-1}(1-f_{\pm} )
+ O(\rho^{6}) 
\label{eq:pertf}
\\
\delta A_{1{\pm}}
&=
\frac13 \rho^{4} (\cos \sigma -1)
\left [
f_{\pm}^{-1} +2 -6(1-f_{\pm})^{-1}-6 f_{\pm}(1-f_{\pm})^{-2} \ln f_{\pm}
\right ]
+ O(\rho^{6}) 
\label{eq:pertA1}
\\
\delta A_{2\pm} &= 2 \rho^{4}
\sin \sigma
f_{\pm}^{-1}(1-f_{\pm})^{2}
+ O(\rho^{6}) 
\,.
\label{eq:pertA2}
}
which indeed is a small perturbation of the (anti-)instanton 
$D_{\pm}$ everywhere on $S^{4}$.

\section{The gradient formula and \texorpdfstring{$S_{\YM}$}{S-YM} on the slow manifold}

We check that the Y-M flow on the slow manifold is a gradient 
flow with respect to the metric induced from the space of 
connections $\mathcal{A}$.

Let $d \Dslow$ be an infinitesimal variation in the slow manifold,
corresponding to variations $d\rho$ and $d\alpha(x)$ of the parameters.
In the metric on $\mathcal{A}$, 
given by equation~\ref{eq:metric},
the length-squared of the variation is
\eq
(ds^{2})^{\slow}
= \frac1{4\pi^{2}}\int_{S^{4}} \tr\left (-d \Dslow{*}d \Dslow\right )
\en
In each hemisphere, to leading order,
\eq
d \Dslow  = 
e^{i \alpha_{\pm} (x) (P-Q)} \,
\left (
d\rho_{\pm} \partialby{\rho_{\pm} } \Dpm
+ \left [\Dpm ,\, -i d\alpha_{\pm} (x)(P-Q)\right ]
\right )\, 
e^{- i \alpha_{\pm} (x) (P-Q)}
\en
with
\aeq{
d\rho_{\pm} \partialby{\rho_{\pm} } \Dpm &= -2 \rho_{\pm} ^{-1}d \rho_{\pm} 
f_{\pm} (1-f_{\pm} )\omega 
\\
\left [\Dpm ,\, -i d\alpha_{\pm} (x)(P-Q)\right ]
&=
-
\partial_{x}d \alpha_{\pm}  dx\,  i (P-Q) 
- 2 f_{\pm} d \alpha_{\pm} \omega_{2}
\,.
}
Using the inner-product formulas given in
Appendix~\ref{app:newbasis},
equation~\ref{eq:innerprod},
we get
\eq
(ds^{2})^{\slow}
=
\int_{-\infty}^{\infty} dx\, R^{2}(x)
\left [
12 \rho_{\pm}^{-2} (d\rho_{\pm})^{2}f_{\pm}^{2}(1-f_{\pm})^{2} + (\partial_{x}d\alpha_{\pm})^{2}
+ 8 f_{\pm}^{2} (d \alpha_{\pm})^{2}
\right ]
\en
We replace the conformal factor $R^{2}(x)$
by its leading order approximation,
equation~\ref{eq:R2approx},
getting
\eq
(ds^{2})^{\slow}
=
\int_{-\infty}^{\infty} dx\,4\rho_{\pm} ^{2} f_{\pm} ^{-1}(1-f_{\pm} ) 
\left [
12 \rho_{\pm}^{-2} (d\rho_{\pm})^{2}f_{\pm}^{2}(1-f_{\pm})^{2} + (\partial_{x}d\alpha_{\pm})^{2}
+ 8 f_{\pm}^{2} (d \alpha_{\pm})^{2}
\right ]
\en
Specializing to the $x\leftrightarrow -x$ symmetric subfamily, 
and again writing $f$ for $f_{+}$, we have
\aeq{
(ds^{2})^{\slow}
&=
2 \int_{0}^{\infty} dx\,4\rho ^{2} f ^{-1}(1-f ) 
\left [
12 \rho^{-2} (d\rho)^{2}f^{2}(1-f)^{2} + (\partial_{x}d\alpha)^{2}
+ 8 f^{2} (d \alpha)^{2}
\right ]
\\
&=
16 \int_{0}^{1} df\,
\left [
3  (d\rho)^{2}(1-f)^{2} + \rho ^{2}(1-f)^{2}(\partial_{f}d\alpha)^{2}
+ 2 \rho ^{2}(d \alpha)^{2}
\right ]
\\
&=
16 (d\rho)^{2} +
16  \rho ^{2}\int_{0}^{1} df\,
\left [
(1-f)^{2}(\partial_{f}d\alpha)^{2}
+ 2 (d \alpha)^{2}
\right ]
\label{eq:slowmetricA}
}
The generator of the Y-M flow, $\dot \rho$, $\dot \alpha$,
has inner product with a general variation
\aeq{
(ds^{2})^{\slow}(\dot\rho ,\dot \alpha;d\rho,d\alpha )
&=
16 \dot \rho  d\rho +
16  \rho ^{2}\int_{0}^{1} df\,
\left [
(1-f)^{2}\partial_{f}\dot \alpha \partial_{f}d\alpha
+ 2 \dot \alpha d \alpha
\right ]
\\
&=
16 \dot \rho  d\rho +
4\dot\sigma  \rho ^{2}\int_{0}^{1} df\,
\partial_{f}\left [
-(1-f)^{2} d\alpha
\right ]
\\
&=
16 \dot \rho  d\rho +
\dot\sigma  \rho ^{2}
d\sigma
\\
&=
16 \rho^{3}(1+2\cos \sigma)  d\rho
- 8 \rho^{2} \sin \sigma  \rho ^{2}
d\sigma
}
so
\eq
(ds^{2})^{\slow}(\dot\rho ,\dot \alpha;d\rho,d\alpha )
=
-d S_{\YM}
\en
with
\eq
S_{\YM} = 2 - 4 \rho^{4}(1+2\cos \sigma) + O(\rho^{6})\,.
\en
This is the gradient formula, equation~\ref{eq:gradient}.
The additive constant in $S_{YM}$ is fixed because
$S_{\YM}=2$ for the twisted pairs at $\rho=0$.

\section{The metric on the slow manifold in 
\texorpdfstring{$\mathcal{A}/\mathcal{G}$}{A/G}}

The metric on the slow manifold in $\mathcal{A}$ is given by 
equation~\ref{eq:slowmetricA}.  To find the metric on the slow 
manifold in $\mathcal{A}/\mathcal{G}$, we need to project on the 
horizontal subspace of the tangent space of $\mathcal{A}$ --- the variations
orthogonal to the infinitesimal gauge transformations.

The infinitesimal gauge transformations are the perturbations 
$d\alpha_{V}(x)$ with $d\alpha_{V}(0)=0$.
A variation $d\alpha_{H}$ is perpendicular to the gauge 
transformations iff,
for all $d\alpha_{V}(x)$ with $d\alpha_{V}(0)=0$, 
\eq
\int_{0}^{1} df\,
\left [
(1-f)^{2}(\partial_{f}d\alpha_{V})(\partial_{f}d\alpha_{H})
+ 2 (d \alpha_{V})(d \alpha_{H})
\right ] = 0
\en
which is to say that $d\alpha_{H}$ satisfies the 
ordinary differential equation
\eq
\left [
-\partial_{f}(1-f)^{2}\partial_{f} +2
\right ]
d \alpha_{H} = 0
\,.
\en
The only solution that vanishes at $f=1$ is
\eq
d \alpha_{H} = \frac14 d\sigma (1-f)
\,.
\en
Substituting in equation~\ref{eq:slowmetricA},
we get the metric on the slow 
manifold in $\mathcal{A}/\mathcal{G}$,
\eq
(ds^{2})^{\slow}_{\mathcal{A}/\mathcal{G}}
=
16 (d\rho)^{2} +
\rho ^{2}(d\sigma )^{2}
\,.
\en
The gradient formula of course holds here as well,
\eq
(ds^{2})^{\slow}_{\mathcal{A}/\mathcal{G}}
(\dot \rho,\dot \sigma; d\rho,d\sigma)
= - d S_{\YM}
\,.
\en

\section{Long time behavior of the flow}

The Y-M flow on the slow manifold,
\eq
\frac{d\rho}{dt} =  \rho^{3}
(1+2\cos\sigma)
\,,\qquad
\frac{d \sigma }{dt}= - 8\rho^{2}\sin\sigma
\,,
\label{eq:slowflowA}
\en
has flow lines given by
\eq
\frac{d\rho}{d\sigma} = -\frac\rho8 \left ( \frac{1+2\cos\sigma}{\sin\sigma} \right )
\en
which integrates to
\eq
\rho^{8}(1-\cos\sigma)\sin\sigma = 4 C
\,.
\en
Changing variable from $\sigma$ to 
\eq
s=\cos\frac\sigma2\,,\qquad s\in [-1,1]\,,
\en
the flow is
\eq
\frac{d\rho}{dt} =  \rho^{3}
(4 s^{2}-1)
\,,\qquad
\frac{d s }{dt}= 8\rho^{2} s(1-s^{2})
\,,
\en
and the flow lines are
\eq
\rho^{8} s (1-s^{2})^{3/2} = C
\,.
\en
On a flow line, say on the side $s\ge 0$ where $C\ge 0$,
the flow equation is
\eq
\frac{d s }{dt}= 8 C^{1/4} (1-s^{2})^{5/8}s^{3/4}
\,.
\en
which integrates to
\eq
2 C^{1/4}t = s_{t}^{\frac14} F(s_{t}^{2}) - s_{0}^{\frac14} F(s_{0}^{2})
\en
where $H(s^{2})$ is the hypergeometric function
\eq
F(z) = F\left ({\textstyle\frac18}, {\textstyle\frac58}; {\textstyle\frac98}; z\right )
\,,\qquad
F(0)=1\,,\quad
F(1-\epsilon) = F(1) -\frac{1}3\epsilon^{\frac38}+O(\epsilon)
\,.
\en
Substituting for $C$, we get
\eq
2 \rho_{0}^{2} (1-s_{0}^{2})^{3/8} 
t = \left (\frac{s_{t}}{s_{0}}\right ) ^{\frac14} F(s_{t}^{2}) - F(s_{0}^{2})
\en
\eq
2 \rho_{t}^{2}  (1-s_{t}^{2})^{3/8} 
t =  F(s_{t}^{2}) - \left (\frac{s_{0}}{s_{t}}\right )^{\frac14} F(s_{0}^{2})
\en
Suppose $t$ large.
If we hold $\rho_{0}$ fixed  and letting $s_{0}$ vary near 0, we
see explicitly from these formulas that
the trajectory moves first towards $\rho=0$,$s=0$, then along 
the $s$-axis to the neighborhood of $\rho=0$, $s=1$, then outward 
to increasing $\rho$ with $s$ near 1.

\section{The outgoing trajectory}

For the symmetric twisted pair, where the instanton and 
anti-instanton are located at opposite poles in the round $S^{4}$,
we can show that the outgoing trajectory at $\sigma=0,2\pi$ ends at 
the flat connection. 
The argument does not work for other 
twisted pairs, whose outgoing trajectories have less symmetry.

The perturbations 
$\delta A_{1,2}$, equations~\ref{eq:pertA1} and \ref{eq:pertA2} 
vanish for $\sigma=0,2\pi$,
so the outgoing trajectory has the full $SO(4)$ symmetry
of the aligned instanton-anti-instanton and of the round geometry on 
$S^{4}$.
The connections on the outgoing 
trajectory are therefore all of the form
\eq
D = d_{\omega}-f\omega
\,,\qquad f(\pm \infty) = 1
\,.
\en
From equation~\ref{eq:fomegaLpm}, the Y-M action is
\eq
S_{\YM} = \int dx \,
\frac32 \left [ (\partial_{x}f)^{2}+ 4f^{2}(1-f)^{2}\right]
\,.
\en
From Appendix~\ref{app:InvConnections},
\aeq{
{*}F_{\pm}&= -\left [\partial_{x}f\pm 2f(1-f)\right ]
\frac12 ({*}\pm 1)dx \omega \\
{*}D{*}F_{\pm}&= -\frac12 \Round(x)^{-1}\left[ \partial_{x} \pm 2(2f-1) \right ]
\left [\partial_{x}f\pm 2f(1-f)\right ]
\omega
}
so the Y-M flow equation is
\eq
\frac{df}{dt} =\Round(x)^{-1}\left [ \partial_{x}^{2}f + 
4f(1-f)(2f-1)\right ]
\,.
\label{eq:YMflowU2}
\en
Let us 
assume that the flow ends at a fixed point.
The fixed point equation is
\eq
\partial_{x}^{2}f + 4f(1-f)(2f-1) = 0
\,.
\en
For any solution $f$ of the fixed point equation,
the quantity
\eq
A = (\partial_{x}f)^{2}- 4f^{2}(1-f)^{2}
\en
is constant,
$\partial_{x}A=0$,
and must vanish because $S_{\YM}<\infty$.
So, for all $x$,
\eq
\partial_{x}f = \pm 2f(1-f)
\,.
\en
The only solution of this equation
compatible with the boundary conditions $f(\pm\infty)=1$, besides the twisted pair,
is $f=1$, the flat connection.
There is no other fixed point where the outgoing trajectory can end.

\section{Stable 2-manifolds of \texorpdfstring{$SU(2)$ and 
$SU(3)$}{SU(2) and SU(3)} gauge fields}

Nontrivial stable 2-spheres of gauge fields might
give 2-d instanton corrections to the space-time quantum field theory
in the lambda model (discussed in
section~\ref{app:lambdamodel} below).
Nontrivial 2-spheres of gauge fields are classified by 
$\pi_{2}(\mathcal{A}/\mathcal{G})$,
which is $\pi_{5}$
of the gauge group.
Potentially interesting examples are
$\pi_{5}SU(2)=\Integers_{2}$ and $\pi_{5}SU(3)=\Integers$.
We describe some partial results
towards constructing stable 2-spheres for
$SU(3)$ and for $SU(2)$ gauge groups.

\subsection{\texorpdfstring{$SU(3)$}{SU(3)}}

Numerical evidence suggests that there is a stable 2-sphere of $SU(3)$ connections
on $S^{4}$ consisting again of zero-size instanton-anti-instanton
twisted pairs \cite{Friedan2009Pisa}.  The numerical calculation is
analogous to the $SU(2)$ calculation reported above (and was actually
done first).  The $SU(3)$ principle bundles over $S^{6}$ are
classified by $\pi_{5}SU(3)=\Integers$.  The homogeneous space $SU(3)\rightarrow
G_{2}\rightarrow S^{6}$ represents a generator of $\pi_{5}SU(3)$
\cite{Chaves1996}.  Pulling back along a suitably chosen map $S^{2}\times
S^{4}\rightarrow S^{6}$ gives a nontrivial 2-sphere of connections in
the trivial $SU(3)$ bundle over $S^{4}$, representing a generator of
$\pi_{2}(\mathcal{A}/\mathcal{G})$.  The south pole of $S^{4}$ is
mapped to the flat connection.  Some of the $G_{2}$ symmetry survives,
so that all of the connections on $S^{4}$ are $SU(2)$-invariant.  An additional
$U(1)$ symmetry acts on the 2-sphere family of connections, rotating the 2-sphere around
its poles.  The north pole of the 2-sphere is left fixed, so the
connection at the north pole has an additional $U(1)$ symmetry.  It also
has a discrete symmetry exchanging $x\rightarrow -x$.  It seems
plausible that this connection flows to an index 2 fixed point whose
two dimensional unstable manifold is a stable 2-sphere.  It also seems
plausible that this connection flows to the connection that minimizes
$S_{\YM}$ among all connections with the same symmetries.  Carrying
out this minimization of $S_{\YM}$ numerically, we find strong
indications that the minimum value is $S_{\YM}=2$, realized by a
twisted pair.

All $SU(3)$ instantons on $S^{4}$ of instanton number $\pm 1$ are 
reducible \cite{Atiyah1978}.
That is, they are $SU(2)$ instantons embedded in 
$SU(3)$.  We identify $SU(2)$ with the upper-left $2\times2$ block in 
$SU(3)$, identifying an element $g\in SU(2)$ with the block matrix
\eq
g\in SU(2) \equiv
\left (
\begin{array}{c|c}
g & 0\\
\hline
0 & 1
\end{array}
\right )
\in SU(3)
\,.
\en
The basic $SU(2)$ instanton $D_{+}$ is now an $SU(3)$ 
instanton.  The general $SU(3)$ instanton --- of given size 
and location --- is $G D_{+} G^{-1}$ for $G\in SU(3)$,
up to the equivalence $G\sim GK(\theta)$,
for $K(\theta)$ in
the $U(1)$ subgroup of $SU(3)$ of elements
that commute with $SU(2)$, which take the block matrix form
\eq
K(\theta) = 
\left (
\begin{array}{c|c}
e^{i\theta} & 0\\
\hline
0 & e^{-2i\theta}
\end{array}
\right )
\,.
\en
The space of orientations of the $SU(3)$ instanton is 
thus $SU(3)/U(1)$.  The space of relative twists of a twisted pair of 
$SU(3)$ instantons is $M_{\tw}^{SU(3)}=U(1)\backslash SU(3)/U(1)$,
which consists of the individual orientations of the instanton and 
anti-instanton, $SU(3)/U(1) \times 
SU(3)/U(1)$, modulo the global $SU(3)$ gauge transformations.
$M_{\tw}^{SU(3)}$ contains nontrivial 2-spheres,
$\pi_{2}M_{\tw}^{SU(3)} \supset \Integers $,
that can represent $\pi_{2}(\mathcal{A}/\mathcal{G})$.

To write a concrete nontrivial 2-sphere of relative twists,
it is convenient to parametrize $SU(3)$
as $SU(2)\times D^{2} \times SU(2)$,
\eq
G(g_{-},u,g_{+}) = g_{-}^{-1} G_{1}(u) g_{+}
\en
where
\eq
G_{1}(u) = \begin{pmatrix}
1 & 0 & 0\\
0 & u & -\sqrt{1-|u|^{2}} \\
0 & \sqrt{1-|u|^{2}} & \bar u
\end{pmatrix}
\,,\qquad
|u|\le 1\,.
\en
The parametrization is faithful for $|u|<1$,
while at the boundary of the 2-disk, $|u|=1$, it gives a redundant 
parametrization of the subgroup
$SU(2)\times U(1)\subset SU(3)$.
The $U(1)$ subgroup of $SU(3)$ acts on the left and right by
\eq
K(\theta) G(g_{-},u,g_{+})K(\theta')^{-1} 
= G( h({\theta+2\theta'}) g_{-},e^{2i\theta-2i\theta'}u, h({2\theta+\theta'}) g_{+})
\en
where
\eq
h(\theta) =
\begin{pmatrix}
e^{i\theta} & 0 \\
0 & e^{-i\theta}
\end{pmatrix}
\,.
\en
The $O(4)=SU(2)\times SU(2)/\{\pm1\}$ group of rotations around the poles of $S^{4}$
acts on the relative twists by $(g_{L},g_{R}): G \mapsto g_{R}^{-1} G 
g_{R}$, the $g_{L}$ all leaving the twisted pair invariant.  In our parametrization
of $SU(3)$, 
the symmetries act by
\eq
G(g_{-},u,g_{+}) \mapsto G(g_{-}g_{R},u,g_{+}g_{R})
\en
We represent the symmetry classes of twists by the $G(1,u,g_{+})$
subject to the gauge equivalence
\eq
G(1,u,g_{+}) \equiv 
G(1,e^{2i\theta}u,h(\theta) g_{+})
=K\left (\frac{2\theta}3\right ) G(1,u,g_{+}) K\left (-\frac\theta3 \right )^{-1}
\en
and a remaining $U(1)$ symmetry
\eq
G(1,u,g_{+}) \mapsto  G(1,u,h({\theta})g_{+}h({\theta})^{-1})
\,.
\en
The symmetry classes of twisted pairs with an additional $U(1)$ invariance are
the $G(1,u,1)$ and also $G(1,0,g_{0})$ with
\eq
g_{0} = 
\begin{pmatrix}
0 & 1 \\
-1 & 0
\end{pmatrix}
\,.
\en
The latter, $G(1,0,g_{0})$, is the $U(2)$-invariant twisted pair 
indicated by the computer calculation.

A 2-sphere family of twisted pairs invariant under $U(1)$ acting by 
rotation around the poles of $S^{2}$ is given by
\eq
G(w) = G(1,w^{2},g_{1}(w))
\,,\qquad
|w|\le 1
\en
where
\eq
g_{1}(w) = 
\begin{pmatrix}
w & -\sqrt{1-|w|^{2}} \\
\sqrt{1-|w|^{2}} & \bar w
\end{pmatrix}
\,.
\en
The $U(1)$ symmetry is
\eq
G(w) \mapsto G(e^{i\theta}w)
\,.
\en
The twisted pair at the north pole of $S^{2}$, $w=0$, is
the $U(2)$-invariant $G(1,0,g_{0})$.
At $|w|=1$,
\eq
G(w) = K(w) \equiv 1
\en
so $|w|=1$ can be identified to the the south pole in $S^{2}$,
which is mapped to the aligned twisted pair, $G=1$.

It should be straightforward to check directly that $w\mapsto G(w)$ represents 
a generator of $\pi_{2}(\mathcal{A}/\mathcal{G})$, by the same 
argument used above to check the nontriviality of the loop of $SU(2)$ 
twisted pairs.  We leave $A_{0}=0$ gauge, making $G(w) D_{+} G(w)^{-1}$ non-singular at the 
south pole of $S^{4}$ by a gauge transformation $\phi(w,x)\in SU(3)$.
The twisted pairs at $|w|=1$ will all be gauge equivalent, giving
a loop in the gauge group, a map $S^{1}\times S^{4} \rightarrow 
SU(3)$.  This will factor through a map $S^{5} \rightarrow SU(3)$,
which we can check is a generator of $\pi_{5}SU(3)$ by
composing with $SU(3)\rightarrow S^{5}=SU(3)/SU(2)$
to get a map $S^{5}\rightarrow S^{5}$ whose index should be $\pm 1$ 
\cite{Eckmann1942thesis}.

A quicker way to check the nontriviality of the 2-sphere $w\mapsto 
G(w)$ is to evaluate the family index \cite{Atiyah1971} of the Dirac operator on 
$S^{4}$ acting on spinors tensored with the defining representation, 
$\mathbf3$, of $SU(3)$.  The chiral zero-modes of the Dirac operator
of each handedness are localized respectively in the instanton and and the anti-instanton.
It is a simple calculation to show that the left-handed zero mode forms a line 
bundle of Chern number 1 over the 2-sphere of twisted pairs,
which must then necessarily be a generator of 
$\pi_{2}(\mathcal{A}/\mathcal{G})=\Integers$.

The Y-M flow on the slow manifold remains to be calculated in order to 
check that that the 2-sphere 
$w\mapsto G(w)$, or some deformation, is locally stable under 
the flow.  
The calculation is the same, in principle, as for the $SU(2)$ twisted 
pairs.
For $SU(3)$, the symmetry classes of twists are described by 3
parameters, analogous to the twist angle $\sigma$ for $SU(2)$ twists.
Unfortunately, there does seem to be any symmetry that singles out a distinguished 
set of representatives of the symmetry classes, closed under the flow,
analogous to the $U(2)$ symmetry for $SU(2)$ twisted pairs.
It might be possible to find a perpendicular slice through the 
symmetry classes, which would be closed under the gradient flow.
Otherwise, it will be necessary to parametrize the slow manifold
by the full 6 parameter space of $SU(3)$ twists, in addition to the 
instanton size $\rho$.  Inverting the instanton laplacian will be 
considerably more work than in the $SU(2)$ case.
In any case, the calculation of the Y-M flow on the slow manifold 
and the check of local stability remain to be done.

\subsection{\texorpdfstring{$SU(2)$}{SU(2)}}

Since $\pi_{5} SU(2) = \Integers_{2}$,
there should be a nontrivial stable 2-sphere of $SU(2)$ gauge fields 
on $S^{4}$.
We do not know of a homogeneous realization of the generator of 
$\pi_{5} SU(2)$
analogous to the bundles
$SU(2)\rightarrow SU(3)\rightarrow S^{5}$ for $\pi_{4}SU(2)$ and $SU(3)\rightarrow 
G_{2}\rightarrow S^{6}$ for $\pi_{5}SU(3)$,
but there is available a realization with enough symmetry to reduce the problem to
minimizing $S_{\YM}$ on the space of connections with a 
certain fixed symmetry group, as in the other two cases.
In this case, the symmetry group is large enough that
numerical minimization is (barely) practical.

We construct a nontrivial 2-sphere family of $SU(2)$ bundles over $S^{4}$, 
each having the symmetry group $(U(1)\times U(1)/\Integers_{2})\times 
\Integers_{2}$.
The bundle at the north pole in $S^{2}$ has 
an extra $\Integers_{2}\times\Integers_{2}$ symmetry.
We attempt to minimize $S_{\YM}$ numerically over connections with 
the enhanced symmetry group $(U(1)\times U(1)/\Integers_{2})\times 
(\Integers_{2})^{3}$, again approximating the space of such 
connections by finite dimensional affine subspaces.
We find $\min(S_{\YM}) < 4.0053$.
The numerical computations are more expensive in processing time and memory
than the previous ones
because the two continuous symmetries reduce $S^{4}$ to a 
2-dimensional domain, instead of the 1-dimensional domain of the 
previous calculations.
We have to minimize $S_{\YM}$ over 
connections that are polynomials in two variables.

The numerical results suggest that, at the
enhanced symmetry point in the 2-sphere family,
$\min(S_{\YM})$ is realized 
by a fixed point of the Y-M flow
that consists of two zero-size instantons and two zero-size 
anti-instantons,
arranged along the $x$ axis in the order $\bar I I \bar I I$.
Writing the sizes of the instantons $r_{+1}=e^{-x_{+1}}$,
$r_{+2}=e^{x_{+2}}$ and the sizes of the anti-instantons 
$r_{-1}=e^{x_{-1}}$, $r_{-2}=e^{-x_{-2}}$,
the zero-size limit is taken with
\eq
x_{-1}\ll x_{+2}\ll 0 \ll x_{-2}\ll x_{+1}
\,.
\en
Each pair of neighbors in the sequence is
maximally twisted.
It seems plausible that repulsion between neighbors will drive
such a configuration of finite-size 
instantons and anti-instantons to this zero-size limit.
In the limit, there is an
an instanton/anti-instanton pair at
each of the poles.

The evidence for $\min(S_{\YM}) = 4$ at the enhanced symmetry point,
realized by the twisted quadruplet of zero-size (anti-)instantons, is good,
though perhaps not as compelling as in the previous
calculations.
The twisted quadruplet has $S_{\YM}=4$,
so $S_{\YM}\le 4$ is a rigorous upper bound at the enhanced symmetry 
point.

The continuous $U(1)\times U(1)/\Integers_{2}$ symmetry restricts the 
relative twists of the instantons to the diagonal $SU(2)$ matrices.
We write explicitly a 2-parameter family of twisted quadruplet 
connections, in the 2-parameter family of bundles.
This family of connections forms a 2-torus, not a 2-sphere.
It remains to calculate the Y-M flow in the slow modes, to check
first that the twisted quadruplet connection at the enhanced symmetry point
has a 2-dimensional unstable manifold,
and then to find the global structure of that unstable manifold,
presumably either a 2-torus of zero area or a 2-sphere of nonzero 
area.
The first possibility would be of interest for the lambda model.

\subsubsection{A nontrivial 2-sphere of \texorpdfstring{$SU(2)$ 
bundles over $S^{4}$}{SU(2) bundles over S4}}
\label{sect:twospherefamily}
The nontrivial element in $\pi_{5}SU(2)$
was originally realized
as the suspension map
$S(h\circ Sh):S^{5}\rightarrow S^{3}$,
where $h:S^{3}\rightarrow S^{2}$ is the Hopf fibration,
and $Sh:S^{4}\rightarrow S^{3}$ is its suspension  \cite{Whitehead1950}.
We write explicitly
\begin{gather}
S(h\circ Sh) : [0,\pi]^{2}\times SU(2) \rightarrow 
SU(2)
\\
S(h\circ Sh)(\beta_{1},\beta_{2},g) =
\left ( g h_{\beta_{2}} g^{-1} \right )^{-1}
h_{\beta_{1}}
\left ( g h_{\beta_{2}} g^{-1}\right )
\end{gather}
where
\eq
h_{\beta} =
\begin{pmatrix}
e^{i\beta} & 0 \\
0 & e^{-i\beta}
\end{pmatrix}
\,.
\en
We make a topologically insignificant modification, defining
\begin{gather}
\Phi_{2} : [0,\pi]^{2}\times SU(2) \rightarrow 
SU(2)
\\
\Phi_{2}(\beta_{1},\beta_{2})(g)
= h_{\beta_{1}}^{-1}S(h\circ Sh)
= h_{\beta_{1}}^{-1}
\left ( g h_{\beta_{2}} g^{-1} \right )^{-1}
h_{\beta_{1}}
\left ( g h_{\beta_{2}} g^{-1}\right )
\,,
\end{gather}
which satisfies
\eq
\Phi_{2}(\beta_{1},0,g) =
\Phi_{2}(\beta_{1},\pi,g) =
\Phi_{2}(0,\beta_{2},g) =
\Phi_{2}(\pi,\beta_{2},g) =
1
\,,
\en
so the boundary of the square $[0,\pi]^{2}$ can be identified to a 
point, the square becoming a 2-sphere,
and $\Phi_{2}$ becoming a nontrivial map $S^{2}\times SU(2)\rightarrow SU(2)$.
For each $(\beta_{1},\beta_{2})\in S^{2}$, we construct an
$SU(2)$ bundle over $S^{4}$
using $g\mapsto \Phi_{2}(\beta_{1},\beta_{2},g)$
as the gluing map at the equator in $S^{4}$.
Thus $\Phi_{2}$ defines
a nontrivial 2-sphere of trivial $SU(2)$ bundles over $S^{4}$.

The group $SO(4)=SU(2)\times SU(2)/\{\pm1\}$ of rotations of $S^{4}$ 
around the polar axis acts by
\eq
\Phi_{2}(\beta_{1},\beta_{2},g_{L}g g_{R}^{-1}) =
g_{L} \left [
h_{1}^{-1}
\left ( gh_{2} g^{-1} \right )^{-1}
h_{1}
\left ( g h_{2} g^{-1}\right )
\right ]
g_{L}^{-1}
\en
where
\eq
h_{1} = g_{L}^{-1}h_{\beta_{1}}g_{L}\,, \qquad
h_{2} = g_{R}^{-1} h_{\beta_{2}} g_{R}
\,.
\en
If $g_{L}$ and $g_{R}$ are both diagonal,
\eq
g_{L}=h_{\alpha'}\,, \qquad g_{R}=h_{\alpha}\,,
\en
then
\eq
\Phi_{2}(\beta_{1},\beta_{2},h_{\alpha'}g h_{\alpha}^{-1})
= h_{\alpha'} \Phi_{2}(\beta_{1},\beta_{2},g )h_{\alpha'}^{-1}
\en
so each of the $SU(2)$ bundles over $S^{4}$ is invariant under the $U(1)\times 
U(1)/\{\pm1\}$ subgroup of diagonal matrices 
$(h_{\alpha'},h_{\alpha})$ modulo $(-1,-1)$.

In addition, the entire 2-sphere family of bundles is invariant under the $\Integers_{2}\times 
\Integers_{2}$ subgroup  generated by $(g_{L},g_{R})=(\mu_{2},\mu_{2})$ and $(\mu_{1},\mu_{3})$
where
\eq
\mu_{1}=
\begin{pmatrix}
0 & i \\
i & 0
\end{pmatrix}
\,,\qquad
\mu_{2}=
\begin{pmatrix}
0 & 1 \\
-1 & 0
\end{pmatrix}
\,,\qquad
\mu_{3}=
\begin{pmatrix}
i & 0 \\
0 & -i
\end{pmatrix}
\,.
\en
This $\Integers_{2}\times \Integers_{2}$ acts on 
the family of bundles by
\aeq{
\Phi_{2}(\beta_{1},\beta_{2},\mu_{2} g\mu_{2}^{-1})
&= \mu_{2} \Phi_{2}(\pi-\beta_{1},\pi-\beta_{2},g )\mu_{2}^{-1} \\
\Phi_{2}(\beta_{1},\beta_{2},\mu_{1} g \mu_{3}^{-1})
&= \mu_{1} \Phi_{2}(\pi-\beta_{1},\beta_{2},g )\mu_{1}^{-1}
\\
\Phi_{2}(\beta_{1},\beta_{2},\mu_{3} g \mu_{1})
&= \mu_{3} \Phi_{2}(\beta_{1},\pi-\beta_{2},g )\mu_{3}^{-1}
\,.
}

Finally, there is a $\Integers_{2}$ symmetry
\eq
\Phi_{2}(\beta_{1},\beta_{2},g )^{-1}
=  h_{\beta_{1}}^{-1} \Phi_{2}(\pi-\beta_{1},\beta_{2},g ) h_{\beta_{1}}
\en
that acts by reflecting $S^{4}$ in the equator, taking 
$\theta\rightarrow \pi-\theta$, $x\rightarrow -x$.
Combining with the discrete symmetry $g\mapsto \mu_{1}g\mu_{3}^{-1}$,
we get a reflection symmetry of each bundle in the family,
\eq
\Phi_{2}(\beta_{1},\beta_{2},g )^{-1}
= h_{\beta_{1}}^{-1} 
\mu_{1}^{-1}\Phi_{2}(\beta_{1},\beta_{2},\mu_{1}g\mu_{3}^{-1} )\mu_{1} h_{\beta_{1}}
\en
so each connection has symmetry group $(U(1)\times 
U(1)/\{\pm1\})\times \Integers_{2}$.

The $SU(2)$ bundle at the midpoint $\beta_{1}=\beta_{2}=\frac\pi2$ 
thus has an extra $\Integers_{2}\times \Integers_{2}$ symmetry.
If we were to choose a 2-sphere family of connections in this 2-sphere
family of $SU(2)$ bundles, respecting the symmetries of the bundles,
then run the Y-M flow on the family of connections, we might expect
that the connection at the midpoint $\beta_{1}=\beta_{2}=\frac\pi2$
would flow to a fixed point with effective Morse index 2 that
minimizes $S_{\YM}$ among all connections with the enhanced symmetry
of the bundle at $\beta_{1}=\beta_{2}=\frac\pi2$.
With this scenario in mind, we attempt to minimize $S_{\YM}$ among the
connections invariant under this
$(U(1)\times U(1)/\{\pm1\})\times (\Integers_{2})^{3}$
group.

\subsubsection{Reduction to 2-dimensions}

The continuous symmetry group $U(1)\times U(1)/\{\pm1\}$ acts on 
$S^{4}$ by
\begin{gather}
g = \begin{pmatrix}
z_{1} & -\bar z_{2} \\
z_{2} & \bar z_{1}
\end{pmatrix}
\mapsto
h_{\alpha'}g h_{\alpha}^{-1}
=
\begin{pmatrix}
z_{1}e^{i\alpha'-i\alpha} & -\bar z_{2}e^{i\alpha'+i\alpha} \\
z_{2}e^{-i\alpha'-i\alpha} & \bar z_{1}e^{-i\alpha'+i\alpha}
\end{pmatrix}
=
\begin{pmatrix}
z_{1}e^{i\alpha_{1}} & -\bar z_{2}e^{-i\alpha_{2}} \\
z_{2}e^{i\alpha_{2}} & \bar z_{1}e^{-i\alpha_{1}}
\end{pmatrix}
\\[2ex]
\alpha'=\frac12(\alpha_{1}-\alpha_{2})
\,,\qquad
\alpha=\frac12(-\alpha_{1}-\alpha_{2})
\,.
\end{gather}
We write
\eq
z_{1}=r_{1}e^{i\theta_{1}}
\,,\qquad
z_{2}=r_{2}e^{i\theta_{2}}
\,,\qquad
r_{1}=\cos \frac12\psi
\,,\qquad
r_{2}=\sin \frac12\psi
\,,\qquad
\psi\in [0,\pi]
\en
and use $\theta, \psi,\theta_{1},\theta_{2}$ as coordinates on 
$S^{4}$.
We can write the coordinate map
\eq
g = h_{\frac12(\theta_{1}-\theta_{2})} g(\psi)
h_{\frac12(-\theta_{1}-\theta_{2})}^{-1}
\en
where
\eq
g(\psi)
=
\begin{pmatrix}
r_{1} & -r_{2}  \\
r_{2} & r_{1}
\end{pmatrix}
= e^{-\frac12 \psi \mu_{2}}
\,.
\en
The coordinate map is redundant at the poles $\theta=0,\pi$ and at 
$\psi=0,\pi$.
All of $S^{4}$ is covered when $\psi$ ranges over $[0,\pi]$, but it is 
useful to think of $\psi$ taking any real value, the coordinate map 
being many-to-one.

The slice $\theta_{1}=\theta_{2}=0$ contains one representative in each 
symmetry class (except at the poles $\theta=0,
\pi$).
A connection on $S^{4}$ invariant under the continuous symmetry 
will reduce to a connection on the slice,
the 2-dimensional domain parametrized by $\theta$ and $\psi$.

A connection in the bundle defined by the patching map 
$\Phi_{2}(\beta_{1},\beta_{2})$ 
consists of a connection in each hemisphere, 
$D_{\pm}=d+A_{\pm}$,
related on the overlap of the hemispheres by
\eq
D_{-} = \Phi_{2} (\beta_{1},\beta_{2}) D_{+} \Phi_{2}(\beta_{1},\beta_{2})^{-1}
\,.
\en
Writing
\eq
\Phi_{2} (\beta_{1},\beta_{2}) = \Phi_{-}^{-1}\Phi_{+}
\en
with
\eq
\Phi_{+} = h_{\beta_{1}} g h_{\beta_{2}} g^{-1}
\,,\qquad
\Phi_{-}= g h_{\beta_{2}} g^{-1} h_{\beta_{1}}
\,,
\en
the patching formula becomes
\eq
\Phi_{-}D_{-}\Phi_{-}^{-1} = \Phi_{+}D_{+}\Phi_{+}^{-1}
\,.
\en

The continuous symmetry group $(U(1)\times U(1)/\{\pm1\})$ acts by
\eq
D_{\pm}(\theta, h_{\alpha'}g h_{\alpha}^{-1})
= h_{\alpha'}D_{\pm}(\theta, g) h_{\alpha'}^{-1}
\en
or, equivalently,
\eq
A_{\pm}(\theta, h_{\alpha'}g h_{\alpha}^{-1})
= h_{\alpha'}A_{\pm}(\theta, g) h_{\alpha'}^{-1}
\,.
\en
We eliminate the dependence on $\theta_{1,2}$
by a gauge transformation
\eq
\tilde D_{\pm} = d+ \tilde A_{\pm} = h_{\frac12(-\theta_{1}+\theta_{2})}
D_{\pm} h_{\frac12(-\theta_{1}+\theta_{2})}^{-1}
\en
\eq
\tilde A_{\pm} = h_{\frac12(-\theta_{1}+\theta_{2})} A_{\pm}h_{\frac12(-\theta_{1}+\theta_{2})}^{-1}
+ \frac12 (d\theta_{1}-d\theta_{2}) \mu_{3}
\en
\eq
\partial_{\theta_{1}} \tilde A_{\pm} = \partial_{\theta_{2}} \tilde A_{\pm}=0
\,.
\en
The patching formula now becomes
\eq
\tilde \Phi_{-}\tilde D_{-}\tilde \Phi_{-}^{-1} = \tilde \Phi_{+}\tilde D_{+}\tilde \Phi_{+}^{-1}
\en
where
\eq
\tilde \Phi_{\pm}
=  h_{\frac12(-\theta_{1}+\theta_{2})} \Phi_{\pm} h_{\frac12(-\theta_{1}+\theta_{2})}^{-1}
\en
also do not depend on $\theta_{1,2}$,
\eq
\tilde \Phi_{+} = h_{\beta_{1}} g(\psi) h_{\beta_{2}} g(\psi)^{-1}
\,,\qquad
\tilde \Phi_{-}= g(\psi) h_{\beta_{2}} g(\psi)^{-1} h_{\beta_{1}}
\,.
\en

Finally, we define
\eq
D = d+A = \tilde \Phi_{-}\tilde D_{-}\tilde \Phi_{-}^{-1} = \tilde \Phi_{+}\tilde D_{+}\tilde \Phi_{+}^{-1}
\en
which is regular everywhere on $S^{4}$ except at the poles,
and which is independent of $\theta_{1,2}$.
At the poles,
\ateq{2}{
D &\rightarrow \tilde \Phi_{-}
\left [ d+\frac12 (d\theta_{1}-d\theta_{2}) \mu_{3}\right ] 
\tilde \Phi_{-}^{-1}
&\qquad &\text{at the north pole, $\theta = 0$.}
\\
D &\rightarrow \tilde \Phi_{+}
\left [ d+\frac12 (d\theta_{1}-d\theta_{2}) \mu_{3}\right ]
\tilde \Phi_{+}^{-1}
&\qquad &\text{at the south pole, $\theta = \pi$.}
}
We have traded the patching condition at the equator and the 
dependence on $\theta_{1,2}$ for boundary 
conditions at $\theta=0,\pi$.  We now can write
\eq
A = A_{\theta}(\theta,\psi)d\theta + A_{\psi}(\theta,\psi)d\psi + A_{\theta_{1}}(\theta,\psi)d\theta_{1}+ 
A_{\theta_{2}}(\theta,\psi)d\theta_{2}
\,.
\en

\subsubsection{Reduction from \texorpdfstring{$SU(2)$ to $U(1)$ by a
$\Integers_{2}$ symmetry at $\beta_{1}=\beta_{2}=\frac\pi2$}{SU(2) to 
U(1) by a Z2 symmetry at beta1=beta2=pi/2}}

The extra $\Integers_{2}$ symmetry at $\beta_{1}=\beta_{2}=\frac\pi2$,
\eq
\Phi_{2}\left (\frac\pi2,\frac\pi2,\mu_{2} g\mu_{2}^{-1} \right )
= \mu_{2} \Phi_{2}\left (\frac\pi2,\frac\pi2,g \right )\mu_{2}^{-1} \\
\en
becomes, on the slice,
\eq
A(\theta,\psi,-\theta_{1},-\theta_{2}) = \mu_{2} A(\theta,\psi,\theta_{1},\theta_{2}) \mu_{2}^{-1}
\en
because
\eq
\mu_{2} g(\psi) \mu_{2}^{-1} = g(\psi)
\,,\qquad
\mu_{2} h_{\frac12(-\theta_{1}+\theta_{2})}  \mu_{2}^{-1} = 
h_{\frac12(\theta_{1}-\theta_{2})} 
\,,
\en
and, at the enhanced symmetry point,
\ateq{3}{
\tilde \Phi_{-}
&=
g(\psi) h_{\frac\pi2} g(\psi)^{-1} h_{\frac\pi2}
&
&= e^{-\frac12 \psi \mu_{2}}\mu_{3}e^{\frac12 \psi \mu_{2}}
\mu_{3} 
&&= - e^{-\psi \mu_{2}}
\\
\tilde \Phi_{+}
&=
h_{\frac\pi2}g(\psi) h_{\frac\pi2} g(\psi)^{-1}
&
&= \mu_{3}e^{-\frac12 \psi \mu_{2}}e^{+\frac12 \psi \mu_{2}}
\mu_{3} 
&&= - e^{\psi \mu_{2}}
\,.
}
The connection on the slice therefore takes the form
\eq
A = a_{\theta} \mu_{2} d\theta + 
a_{\psi} \mu_{2} d\psi
+ (v_{1}\mu_{+}-\bar v_{1}\mu_{+}^{\dagger}) d\theta_{1}
+ (v_{2}\mu_{+}-\bar v_{2}\mu_{+}^{\dagger}) d\theta_{2}
\en
where
\eq
\mu_{+} = \frac12(\mu_{3}+i\mu_{1})
\,,\qquad
\mu_{2}\mu_{+}\mu_{2}^{-1} = -\mu_{+}
\,,\qquad
[\mu_{2},\,\mu_{+}] = 2i\mu_{+}
\en
and where the components $a_{\theta}$, $a_{\psi}$, and $v_{1,2}$ are 
functions only of $\theta$ and $\psi$.
Thus the invariant $SU(2)$ connection reduces to a $U(1)$ connection
on the slice, plus the two additional fields $v_{1,2}$.

We write the $U(1)$ connection as
\eq
D_{r} = d + iA_{r}
=d\theta D_{r,\theta}+d\psi D_{r,\psi}\,,\qquad 
A_{r}=a_{\theta}d\theta + a_{\psi}d\psi
\,.
\en
Its curvature 2-form is
\eq
F_{r} = dA_{r}= F_{r,\theta\psi}d\theta d\psi
\,,\qquad F_{r,\theta\psi} = \partial_{\theta}a_{\psi}- 
\partial_{\psi}a_{\theta}
\en

\subsubsection{\texorpdfstring{$S_{\YM}$}{S-YM}}
The curvature 2-form of $D$ is
\eq
F = 
\left [F_{r}+i(v_{1}\bar v_{2}-\bar v_{1} v_{2})d\theta_{1}d\theta_{2} \right ]\mu_{2}
+
\left ( D_{r}v_{1}d\theta_{1} + D_{r}v_{2}d\theta_{2}\right ) \mu_{+}
-
\left ( \overline{D_{r}v_{1}}d\theta_{1} + \overline{D_{r}v_{2}}d\theta_{2}\right )\mu_{+}^{\dagger}
\en
where the covariant derivatives of the fields $v_{1,2}$ are given by
\eq
D_{r} v_{k} = (d + 2iA_{r})v_{k}
\,,\qquad
D_{r,\theta} v_{k} = (\partial_{\theta} + 2ia_{\theta})v_{k}
\,,\quad
D_{r,\psi} v_{k} = (\partial_{\psi} + 2ia_{\psi})v_{k}
\,.
\en
The round metric on $S^{4}$ is
\eq
(ds)_{S^{4}}^{2}=  (d\theta)^{2} + (\sin \theta)^{2}\frac14 (d\psi)^{2}
+ (\sin \theta)^{2} \frac12(1+\cos\psi)(d\theta_{1})^{2}
+ (\sin \theta)^{2} \frac12(1-\cos\psi)(d\theta_{2})^{2}
\,.
\en
The volume element of $S^{4}$ reduced to the 2-dimensional domain is
\eq
\frac1{8\pi^{2}}\int_{0}^{2\pi}\int_{0}^{2\pi}d\theta_{1}d\theta_{2}
\;\mathrm{dvol}_{S^{4}} =
\frac18 d\theta d\psi \,  (\sin\theta)^{3}\sin\psi
\,.
\en
The Yang-Mills action is most neatly written in terms of a certain metric on the 2-dimensional 
domain
\eq
(ds)_{2}^{2}  =  (2 \sin\psi)^{-1} \left [ 4(\sin\theta)^{-2}(d\theta)^{2} + (d\psi)^{2} 
\right ]
=  (2 \sin\psi)^{-1} [4(dx)^{2} + (d\psi)^{2}]
\en
whose area element is
\eq
\mathrm{dvol}_{2}= d\theta d\psi \, (\sin\theta \sin\psi)^{-1}  = dx d\psi \, (\sin\psi)^{-1}
\en
and in terms of hermitian forms on the line bundles
\eq
\lVert v_{1}\rVert_{2}^{2} = h_{\bar 11}|v_{1}|^{2}
\,,\qquad
\lVert v_{2}\rVert_{2}^{2} = h_{\bar 22}|v_{2}|^{2}
\,,\qquad
h_{\bar 11} = \tan\frac12\psi \,,\qquad
h_{\bar 22} = \cot\frac12\psi = h^{1\bar 1}
\,.
\en
That is, $v_{1}$ is a section of a line bundle $L_{1}$
with hermitian form $h$,
and $v_{2}$ is a section of $L_{2}= \bar L_{1}^{-1}$.
Then $S_{\YM}$ is given by the covariant formula
\eq
S_{\YM} = 
\int \mathrm{dvol}_{2}  \;\left (
\frac12 \lVert F \rVert_{2}^{2}
+ \lVert v_{1}\bar v_{2}-\bar v_{1} v_{2}\rVert_{2}^{2}
+  \lVert D_{r}v_{1} \rVert_{2}^{2}
+  \lVert D_{r}v_{2} \rVert_{2}^{2}
\right )
\en
where
\eq
\lVert F \rVert_{2}^{2}
=
F_{r,ab}F_{r}^{ab}
\,,\quad
\lVert v_{1}\bar v_{2}-\bar v_{1} v_{2}\rVert_{2}^{2}
=
h_{\bar 11}h_{\bar 22}|v_{1}\bar v_{2}-\bar v_{1} v_{2}|^{2}
\,,\quad
\lVert D_{r}v_{k} \rVert_{2}^{2}
=
h_{\bar kk} \overline{D_{r,a}v_{k}} D_{r}^{a}v_{k}
\,.
\en

\subsubsection{Discrete symmetries and boundary conditions}
The remaining $\Integers_{2}\times \Integers_{2}$ symmetries are:
\ateq{2}{
a_{\theta}(\pi-\theta,\psi) &= a_{\theta}(\theta,\psi)
\qquad
&
v_{1}(\pi-\theta,\psi) &= \bar v_{1}(\theta,\psi)
\\
a_{\psi}(\pi-\theta,\psi) &= -a_{\psi}(\theta,\psi)
&v_{2}(\pi-\theta,\psi) &= \bar v_{2}(\theta,\psi)
}
\ateq{2}{
a_{\theta}(\theta,\pi-\psi) &= -a_{\theta}(\theta,\psi)
\qquad
&v_{1}(\theta,\pi-\psi) &= -\bar v_{2}(\theta,\psi)
\\
a_{\psi}(\theta,\pi-\psi) &= a_{\psi}(\theta,\psi)
&v_{2}(\theta,\pi-\psi) &= -\bar v_{1}(\theta,\psi)
}
In addition, if we regard the connection as a function of $\psi\in 
\Reals$, we have
\ateq{2}{
a_{\theta}(\theta,-\psi)& = -a_{\theta}(\theta,\psi)
\qquad
&v_{1}(\theta,-\psi) &= \bar v_{1}(\theta,\psi)
\\
a_{\psi}(\theta,-\psi) &= a_{\psi}(\theta,\psi)
&v_{2}(\theta,-\psi)& = \bar v_{2}(\theta,\psi)
\,.
}
Combined with the $\psi\rightarrow \pi-\psi$ symmetry,
this gives
\ateq{2}{
a_{\theta}(\theta,\psi+\pi) &= a_{\theta}(\theta,\psi)
\qquad
&
v_{1}(\theta,\psi+\pi) &= -v_{2}(\theta,\psi)
\\
a_{\psi}(\theta,\psi+\pi) &= a_{\psi}(\theta,\psi)
&v_{2}(\theta,\psi+\pi) &= - v_{1}(\theta,\psi)
}
so the $U(1)$ connection $D_{r}$ lives on the 2-sphere
parametrized by $\theta,\psi$ with the identification $\psi \sim \psi 
+ \pi$.
The fields $v_{1,2}$ live on a 2-sheeted covering of this 2-sphere.

The  boundary conditions at $\theta=0,\pi$ are:
\ateq{4}{
a_{\theta}(0,\psi)&= 0
\,,\qquad
&a_{\psi}(0,\psi) &= 1
\,,\qquad
&v_{1}(0,\psi) &= \frac12e^{-2i\psi}
\,,\qquad
&v_{2}(0,\psi) &= - \frac12e^{-2i\psi}
\,,
\\
a_{\theta}(\pi,\psi)&= 0
&a_{\psi}(\pi,\psi) &= -1
&v_{1}(\pi,\psi) &= \frac12e^{2i\psi}
&v_{2}(\pi,\psi) &= -\frac12e^{2i\psi}
\,.
}
At $\psi=0$ and at $\psi=\pi$,
the $U(1)\times U(1)/\{\pm 1\}$ continuous symmetry
degenerates to  $U(1)$,
giving rise to boundary conditions at $\psi=0,\pi$
\ateq{4}{
a_{\theta}(\theta,0) &= 0
\,,\qquad
&v_{1}(\theta,0) &= \bar v_{1}(\theta,0)
\,,\qquad
&v_{2}(\theta,0) &=-\frac12
\,,\qquad
&D_{r,\psi}v_{2}(\theta,0) &= 0
\,,
\\
a_{\theta}(\theta,\pi) &= 0
&v_{2}(\theta,\pi) &= \bar v_{2}(\theta,\pi)
&v_{1}(\theta,\pi) &=\frac12
&D_{r,\psi}v_{1}(\theta,\pi) &= 0
\,.
}
Because of the degeneration of the continuous symmetry group,
the $U(1)$ gauge transformations must act trivially at $\psi=0,\pi$.
The boundary conditions on $v_{1,2}$ at $\psi=0,\pi$ are gauge 
invariant for this restricted group of gauge transformations.

Over the 2-sphere $\psi\sim \psi+\pi$,
\eq
\frac1{2\pi}\int 2 F_{r} = -2
\en
so the fields $v_{1,2}$ live in the $U(1)$ bundle of Chern number $-2$.

\subsubsection{Numerical computations}
\label{sect:SU2numerics1}

We try to minimize $S_{\YM}$ numerically in this two dimensional 
setting by the same technique as in the previous 
one dimensional problems, approximating the space of connections by 
increasing finite dimensional affine subspaces of 
polynomial connections.
We let the fields be polynomials of finite degree,
whose coefficients are real variables.
If there are $N$ of these real variables,
we are approximating the space of connections by an affine subspace 
of dimension $N$.
We use mathematical software \cite{sage} to evaluate
$S_{\YM}$ as a quartic polynomial in these $N$ real 
variables, and then to minimize it.

First, we design the polynomial approximation so that the 
evaluation of $S_{\YM}$ requires only multiplication of polynomials
(to conserve computational resources).
We use as coordinates
\eq
t=\cos \theta
\,,\qquad
s=\cos \psi
\,,
\en
and write
\eq
a_{\theta} = Q_{\theta}(t,s) \sin \theta \sin\psi
\,,\qquad
a_{\psi}  = t + (1-t^{2}) Q_{\psi}(t,s)
\en
\eq
v_{1}= v_{11} + i \tilde v_{12}\sin \psi
\qquad
v_{2}= v_{21} + i  \tilde v_{22}\sin \psi
\en
\aeq{
v_{11} &=-\frac12 + s^{2}+ (1-t^{2})(1+s) P_{11}(t,s)
\\
v_{21} &= \frac12 -s^{2} -(1-t^{2})(1-s) P_{11}(-t,-s)
\\
\tilde v_{12} &= - t s +(1-t^{2}) P_{12}(t,s)
\\
\tilde v_{22} &= t s -(1-t^{2}) P_{12}(-t,-s)
}
where $Q_{\theta}$, $Q_{\psi}$, $P_{11}$, and $P_{12}$ are 
polynomials in $t$ and $s$ obeying the 
symmetry conditions
\eq
Q_{t}(t,s) = -Q_{t}(t,-s) = Q_{t}(-t,s)
\en
\eq
Q_{\psi}(t,s) = Q_{\psi}(t,-s) = -Q_{\psi}(-t,s)
\en
\eq
P_{11}(t,s) = P_{11}(-t,s)
\,,\qquad
P_{12}(t,s) = -P_{12}(-t,s)
\,.
\en
All of the discrete symmetries are automatically satisfied,
as are all of the boundary conditions except the boundary conditions
on $D_{r,\psi}v_{1,2}$.
These last conditions are solved by
\eq
P_{12}(t,s) = Q_{\psi}(t,s) + (1+s) Q_{v}(t,-s)
\en
where $Q_{v}$ is a new polynomial with symmetry
\eq
Q_{v}(t,s) = -Q_{v}(-t,s)
\,.
\en
The connection is now specified by the polynomials
$P_{11}$, $Q_{\theta}$, $Q_{\psi}$, and $Q_{v}$,
which obey the various symmetries written above,
but are otherwise arbitrary.

For simplicity in the computer program, each of the four polynomials is written so as to 
contain the first $n_{t}$ powers 
of $t$ and the first $n_{s}$ powers of $s$ consistent with the 
symmetries, so each polynomial contains $n_{t}n_{s}$ 
coefficients, so the total number of coefficients is $N=4n_{t}n_{u}$.
We are approximating the space of connections by an affine subspace 
of dimension $N$.

Numerical results are shown in
Table~\ref{table:pi5su2-1}.
\begin{table}[t]
\centering
\begin{tabular}{cccl}
$n_{t}$ & $n_{s}$ & $N$ & $\min(S_{\mathit{YM}})$ \\[0.5ex]
2 &  2 & 16 & 5.23\\
3 &  3 & 36 & 4.91\\
4 &  4 & 64 & 4.73\\
5 &  5 & 100& 4.60\\
7 &  3 & 84& 4.48
\end{tabular}
\caption{Numerical minimization of $S_{\YM}$ for
the reduced 2-dimensional $U(1)$ system.}
\label{table:pi5su2-1}
\end{table}
They were obtained using the Sage mathematics software \cite{sage}.
The calculations became too time-consuming for $N>100$.
Nothing is especially suggested by the values of 
$\min(S_{\mathit{YM}})$,
besides insufficiency of the computing resources.

\begin{figure}
\centering
\includegraphics[width=3.5in]{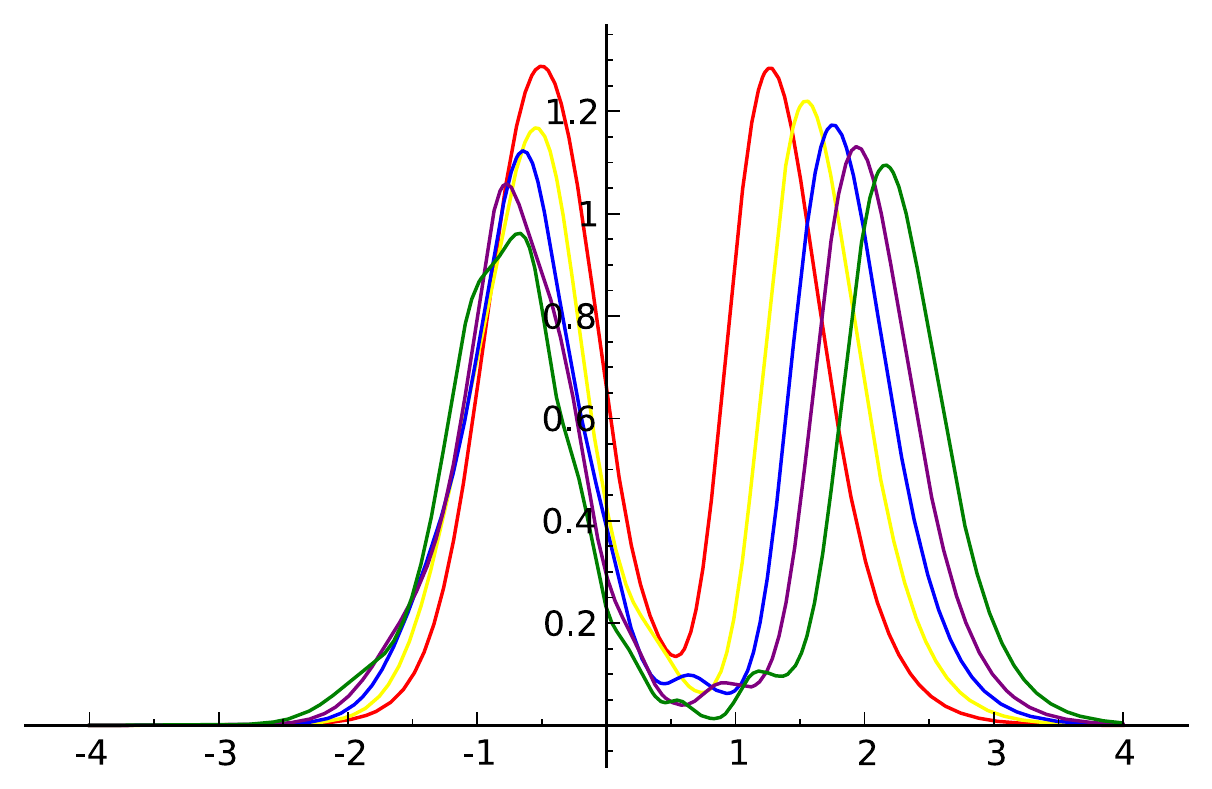}
\caption{Plots of $L_{+}(x)$ for the connections numerically 
minimizing $S_{\YM}$.
The bumps at positive $x$ move away from 
the origin as $\min(S_{\mathit{YM}})$ decreases (becomes a better 
upper bound).
The graphs of $L_{-}(x)$ are given by reflecting $x\rightarrow -x$.}
\label{fig:pi5su2Lplus}
\end{figure}

Slightly more suggestive are the graphs of the chiral action density
\eq
\frac1{8\pi^{2}} 
\tr\left (-F\frac12 (1+{*})F\right ) = dt ds \, L_{+}(t,s)
\en
or rather, of its projection onto the $t$ or $x$ coordinate
\eq
dx\, L_{+}(x) = dt \int_{-1}^{1} ds \, L_{+}(t,s)
\,.
\en
Recall that
\eq
x= \ln \tan \frac12 \theta 
\,,\qquad
t = \cos \theta = - \tanh x
\,.
\en
The graphs are shown in Figure~\ref{fig:pi5su2Lplus}.
There appears to be a separation into four lumps of alternating 
topological charges (the two lumps of positive charge are shown in 
the graphs), though there is no indication that the 
topological charges are quantized.
Still, we can guess that the outer lumps will travel to $x=\pm 
\infty$, the lump going to $x=\infty$ resolving into a zero-size 
instanton and the lump going to $x=-\infty$ resolving into a
zero-size anti-instanton,
the argument being that there seems to be nothing to stop this happening.
A better method of approximation is needed
that could give more convincing numerical evidence
in support of this extrapolation.

\subsubsection{Assume zero-size (anti-)instantons at the poles}

We now assume that a zero-size instanton has gone to the south pole and a zero-size 
instanton to the north pole.  These provide new boundary conditions 
at $\theta=0,\pi$ for the connection away from the poles.  We
minimize $S_{\YM}$ numerically using the new boundary conditions.

An instanton of size $r_{+}=e^{-x_{+}}$ at the south pole is the 
connection
\eq
D_{+} = d + (1-f_{+}) \omega
\,,\qquad
\omega = g d(g^{-1})
\,,\qquad
f_{+}(x) = \frac1{1-e^{-2(x-x_{+})}}
\,.
\en
In the limit of zero size, the instanton becomes
\eq
D_{+} = d +\omega = g d g^{-1}
\en
which will provide the new boundary condition at the south pole,
$\theta = \pi$, $x=\infty$.

Going to the slice $\theta_{1,2}=0$, the zero-size instanton becomes
\aeq{
\tilde D_{+} &= h_{\frac12(-\theta_{1}+\theta_{2})}
D_{+} h_{\frac12(-\theta_{1}+\theta_{2})}^{-1}
\\
&= 
g(\psi) h_{\frac12(-\theta_{1}-\theta_{2})}^{-1} d \,
h_{\frac12(-\theta_{1}-\theta_{2})}
g(\psi)^{-1} 
\\
&=
g(\psi) \left [ d - \frac12 (d\theta_{1}+d\theta_{2})\mu_{3} \right ]
g(\psi)^{-1}
\\
&=
e^{-\frac12 \psi \mu_{2}} \left [ d - \frac12 (d\theta_{1}+d\theta_{2})\mu_{3} \right ]
e^{\frac12 \psi \mu_{2}}
\\[2ex]
D &= \tilde \Phi_{+} \tilde D_{+} \tilde \Phi_{+}^{-1}
\\
&= e^{\psi \mu_{2}} \tilde D_{+} e^{-\psi \mu_{2}}
\\
&= e^{\frac12 \psi \mu_{2}} \left [ d - \frac12 (d\theta_{1}+d\theta_{2})\mu_{3} \right ]
e^{-\frac12 \psi \mu_{2}}
\\
&=
d -\frac12 d\psi \mu_{2} - \frac12 
(d\theta_{1}+d\theta_{2})(e^{i\psi}\mu_{+}-e^{-i\psi}\mu_{+}^{\dagger})
\,.
}
At the north pole we put the reflected connection, respecting the 
$\theta\rightarrow \pi-\theta$ symmetry:
\eq
D = d +\frac12 d\psi \mu_{2} - \frac12 
(d\theta_{1}+d\theta_{2})(e^{-i\psi}\mu_{+}-e^{i\psi}\mu_{+}^{\dagger})
\,.
\en
Working backwards,
\aeq{
D &=  e^{-\frac12 \psi \mu_{2}}
\left [ d - \frac12 (d\theta_{1}+d\theta_{2})\mu_{3} \right ]
e^{\frac12 \psi \mu_{2}}
\\
&= \tilde \Phi_{-}
\mu_{3}
g(\psi)
\mu_{3}^{-1}
\left [ d - \frac12 (d\theta_{1}+d\theta_{2})\mu_{3} \right ]
\mu_{3}
g(\psi)^{-1}
\mu_{3}^{-1}
\tilde \Phi_{-}^{-1}
}
so the zero-size anti-instanton is given by
\eq
D_{-} = \mu_{3} (g d g^{-1}) \mu_{3}^{-1}
\en
which is the maximally twisted zero-size anti-instanton.

The new boundary conditions at $\theta=0,\pi$ are:
\ateq{4}{
a_{\theta}(0,\psi)&= 0
\,,\qquad
&a_{\psi}(0,\psi) &= \frac12
\,,\qquad
&v_{1}(0,\psi) &= -\frac12e^{-i\psi}
\,,\qquad
&v_{2}(0,\psi) &= - \frac12e^{-i\psi}
\,,
\\
a_{\theta}(\pi,\psi)&= 0
&a_{\psi}(\pi,\psi) &= -\frac12
&v_{1}(\pi,\psi) &= -\frac12e^{i\psi}
&v_{2}(\pi,\psi) &= -\frac12e^{i\psi}
\,.
}
The fields $v_{1,2}$ now live in the $U(1)$ bundle of Chern number 
$-1$.

\subsubsection{Numerical calculations with the new boundary conditions}

\begin{figure}
\centering
\includegraphics[width=3.5in]{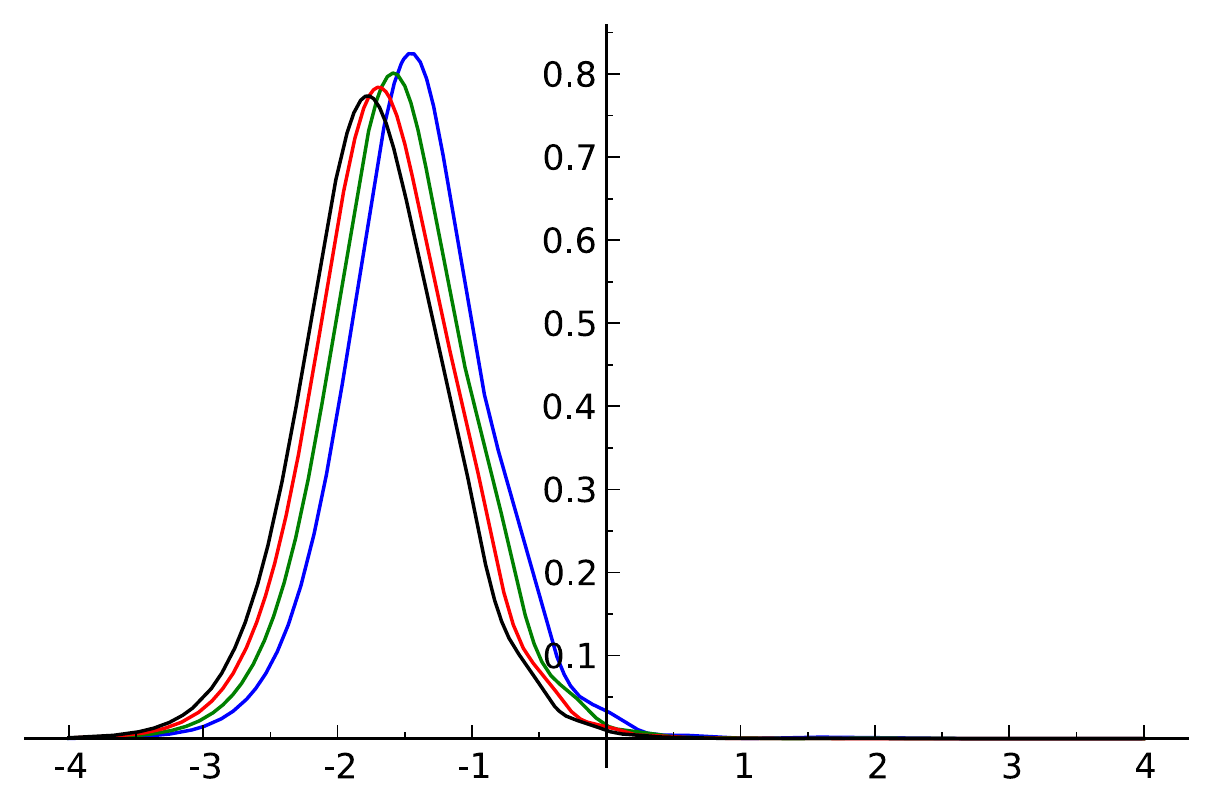}
\caption{Plots of $L_{+}(x)$ for the connections numerically 
minimizing $S_{\YM}$, with a zero-size instanton at the south pole 
and a zero-size anti-instanton at the north pole.
The bumps move away from 
the origin as $\min(S_{\mathit{YM}})$ decreases.
The graphs of $L_{-}(x)$ are given by the reflection $x\rightarrow -x$.}
\label{fig:pi5su2Lplusinst}
\end{figure}

We use the same technique to minimize $S_{\YM}$ with the zero-size 
instanton at the south pole and the zero-size anti-instanton at the 
north pole, using the same Sage program, making only the changes needed to implement the new 
boundary conditions.  The numerical results are shown in 
Table~\ref{table:pi5su2-inst}.
\begin{table}[t]
\centering
\begin{tabular}{cccl}
$n_{t}$ & $n_{s}$ & $N$ & $\min(S_{\mathit{YM}})$ \\[0.5ex]
4 &  4 & 64 & 2.0174\\
5 &  5 & 100& 2.0109\\
6 &  3 &  72& 2.0073\\
7 &  3 &  84& 2.0053
\end{tabular}
\caption{Numerical minimization of $S_{\YM}$ for
the reduced 2-dimensional $U(1)$ system,
with a zero-size instanton at the south pole
and a zero-size anti-instanton at the north pole.
$S_{\YM}$ here does not include the contribution of $2$ units from the instantons.
$N$ is the dimension of the affine subspace of connections on which
$S_{\YM}$ is minimized.}
\label{table:pi5su2-inst}
\end{table}
Counting the two units of action from the instantons,
we now have an upper bound $\min(S_{\mathit{YM}})<4.0053$.
Graphs of $L_{+}(x)$ are shown in Figure~\ref{fig:pi5su2Lplusinst}.
It seems clear that an instanton is moving towards $x=\infty$ and
an anti-instanton towards $x=-\infty$.

Given a widely separated sequence of instantons, $\bar I I \bar I I$,
the symmetry conditions at the enhanced symmetry point
force all three of the neighboring pairs to be maximally twisted.

The zero-size limit of such a sequence, $\bar I I \bar I I$, of 
instantons and anti-instantons, widely separated 
in $x$, seems a plausible candidate for the enhanced symmetry fixed 
point connection with effective index $2$.
It seems at least worth trying to check by calculating the Yang-Mills flow on the slow 
manifold.

It is worrisome that the first numerical minimization did not get
closer to $\min(S_{\mathit{YM}})=4$.  We would naively expect the
sequence of four separated (anti-)instantons to appear quickly,
leaving only the sizes as slow modes.  Perhaps there is a competing
process.  Or perhaps there is an error in the computer program.  Most
likely, the polynomials are not of high enough degree in $t$ to
sufficiently resolve the region near $t=\pm 1$.

There is no compelling evidence from the numerical calculations
that $\min(S_{\mathit{YM}})=4$ at the enhanced 
symmetry point.  There could still be a smooth fixed point with 
$S_{\mathit{YM}}<4$, or a hybrid connection containing a zero-size 
instanton and a zero-size anti-instanton plus a smooth part,
with total action $2< S_{\mathit{YM}}<4$.
We might note that this could not be 
the Sibner-Sibner-Uhlenbeck \cite{Sibner1989} solution of the Yang-Mills equation with $S_{\YM}<4$.
Their fixed point must have
an unstable manifold of dimension $\ge 3$.
The Sibner-Sibner-Uhlenbeck construction presupposes a certain $U(1)$ 
symmetry, and produces a connection with a 1 dimensional unstable 
manifold in the space of $U(1)$-invariant connections.
Any other unstable directions would have to come in doublets of the 
$U(1)$ symmetry group (two dimensional 
real representations).
So there is no possibility of a two dimensional unstable manifold.
A smooth fixed-point must have at least two unstable directions \cite{Taubes1983},
so the Sibner-Sibner-Uhlenbeck connection with $S_{\YM}<4$ must have 
at least 3 unstable directions.

\subsubsection{Improved numerical results}

The first calculation, described in 
Section~\ref{sect:SU2numerics1} above,
can be re-done with improved resolution near the poles by a trivial 
modification, simply rescaling 
$x\rightarrow x/x_{1}$, redefining $t=-\tanh(x/x_{1})$,
for appropriate values of $x_{1}$ that are determined empirically.
The only change to the computer program is a rescaling of
each term in the Yang-Mills action by a power of $x_{1}$.

\begin{figure}[ht]
\centering
\includegraphics[width=3.5in]{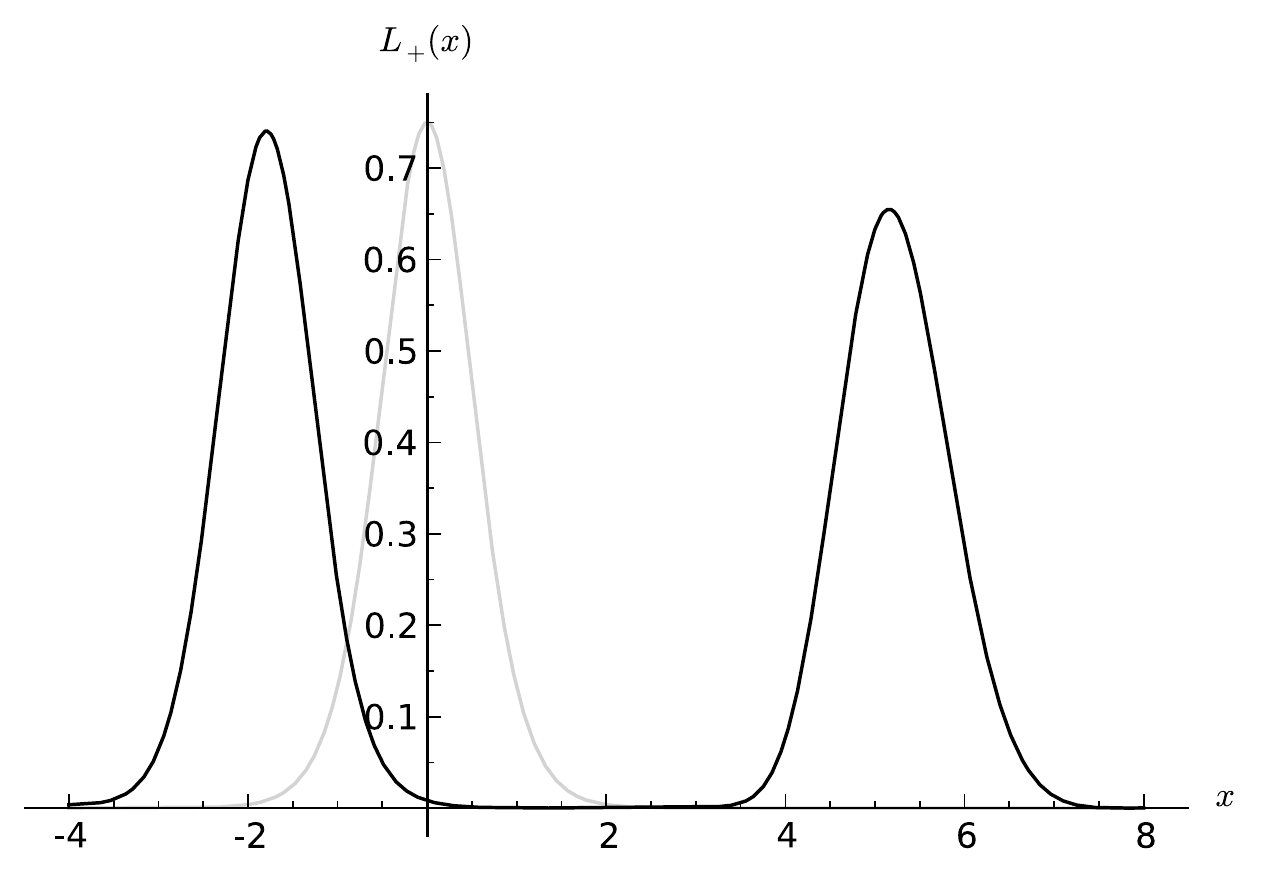}
\caption{$L_{+}(x)$ for the connection
minimizing $S_{\YM}$ at $4.04$ with the improved resolution,
the last run in Table~\ref{table:pi5su2-improved}.
For comparison,
the curve centered at $x=0$ is $L_{+}(x)$ for an instanton.}
\label{fig:pi5su2Lplus-improved}
\end{figure}

\begin{figure}[ht]
\centering
\includegraphics[width=3.5in]{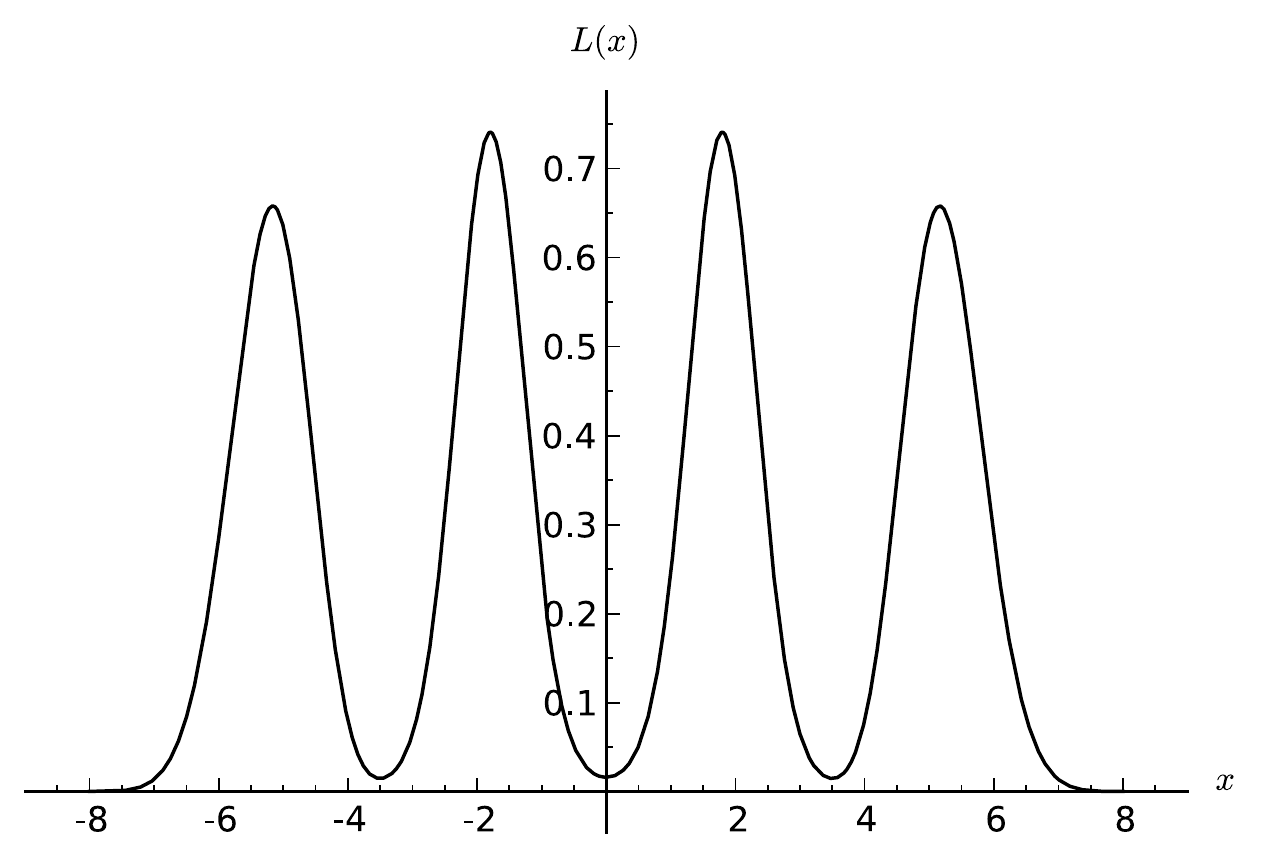}
\caption{$L(x)$ for the connection
minimizing $S_{\YM}$ at $4.04$ with the improved resolution,
the last run in Table~\ref{table:pi5su2-improved}.
The separation into four small (anti-)instantons is more apparent.}
\label{fig:pi5su2L-improved}
\end{figure}

Some results are shown in Table~\ref{table:pi5su2-improved}.
\begin{table}[t]
\centering
\begin{tabular}{ccccl}
$n_{t}$ & $n_{s}$ & $N$ & $x_{1}$ & $\min(S_{\mathit{YM}})$ \\[0.5ex]
3 &  5 & 60 & 3.0 & 4.34\\
4 &  2 & 32 & 3.0 & 4.17\\
5 &  2 & 40 & 3.5 & 4.13\\
5 &  5 & 100 & 4.0 & 4.13\\
6 &  2 & 48 & 4.5 & 4.08\\
8 &  2 & 64 & 3.5 & 4.05\\
10 &  2 & 80 & 4.0 & 4.04\\
12 &  2 & 96 & 4.0 & 4.04
\end{tabular}
\caption{Numerical minimization of $S_{\YM}$ with
coordinate re-scaling $x\rightarrow x/x_{1}$,
choosing $x_{1}$ to obtain the best minimum (roughly).
}
\label{table:pi5su2-improved}
\end{table}
The evidence for a local minimum at $S_{\YM}=4$ is much better.
Figure~\ref{fig:pi5su2Lplus-improved}
shows the chiral action density $L_{+}(x)$.
Figure~\ref{fig:pi5su2L-improved}
shows the full action density $L(x)=L_{+}(x)+L_{-}(x)$.
The evidence is stronger for separation into a quadruplet of 
zero-size (anti-)instantons at the minimum.

\subsubsection{A 2-torus family of twisted quadruplets}

We write an explicit 2-parameter family of twisted quadruplets of zero-size 
(anti-)instantons living in the 2-parameter family of bundles 
constructed in Section~\ref{sect:twospherefamily} above,
parametrized by $(\beta_{1},\beta_{2})\in [0,\pi]^{2}$.
Although the family of bundles forms a 2-sphere,
the boundary of the square being identified to a point,
the 2-parameter family of connections forms a 2-torus,
opposite sides of the square being identified
with each other.

Write the basic instanton as
\eq
D_{I}(x-x_{+}) = d + \omega - f_{+} \omega
\,,\qquad
f_{+} = \frac1{1+e^{-2(x-x_{+})}}
\,.
\en
The basic anti-instanton is
\eq
D_{\bar I}(x-x_{-}) = D_{I}(-x+x_{-})
\,.
\en
The twisted quadruplet is constructed
from an instanton $D_{I}(x-x_{+})$,
an anti-instanton $D_{\bar I}(x-x_{-})$,
and their reflections under $x\rightarrow -x$,
the instanton $D_{I}(x+x_{-})=D_{\bar I}(-x-x_{-})$,
and the anti-instanton $D_{\bar I}(x+x_{+})=D_{I}(-x-x_{+})$,
in the limit
\eq
x_{-}\rightarrow \infty
\,,\qquad
x_{+}-x_{-}\rightarrow \infty
\en
twisted as follows,
\eq
D(\beta_{1},\beta_{2}) = \left \{
\begin{array}{crrll}
\Phi_{-}h_{\beta_{1}}^{-1}
D_{\bar I}(x + x_{+})
h_{\beta_{1}}\Phi_{-}^{-1}
&\qquad & &x& < -\frac12 (x_{+}+x_{-})
\\
D_{I}(x+x_{-})
&& -\frac12 (x_{+}+x_{-}) < & x& < 0
\\
h_{\beta_{1}} D_{\bar I}(x-x_{-}) h_{\beta_{1}}^{-1}
&& 0 < &x& < \frac12 (x_{+}+x_{-})
\\
\Phi_{+} D_{I}(x-x_{+}) \Phi_{+}^{-1}
&& \frac12 (x_{+}+x_{-})< &x&
\end{array}
\right .
\en
Recall that the patching map for the bundle is
\eq
\Phi_{2} (\beta_{1},\beta_{2}) = \Phi_{-}^{-1}\Phi_{+}
\,,\qquad
\Phi_{+} = h_{\beta_{1}} g h_{\beta_{2}} g^{-1}
\,,\qquad
\Phi_{-}= g h_{\beta_{2}} g^{-1} h_{\beta_{1}}
\,.
\en
In the limit,
the instantons and anti-instantons
agree at the junctions,
taking the values
\eq
D = \left \{
\begin{array}{crll}
g d g^{-1}&\qquad & x = -\frac12 (x_{+}+x_{-})
\\
d
&&  x = 0
\\
h_{\beta_{1}} g d g^{-1} h_{\beta_{1}}^{-1}
&& x = \frac12 (x_{+}+x_{-})
\end{array}
\right .
\en
At the north pole, $x=-\infty$,
\eq
D = \Phi_{-}h_{\beta_{1}}^{-1}
d
h_{\beta_{1}}\Phi_{-}^{-1}
=
\Phi_{-}
d
\Phi_{-}^{-1}
\en
and at the south pole, $x=\infty$,
\eq
D = \Phi_{+} d \Phi_{+}^{-1}
\en
so $D$ satisfies the boundary conditions defining the bundle.
Alternatively, we have connections on the two hemispheres, 
nonsingular at the poles,
\eq
D_{-} = \left \{
\begin{array}{crrll}
h_{\beta_{1}}^{-1}
D_{\bar I}(x + x_{+})
h_{\beta_{1}}
&\qquad & &x& < -\frac12 (x_{+}+x_{-})
\\
\Phi_{-}^{-1}D_{I}(x+x_{-})\Phi_{-}
&& -\frac12 (x_{+}+x_{-}) < & x& < 0
\end{array}
\right .
\en
\eq
D_{+} = \left \{
\begin{array}{crrll}
 \Phi_{+}^{-1}h_{\beta_{1}} D_{\bar I}(x-x_{-}) h_{\beta_{1}}^{-1}\Phi_{+}
&& 0 < &x& < \frac12 (x_{+}+x_{-})
\\
 D_{I}(x-x_{+})
&& \frac12 (x_{+}+x_{-})< &x&
\end{array}
\right .
\en
whose values at $x=0$ are
\eq
D_{-} = \Phi_{-}^{-1}d\Phi_{-}
\,,\qquad
D_{+} =  \Phi_{+}^{-1}h_{\beta_{1}}d h_{\beta_{1}}^{-1}\Phi_{+}
= \Phi_{+}^{-1} d \Phi_{+}
\en
which are related by the patching map defining the bundle
\eq
D_{-} =\Phi_{2} (\beta_{1},\beta_{2}) D_{+} \Phi_{2} 
(\beta_{1},\beta_{2})^{-1}
\,.
\en
So the connection $D (\beta_{1},\beta_{2})$ lives in the bundle 
defined by the patching map $\Phi_{2} (\beta_{1},\beta_{2})$.
Moreover, it can be checked that all the symmetry conditions of the family of bundles are satisfied 
by the family of connections $D (\beta_{1},\beta_{2})$.

On the boundary of the square, 
$\beta_{1,2}=0,\pi$, the patching map is trivial, $\Phi_{2} 
(\beta_{1},\beta_{2})=1$,
but the connections $D (\beta_{1},\beta_{2})$ are not all gauge 
equivalent on the boundary.
At $\beta_{1}=0,\pi$,
\eq
D_{\mp} = \left \{
\begin{array}{crrll}
D_{\bar I}(x + x_{+})
&\qquad & &x& < -\frac12 (x_{+}+x_{-})
\\
\left ( g h_{\beta_{2}} g^{-1}\right )^{-1}D_{I}(x+x_{-})\left ( g h_{\beta_{2}} g^{-1}\right )
&& -\frac12 (x_{+}+x_{-}) < & x& < 0
\\
\left ( g h_{\beta_{2}} g^{-1}\right )^{-1} D_{\bar I}(x-x_{-})\left ( g h_{\beta_{2}} g^{-1}\right )
&& 0 < &x& < \frac12 (x_{+}+x_{-})
\\
 D_{I}(x-x_{+})
&& \frac12 (x_{+}+x_{-})< &x&
\end{array}
\right .
\en
At $\beta_{2}=0,\pi$,
\eq
D_{\mp} = \left \{
\begin{array}{crrll}
h_{\beta_{1}}^{-1}
D_{\bar I}(x + x_{+})
h_{\beta_{1}}
&\qquad & &x& < -\frac12 (x_{+}+x_{-})
\\
h_{\beta_{1}}^{-1}
D_{I}(x+x_{-})h_{\beta_{1}}
&& -\frac12 (x_{+}+x_{-}) < & x& < 0
\\
D_{\bar I}(x-x_{-})
&& 0 < &x& < \frac12 (x_{+}+x_{-})
\\
 D_{I}(x-x_{+})
&& \frac12 (x_{+}+x_{-})< &x&
\end{array}
\right .
\en
The boundary of the square cannot be identified to a single point to 
give a 2-sphere family of connections.
Rather,
the opposite sides of the square are identified, giving a 2-torus family of 
connections,
as might have been expected from the symmetry conditions on the family.

It remains to calculate the Y-M flow near this family of connections,
first to check that the connection at the enhanced symmetry point
$\beta_{1}=\beta_{2}=\frac\pi2$ has a two-dimensional unstable
manifold, then to trace the global shape of that 2-manifold.
We see two possibilities, depending on details of the flow on the 
slow manifold, yet to be calculated.
The 2-torus of twisted quadruplets could be connected to the flat 
connection by a single outgoing trajectory leaving from the 
distinguished point $(\beta_{1},\beta_{2})=(0,0)$
(identified with the other 3 corners of the square).
This outgoing trajectory in $\mathcal{A}/\mathcal{G}$ 
would lift to a 2-cylinder in $\mathcal{A}$,
each point on the trajectory in $\mathcal{A}/\mathcal{G}$ 
lifting to a nontrivial loop in the group of gauge transformations, 
$\mathcal{G}$.
In this scenario,
the Y-M flow would have a stable 2-torus of zero area which might,
in the lambda model,
produce non-canonical low energy interactions in $SU(2)$ gauge theory.
A second, less attractive, scenario would have outgoing trajectories leaving from each point on 
the boundary of the square, travelling to the flat connection.
The unstable manifold would form a 2-sphere stable under the 
Y-M flow, with non-zero area.

\section{Questions and comments}

\subsection{Does the outgoing trajectory end at the flat connection?}

The topology of the Y-M flow is the same 
whatever the locations of the twisted pair in $S^{4}$ and whatever 
the geometry on $S^{4}$.  There is always an outgoing trajectory from 
the aligned twisted pair.
Does that outgoing trajectory always end at the flat connection?

Euclidean $\Reals^{4}$ is the setting of interest
for the possible physics application (discussed in section~\ref{app:lambdamodel} below).
The twisted pair in euclidean $\Reals^{4}$ can be obtained as the limit
of twisted pairs in $S^{4}$ in which the instanton and anti-instanton 
are brought together
while the metric on $S^{4}$ is scaled so that the distance of 
separation remains constant.
Scaling the metric on $S^{4}$ is equivalent to scaling the Y-M flow time,
so the outgoing trajectory for twisted pairs in $\Reals^{4}$ is the 
same as the limiting trajectory for twisted pairs in $S^{4}$ as the 
instanton and anti-instanton approach each other.
Even if the outgoing trajectory ends at the flat connection for all 
twisted pairs in $S^{4}$, 
there would still 
remain
the possibility of cross-over to another fixed point as the 
locations of the instanton and anti-instanton approach each other,
which would govern the outgoing trajectory for twisted pairs in 
$\Reals^{4}$.
Does such a cross-over take place?  or does the outgoing trajectory 
end at the flat connection for twisted pairs in euclidean 
$\Reals^{4}$?

Whatever the fixed point at the end of the outgoing trajectory,
it will have $S_{\YM}<2$, so it cannot be singular.  If it is not the 
flat connection, then Taubes' theorem 
\cite{Taubes1983} says it must have at least a two 
dimensional unstable manifold.
It would seem extraordinary for the outgoing trajectory from the
twisted pairs to end exactly on a twice unstable fixed point.

\subsection{Asymptotic behavior of the outgoing trajectory?}

For the possible application to physics,
we would like to know
how the outgoing trajectory approaches the flat connection at large 
time, especially for twisted pairs in euclidean $\Reals^{4}$  (presuming the 
outgoing trajectory does end at the flat connection).
Near its end, the trajectory $A_{t}$ will approach the flat connection as a decaying perturbation
whose Fourier transform in $\Reals^{4}$ takes the form
\eq
\tilde A_{t}(p) = e^{-t p^{2}} \tilde A_{0}(p)
\,.
\en
The amplitudes $\tilde A_{0}(p)$ will depend on the locations and 
twist of the twisted pair and will presumably control the observable 
properties of the hypothetical physical states associated with the twisted pair.

We might explore for clues to the outgoing trajectories by looking at 
the outgoing trajectory for the $U(2)$-invariant twisted pair on
$S^{4}$, given by equation~\ref{eq:YMflowU2},
\eq
\frac{df}{dt} =\Round(x)^{-1}\left [ \partial_{x}^{2}f + 
4f(1-f)(2f-1)\right ]
 =(\cosh x)^{2}\left [ \partial_{x}^{2}f + 
4f(1-f)(2f-1)\right ]
\,.
\label{eq:outgo}
\en
The flow equation for the outgoing trajectory in the slow manifold
is given by equation~\ref{eq:slowflowA} at $\sigma=0$ or $\sigma=2\pi$,
\eq
\frac{d\rho}{dt} = 3 \rho^{3}
\,.
\en
It determines the asymptotic initial conditions
in the far past, $t\rightarrow -\infty$,
for the outgoing trajectory,
\eq
f_{t}(x) \rightarrow 
f_{+}(|x|)
\,,\qquad
f_{+}(x)=\frac1{1+e^{-2(x-x_{+}(t))}}
\,,\qquad
x_{+}(t) = \frac12 \ln(-6t) + O(t^{-1})
\,.
\en
If, instead of the round metric on $S^{4}$, we were to use the 
cylindrical metric,
$\Round(x)=1$, the outgoing trajectory
would be given by
\eq
\frac{df}{dt} =  \partial_{x}^{2}f + 4f(1-f)(2f-1)
\label{eq:nonlineardiff}
\en
which is a nonlinear diffusion 
or reaction-diffusion equation known as the Newell-Whitehead-Segel
equation, a special case of the 
Kolmogorov-Petrovsky-Piskounov/FitzHugh-Nagumo
equation.
It is apparently not integrable, but some exact solutions are known
(see for example \cite{Nucci199249}).  Given the asymptotic
$t\rightarrow -\infty$ conditions we need for the outgoing
trajectory,
the methods by which the exact
solutions were produced do not seem applicable \cite{VCalian2009}.
Still, the possibility of an exact solution
for the outgoing trajectory is tantalizing.

For twisted pairs in $\Reals^{4}$, the only symmetry of the outgoing trajectory is
the $SO(3)$ group of rotations around the 
axis that passes through the locations of the instanton and anti-instanton.
The unstable trajectory is given by a set of nonlinear 
diffusion equations in two spatial dimensions.
Numerical integration might be the only way to find its long time 
asymptotic behavior.

%

\subsection{The lambda model}
\label{app:lambdamodel}

The lambda model \cite{Friedan2003} is a two-dimensional nonlinear model
whose target space is the
manifold of spacetime fields: gauge fields, fermion fields, scalar 
fields, and the spacetime metric of general relativity.
The functional integral of the lambda model is
\eq
\int \mathcal{D}\lambda \; \exp\left [ {-\int d^{2}z\; \frac1{g^{2}} 
G_{ij}(\lambda) \partial\lambda^{i} \bar \partial \lambda^{j}} \right ]
\en
\eq
\mathcal{D}\lambda = \prod_{z,\bar z} d\rho(\lambda(z,\bar z))
\,.
\en
The field $\lambda(z,\bar z)$ maps a two-dimensional domain, 
parametrized by a complex coordinate $z$,
to the manifold of spacetime fields.
The $\lambda^{i}$ are coordinates on the manifold of spacetime fields (e.g., 
their momentum modes),
$G_{ij}(\lambda)d\lambda^{i}d\lambda^{j}$ is the natural metric on the manifold of spacetime 
fields,  $g$ is the coupling constant, and $d\rho(\lambda)$ is a 
measure on the spacetime fields, the 
\emph{a priori} measure of the nonlinear model.
The lambda model differs from the standard two dimensional nonlinear 
model in that the fields $\lambda^{i}(z,\bar z)$ are not precisely
dimensionless, but change with the two-dimensional scale $\Lambda$
according to the gradient flow
\eq
\Lambda \partialby{\Lambda} \lambda^{i} = - \nabla ^{i} 
 S(\lambda)
\,,\qquad
\nabla ^{i} S  = g^{2} G^{ij}\partialby{\lambda^{j}}
\left (\frac1{g^{2}} S \right )
\en
where $\frac1{g^{2}} S $ is the classical action functional 
of the spacetime field theory.  For small fluctuations, the dimension 
of the mode $\lambda^{i}$ is 
$p(i)^{2}$ where $p(i)$ is the spacetime momentum of 
the mode.

The \emph{a priori} measure is produced by the fluctuations of the 
lambda fields at short 2-d distances, acting in combination with
the gradient flow.
The mechanism of production is expressed by
the renormalization group equation for the \emph{a priori} measure,
which at leading order is the driven diffusion equation
\eq
\Lambda \partialby{\Lambda}d\rho =  \nabla_{i}\, g^{2} G^{ij}
\left [ \nabla_{j} + \partial_{j} \left (\frac1{g^{2}} S \right )
\right ] d\rho
\en
which equilibrates at
\eq
d\rho = d\lambda\; e^{-\frac1{g^{2}} S }
\,,
\en
$d\lambda$ being the metric volume element.
We recognize the \emph{a priori} measure produced 
by the small fluctuations as the functional measure 
of the canonically quantized spacetime quantum field theory.

The exact \emph{a priori} measure produced by the lambda model is a quantum field
theory in spacetime that might not be identical to the canonically 
quantized field theory.  We are pursuing the possibility that
non-canonical corrections to the canonical quantum field theory
might be produced by large
two-dimensional fluctuations in the lambda model.
At weak coupling, large fluctuations in a nonlinear model show up
as winding modes, associated with $\pi_{1}$ of the target manifold,
and as 2-d instantons, associated with $\pi_{2}$ of the target manifold.
The winding modes might provide
weakly interacting states not present in 
the canonical quantum $SU(2)$ gauge field theory.
The 2-d instantons might provide interactions
not present in 
the canonical quantum $SU(2)$ and $SU(3)$ gauge field theories.

The evolution of the nonlinear model in the two dimensional scale can be
represented by the radial quantization, in which the wave functions
live on the loop space of the target manifold and the hamiltonian is
the 2-d dilation operator.  The dilation operator of the lambda model
is the ordinary dilation operator of the nonlinear model combined with
the gradient flow.  The winding modes are wave functions on the
nontrivial component(s) of the loop space.  In a normal nonlinear
model, the low-lying winding modes are concentrated on the nontrivial 
loops of minimal length.  In the lambda model, the gradient flow attempts to
concentrate the wave function on the stable nontrivial loops.  The two processes
compete, in principle, so searching only for loops stable under the
gradient flow was ill-conceived.  Finding a stable loop of zero length
was pure luck.

It remains to quantize the stable loop of $SU(2)$ gauge fields.  The
ground state or states will be concentrated on the nontrivial loop at
the tip of the cone in Figure~\ref{fig:flow_lines_2}.  Regarding the
cone as an orbifold of the plane, the states of the stable loop are
the twist states for the orbifold.  The classical ground state energy
is zero because the length of the loop is zero.  But, if the stable
loop is to provide any low energy states, the quantum corrections to
the ground state energy must also vanish, at least to many orders in
the coupling constant.  We assume a high fundamental spacetime energy
scale in the lambda model.  Fermion zero-modes localized in the
zero-size instanton and anti-instanton of the stable loop will provide
degenerate ground states that offers at least a possibility of
canceling the quantum corrections to the ground state energy, but to
get cancellation to many orders or to all orders, we will probably
need perturbative spacetime supersymmetry.  The hypothetical
non-canonical $SU(2)$ gauge theory states would only be visible at low
energy in theories with perturbative supersymmetry.

It also remains to figure out the spacetime interpretation of the
states of the stable loop.  Do they appear as additional fields in
the quantum field theory, or as extra states, in addition to the
quantum field theory?  In any case, they will presumably be bi-local
objects, depending on the two instanton locations parametrizing the
loop of twisted pairs.

For consistency in the 2-d quantum field theory, twist fields have to
be accompanied by a projection that eliminates all states with
nontrivial monodromy around the twist field.  Here, nontrivial
monodromy means change of sign under nontrivial $SU(2)$ gauge
transformations.  Such states arise in canonically quantized gauge
theories that have a global $SU(2)$ anomaly \cite{Witten1982}.  By
projecting out the anomalous states, and adding new non-canonical
states, the lambda model might produce a non-anomalous quantization of
such gauge theories.

We expect that interactions with the ordinary modes of the gauge field theory 
will be determined by the outgoing trajectory that leads from the 
twisted pairs to the flat connection.
We picture two twist fields --- two loops of twisted pairs --- 
merging in the two dimensional domain.
Any closed curve surrounding them will be a topologically trivial loop in 
the twisted pairs.
In the radial quantization,
it will be driven down the outgoing trajectory to the flat connection,
where it can join to the rest of the two dimensional domain 
where there are only small fluctuations around the flat connection.
The interactions of the new states associated with the stable loop
should then be determined by how the outgoing trajectory approaches the flat 
connection.

We picture a 2-d instanton in the lambda model
to consist of a point-like core that is the nontrivial 2-sphere of 
twisted pairs of $SU(3)$ gauge fields (or twisted quadruplets of 
$SU(2)$ gauge fields),
evolving outwards in the radial quantization,
down the outgoing trajectory towards the flat connection.
Again, the effects on the ordinary states will be determined by how 
the outgoing trajectory approaches the flat connection.
On the two dimensional domain,
the instanton will look like a defect, around which the 
nearly flat gauge field winds by a nontrivial loop in the group of 
gauge transformations.

So far, we have only such vague speculations. 
The task now is to figure out how to calculate in 
the lambda model with the stable loop and 
the stable 2-spheres of gauge fields.

\begin{appendices}

\section{\texorpdfstring{$U(2)$-invariant connections on 
$S^{4}$}{U(2)-invariant connections on S4}}
\label{app:InvConnections}

\subsection{\texorpdfstring{$S^{3}$, $SU(2)$, $U(2)$, $SO(4)$, 
$S^{4}$}{S3, SU(2), U(2), SO(4), S4}}
$S^{3}$:
\eq
\zb=
\begin{pmatrix}
z_{1}\\
z_{2}
\end{pmatrix}
\,,
\qquad
\zb^{\dagger}\zb = \bar z_{1}z_{1}+ \bar z_{2} z_{2} =1
\,,
\qquad
P(\zb) = \zb \zb^{\dagger}\,, \qquad Q(\zb) =1 - P(\zb)
\,,
\en
\eq
\mathrm{dvol}_{S^{3}} = - \frac12 (\zb^{\dagger}d\zb) 
(d\zb^{\dagger}d\zb)\,,
\qquad
\int_{S^{3}} \mathrm{dvol}_{S^{3}}= 2\pi^{2}
\,.
\en

\noindent $SU(2)$:
\eq
g(\zb) = \begin{pmatrix}
z_{1} & -\bar z_{2} \\
z_{2} & \bar z_{1}
\end{pmatrix}
\,.
\en
\noindent $U(2)$:
\eq
g(U\zb) = U g(\zb) 
\begin{pmatrix}
1 & 0 \\
0 & (\det U)^{-1}
\end{pmatrix}
\,.
\en
\noindent $SO(4)=SU(2)\times SU(2)/\Integers_{2}$:
\eq
g(O (\zb)) = g_{L}g(\zb) g_{R}^{-1}
\,.
\en
\noindent $S^{4}$:
\eq
\vec y = (\cos \theta, \zb \sin \theta)
\,,\qquad
x = \ln \tan \left ( \frac{\theta}2 \right )\,,\qquad -\infty\le x \le 
\infty
\,.
\en
\eq
(ds)_{S^{4}}^{2}= 
\Round(x) \left [ (dx)^{2} + d\zb^{\dagger}d\zb \right ]
\,,\qquad
\Round(x) = (\cosh x)^{-2}\,.
\en

\subsection{\texorpdfstring{$U(2)$-invariant $su(2)$-valued 1-forms 
on $S^{3}$}{U(2)-invariant su(2)-valued 1-forms on S3}}
\label{app:invariantformsS3}
\begin{alignat}{2}
\eta &= -PdP
   &\quad&= - ( \zb^{\dagger}d \zb)\zb  \zb^{\dagger} - \zb d \zb^{\dagger} \\
\eta^{\dagger} &= -dP P &&= (\zb^{\dagger} d\zb)  \zb \zb^{\dagger}-d \zb \zb^{\dagger}
\\
\eta_{3} &= (\zb^{\dagger}d \zb)(P-Q) 
        &&= (\zb^{\dagger}d \zb) (2\zb\zb^{\dagger}-1)
\end{alignat}
\eq
\eta(\eb) =
\begin{pmatrix}
0 & -d\bar z_{2} \\
0 & 0
\end{pmatrix}
\qquad
\eta^{\dagger}(\eb) =
\begin{pmatrix}
0 & 0 \\
-d z_{2} & 0
\end{pmatrix}
\qquad
\eta_{3}(\eb) =
\begin{pmatrix}
dz_{1} & 0 \\
0 & -dz_{1}
\end{pmatrix}
\en
\eq
\eta^{2}=(\eta^{\dagger})^{2}=\eta_{3}^{2}=0
\en
\eq
d_{\omega}\omega = d\omega+\{\omega,\,\omega\} = \omega^{2}
\en
\eq
d_{\omega}(-\eta+\eta^{\dagger}-\eta_{3}) = \{\eta_{3},\eta\} - 
\{\eta_{3},\eta^{\dagger}\} - \{\eta,\eta^{\dagger}\}
\en
so, by $U(1)$-covariance,
\eq
d_{\omega}\eta = -\{\eta_{3},\eta\}
\qquad
d_{\omega}\eta^{\dagger} = - \{\eta_{3},\eta^{\dagger}\}
\qquad
d_{\omega}\eta_{3} = \{\eta,\eta^{\dagger}\}
\en

\subsection{\texorpdfstring{$U(2)$-invariant $su(2)$-valued 2-forms 
on $S^{3}$}{U(2)-invariant su(2)-valued 2-forms}}
Define
\begin{alignat}{4}
\sigma &= \frac12 d_{\omega}\eta &&=  -\frac12 \{\eta_{3},\,\eta\}
&&=\zb^{\dagger}d\zb PdP && = - \zb^{\dagger}d\zb\,\eta\\
\sigma^{\dagger} &= \frac12 d_{\omega}\eta^{\dagger} &&=-\frac12 
\{\eta_{3},\,\eta^{\dagger}\}
&&=-\zb^{\dagger}d\zb dPP && = \zb^{\dagger}d\zb\,\eta^{\dagger}\\
\sigma_{3}   &= \frac12 d_{\omega}\eta_{3} &&=\frac12 \{\eta,\,\eta^{\dagger}\}
& &=\frac12 (dP)^{2} &&= \frac12 d\zb^{\dagger}d\zb (P-Q)  
\end{alignat}
\eq
\sigma(\eb) =
\begin{pmatrix}
0 & dz_{1}d\bar z_{2} \\
0 & 0
\end{pmatrix}
\qquad
\sigma^{\dagger}(\eb) =
\begin{pmatrix}
0 & 0 \\
d\bar z_{1}dz_{2} & 0
\end{pmatrix}
\qquad
\sigma_{3}(\eb) =
\frac12 d\bar z_{2}dz_{2}
\begin{pmatrix}
1 & 0 \\
0 & -1
\end{pmatrix}
\en

\subsection{\texorpdfstring{$U(2)$-invariant $su(2)$-valued forms on 
$S^{4}$}{U(2)-invariant su(2)-valued forms on S4}}
The volume form on $S^{3}$ is
\eq
\mathrm{dvol}_{S^{3}}(\eb) = -\frac12 dz_{1} d\bar z_{2} dz_{2}
\qquad
\mathrm{dvol}_{S^{3}} = - \frac12 (\zb^{\dagger}d\zb) (d\zb^{\dagger}d\zb)
\qquad
\int_{S^{3}} \mathrm{dvol}_{S^{3}}= 2\pi^{2}
\,.
\en
In $S^{4}$, at each $x$,
we have a basis for the $U(2)$-invariant $su(2)$-valued forms
\aeq{
\text{0-forms:} \qquad & i(P-Q)\,,\\
\text{1-forms:} \qquad & idx (P-Q)\,,
\eta\,,  \eta^{\dagger} \,,  \eta_{3}\,,\hfill\\
\text{2-forms:} \qquad &
\sigma\,, \sigma^{\dagger}\,, \sigma_{3}\,,
dx\,\eta\,, dx\,\eta^{\dagger}\,,  dx\,\eta_{3}\,,
\\
\text{3-forms:} \qquad &
i \mathrm{dvol}_{S^{3}} (P-Q)\,,
dx\,\sigma\,, dx\,\sigma^{\dagger}\,, dx\,\sigma_{3}
\,.
}

\subsection{Hodge \texorpdfstring{$*$}{*}}
At $(x,\eb)\in S^{4}$, the Hodge $*$ operator acts on

1-forms and 3-forms:
\ateq{3}{
{*}^{2}&=-1 \qquad
&{*}dx &=R^{2}(x) \mathrm{dvol}_{S^{3}}\qquad
&{*}dz_{1} &= - R^{2}(x) dx d\bar z_{2}dz_{2}
\\
&
&{*}dz_{2} &= -R^{2}(x) dx dz_{1}dz_{2}\quad
&\quad {*}d\bar z_{2}& = R^{2}(x) dx dz_{1}d\bar z_{2}
}
\indent 2-forms:
\eq
{*}^{2}=1
\qquad
{*}(dxdz_{1}) = \frac12 d\bar z_{2}dz_{2}
\qquad
{*}(dxdz_{2}) = dz_{1}dz_{2}
\qquad
{*}(dxd\bar z_{2}) = -d z_{1}d\bar z_{2}
\en

\noindent
So Hodge ${*}$ acts on the $U(2)$-invariant $su(2)$-valued forms by:

1-forms and 3-forms:
\eq
{*}idx (P-Q)= R^{2}(x) \mathrm{dvol}_{S^{3}} i(P-Q)
\en
\eq
{*}\eta=-R^{2}(x)dx\sigma \qquad {*}\eta^{\dagger}=-R^{2}(x)dx\sigma^{\dagger}
\qquad {*}\eta_{3}=-R^{2}(x)dx\sigma_{3}
\en
\indent 2-forms:
\eq
{*}(dx \eta) = \sigma
\qquad
{*}(dx \eta^{\dagger}) = \sigma^{\dagger}
\qquad
{*}(dx \eta_{3}) = \sigma_{3}
\en

\subsection{\texorpdfstring{$F$, $F_{\pm}$}{F, F+-}}
\label{app:FFpm}
Recall
\eq
d+A =  d_{\omega}+\Delta A
\en
\eq
\Delta A = f(x)\eta-\bar f(x)\eta^{\dagger}+f_{3}(x) \eta_{3}\,,
\qquad  f_{3}= \bar f_{3}
\en
\eq
F = (d_{\omega}+\Delta A)^{2} = d_{\omega}\Delta A + (\Delta A)^{2}
\en
\eq
F_{\pm} = \frac12(1\pm{*})F
\,.
\en
Calculate
\eq
F = \partial_{x}f dx\eta-\partial_{x}\bar f dx\eta^{\dagger}+\partial_{x}f_{3}dx \eta_{3}
+2f\sigma-2\bar f \sigma^{\dagger}+2f_{3} \sigma_{3}
-2 f_{3}f\sigma +2 f_{3}\bar f \sigma^{\dagger} -2f\bar f\sigma_{3}
\en
\aeq{
F_{\pm} 
&= 
[\partial_{x}f\pm 2(1 -f_{3})f)] 
\,\frac12(1\pm{*})dx\eta 
- [\partial_{x}\bar f\pm 2(1 -f_{3})\bar f)] 
\,\frac12(1\pm{*})dx\eta^{\dagger} \\
&\quad{}+  [\partial_{x}f_{3}\pm 2(f_{3} -f \bar f)] 
\,\frac12(1\pm{*})dx\eta_{3}
}

\subsection{\texorpdfstring{$L_{\pm}$}{L+-}}
\eq
\frac14\tr(- F_{\pm}{*}F_{\pm}) = \mp\frac14 \tr(F_{\pm}^{2})
=L_{\pm}(x) dx \,\mathrm{dvol}_{S^{3}}
\en
\aeq{
\mp\frac14 \tr(F_{\pm}^{2})
&= \mp\frac14 \left [\partial_{x}f_{3}\pm 2\left (f_{3} -|f|^{2}\right )\right ]^{2} 
\,\frac12 \tr(\pm dx\eta_{3}\sigma_{3})  
\\
&\qquad {}\mp\frac14 \left |\partial_{x}f\pm 2\left (1 -f_{3}\right )f)\right |^{2}
\,\frac12 \tr(\mp dx\eta\sigma^{\dagger}
\mp dx\eta^{\dagger}\sigma) 
\nonumber\\
&=
\frac14\left [\partial_{x}f_{3}\pm 2\left (f_{3} -|f|^{2}\right 
)\right ]^{2}
dx \,\mathrm{dvol}_{S^{3}}
+\frac12\left |\partial_{x}f\pm 2\left (1 -f_{3}\right )f)\right |^{2}
dx \,\mathrm{dvol}_{S^{3}}
}
\eq
L_{\pm} =
\frac14
\left [\partial_{x}f_{3}\pm 2\left (f_{3} -|f|^{2}\right )\right ]^{2} 
+\frac12 \left |\partial_{x}f\pm 2\left (1 -f_{3}\right )f)\right |^{2}
\en

\subsection{Products of 1-forms and 2-forms}
The nonzero products of 1-forms and 2-forms are
\eq
\sigma \eta^{\dagger}=\eta \sigma^{\dagger} = 2 \mathrm{dvol}_{S^{3}} P
\qquad
\sigma^{\dagger} \eta=\eta^{\dagger} \sigma = 2 \mathrm{dvol}_{S^{3}} Q
\qquad
\eta_{3}\sigma_{3} = \sigma_{3}\eta_{3} = - \mathrm{dvol}_{S^{3}} 
\mathbf1
\en

\noindent
So
\eq
[\omega,\,\sigma] = [\omega,\,\sigma^{\dagger}] =  -2 
\mathrm{dvol}_{S^{3}} (P-Q)
\qquad
[\omega,\,\sigma_{3}] = 0
\en

\subsection{Inner products}
The non-zero inner products are:

\noindent
1-forms:
\eq
\tr \left [- idx (P-Q){*}idx (P-Q) \right ]
=\tr \left ( \eta {*}\eta^{\dagger} \right )
=\tr \left ( \eta^{\dagger} {*}\eta \right )
=\tr \left ( - \eta_{3} {*}\eta_{3} \right )
=2 R^{2}(x) dx\mathrm{dvol}_{S^{3}}
\en

\noindent
2-forms:
\eq
\tr(-dx\eta_{3}\sigma_{3})
=\tr (dx\eta\sigma^{\dagger}) = \tr(dx\eta^{\dagger}\sigma)
= 2 dx \,\mathrm{dvol}_{S^{3}}
\en

\subsection{New basis for the \texorpdfstring{$U(2)$}{U(2)}-invariant forms}
Change basis for the $U(2)$-invariant 
$su(2)$-valued 1-forms and 2-forms on $S^{3}$ to
\ateq{3}{
\omega &= -\eta+\eta^{\dagger}-\eta_{3}
\qquad &\omega_{1} &= \eta-\eta^{\dagger}-2\eta_{3}
\qquad &\omega_{2} &= -i(\eta+\eta^{\dagger})
\\
\chi &= -\sigma+\sigma^{\dagger}-\sigma_{3}
&\chi_{1} &= \sigma-\sigma^{\dagger}-2\sigma_{3}
&\chi_{2} &= -i(\sigma+\sigma^{\dagger})
}
Correspondingly, on $S^{4}$,
\aeq{
\text{1-forms:} \qquad & idx (P-Q)\,,
\omega\,,  \omega_{1} \,,  \omega_{2}\hfill\\
\text{2-forms:} \qquad &
\chi\,, \chi_{1}\,, \chi_{2}\,,
\\
&dx\,\omega\,, dx\,\omega_{1}\,,  dx\,\omega_{2}
\\
\text{3-forms:} \qquad &
i \mathrm{dvol}_{S^{3}} (P-Q)\,,
dx\,\chi\,, dx\,\chi_{1}\,, dx\,\chi_{2}
\,.
}

\subsection{Formulas in the new basis}
\label{app:newbasis}
\ateq{3}{
d_{\omega} \omega &= 2\chi
&d_{\omega} \omega_{1} &= 2\chi_{1}
& d_{\omega} \omega_{2} &= 2\chi_{2}
\\
\{\omega,\,\omega\}& = 4\chi
&\{\omega,\,\omega_{1}\} &= -2\chi_{1}
&\{\omega,\,\omega_{2}\} &= 2 \chi_{2}
\\
\Dpm \omega &= (2-4f_{\pm})\chi
\quad&\Dpm \omega_{1} &= (2+2f_{\pm})\chi_{1}
\quad&\Dpm \omega_{2}& = (2-2f_{\pm})\chi_{2}
}
\eq
[\omega,\,\chi] =[\omega,\,\chi_{1}] = 0
\qquad
[\omega,\,\chi_{2}] = 4i \mathrm{dvol}_{S^{3}} (P-Q)
\en

\noindent
Hodge ${*}$ on 1-forms:
\eq
{*} idx  (P-Q) = \Round(x) \mathrm{dvol}_{S^{3}} i(P-Q)
\en
\eq
{*}\omega =- \Round(x) dx \chi
\qquad
{*}\omega_{1} =- \Round(x) dx \chi_{1}
\qquad
{*}\omega_{2} =- \Round(x) dx \chi_{2}
\en
Hodge ${*}$ on 2-forms:
\eq
{*}\chi =  dx \omega
\qquad
{*}\chi_{1} = dx \omega_{1}
\qquad
{*}\chi_{2} = dx \omega_{2}
\en

\noindent
Non-zero inner products of 1-forms:
\eq
\tr \left [- idx (P-Q){*}idx (P-Q) \right ]
=2 R^{2}(x) dx\mathrm{dvol}_{S^{3}}
\en
\eq
\label{eq:innerprod}
\frac 16 \tr(-\omega{*}\omega) = 
\frac18 \tr(-\omega_{1}{*}\omega_{1}) = 
\frac14 \tr(-\omega_{2}{*}\omega_{2}) = R^{2}(x) dx\mathrm{dvol}_{S^{3}}
\en

\noindent
Non-zero inner products of 2-forms:
\eq
\frac 16 \tr(-dx \omega{*}dx \omega) = 
\frac18 \tr(-dx \omega_{1}{*}dx \omega_{1}) = 
\frac14 \tr(-dx \omega_{2}{*}dx \omega_{2}) = dx\mathrm{dvol}_{S^{3}}
\en

\subsection{Instanton covariant derivatives}

Using the formulas in Appendix~\ref{app:invariantformsS3}, we 
calculate the instanton covariant derivatives of the 1-forms
\eq
\Dpm  idx(P-Q) = -2f_{\pm} dx \omega_{2}
\qquad
\Dpm \omega = \lambda\chi
\qquad
\Dpm \omega_{1} = \lambda_{1} \chi_{1}
\qquad
\Dpm \omega_{2} = \lambda_{2} \chi_{2}
\en
where
\eq
\lambda = 2(1-2f_{\pm}) \qquad \lambda_{1} = 2(1+f_{\pm})
\qquad \lambda_{2} = 2(1-f_{\pm})
\,.
\en
The covariant derivatives of the 2-forms are
\eq
\Dpm (dx\,\omega) = - \lambda dx\, \chi
\qquad
\Dpm (dx\,\omega_{1}) = - \lambda_{1} dx\, \chi_{1}
\qquad
\Dpm (dx\,\omega_{2}) = - \lambda_{2} dx\,\chi_{2}
\en
\eq
\Dpm \chi = 0
\qquad
\Dpm \chi_{1} = 0
\qquad
\Dpm \chi_{2} = -4  f_{\pm}  \mathrm{dvol}_{S^{3}} i(P-Q)
\en
Their Hodge duals are
\eq
{*}\Dpm (dx\,\omega) = - R^{2}(x)^{-1} \lambda \omega
\qquad
{*}\Dpm (dx\,\omega_{1}) = - R^{2}(x)^{-1} \lambda_{1} \omega_{1}
\en
\eq
{*}\Dpm (dx\,\omega_{2}) = - R^{2}(x)^{-1}\lambda_{2} \omega_{2}
\en
\eq
{*}\Dpm \chi = 0
\qquad
{*}\Dpm \chi_{1} = 0
\qquad
{*}\Dpm \chi_{2} = 4  R^{2}(x)^{-1}f_{\pm}i dx (P-Q)
\en

\subsection{The instanton laplacian}
\label{app:laplacianformulas}
\aeq{
\Dpm \delta A_{0\pm}(x) idx(P-Q) &= -\delta A_{0\pm} 2f_{\pm} dx \omega_{2}\\
({*}\mp 1)\Dpm \delta A_{0\pm}(x) idx(P-Q) &=-2f_{\pm} \delta A_{0\pm}
(\chi_{2}\mp dx \omega_{2})\\
\Dpm ({*}\mp 1)\Dpm \delta A_{0\pm}(x) idx(P-Q) &=
-2\partial_{x}(f_{\pm}\delta  A_{0\pm})
dx\chi_{2}
-2f_{\pm} \delta A_{0\pm}
(\Dpm \chi_{2}\mp \Dpm dx \omega_{2})\\
{*}\Dpm ({*}\mp 1)\Dpm \delta A_{0\pm}(x) idx(P-Q) &=
-2\partial_{x}(f_{\pm} \delta A_{0\pm})\Round(x)^{-1}\omega_{2}
\\
&\qquad{}
-2 f_{\pm} \delta A_{0\pm}
\left [ 4 \Round(x)^{-1}  f_{\pm}i dx (P-Q) \right .
\\&\qquad\qquad\qquad\qquad\qquad{}
\left . \mp (-\lambda_{2}) \Round(x)^{-1}\omega_{2}
\right ]
\\
&=
-2\Round(x)^{-1}
(\partial_{x}\pm\lambda_{2})(f_{\pm} \delta A_{0\pm})
\omega_{2}
\\
&\qquad{}
-8 \Round(x)^{-1}  f_{\pm}^{2} \delta A_{0\pm} i dx (P-Q)
}
\aeq{
\Dpm \delta f_{\pm}(x) \omega &= \partial_{x}\delta f_{\pm}dx \omega
+ \delta f_{\pm}\lambda\chi
\\
({*}\mp 1)\Dpm \delta f_{\pm}(x) \omega &=(\partial_{x}\mp \lambda)\delta f_{\pm}
(\chi\mp dx \omega)\\
\Dpm ({*}\mp 1)\Dpm \delta f_{\pm}(x) \omega &=
\partial_{x}(\partial_{x}\mp \lambda)\delta f_{\pm} dx\chi
+(\partial_{x}\mp \lambda)\delta f_{\pm}(\Dpm \chi\mp \Dpm dx \omega)
\\
{*}\Dpm ({*}\mp 1)\Dpm \delta f_{\pm}(x) \omega &=
\Round(x)^{-1}(\partial_{x}\pm\lambda)(\partial_{x}\mp \lambda)\delta f_{\pm} \omega
}
\aeq{
\Dpm \delta  A_{1\pm}(x) \omega_{1} &= \partial_{x}\delta 
A_{1\pm} dx \omega_{1}
+ \delta A_{1\pm}\lambda_{1}\chi_{1}
\\
({*}\mp 1)\Dpm \delta A_{1\pm}(x) \omega_{1} &=(\partial_{x}\mp 
\lambda_{1})\delta A_{1\pm}
(\chi_{1}\mp dx \omega_{1})\\
\Dpm ({*}\mp 1)\Dpm \delta A_{1\pm}(x) \omega_{1} &=
\partial_{x}(\partial_{x}\mp \lambda_{1})\delta A_{1\pm} dx\chi_{1}
+(\partial_{x}\mp \lambda_{1})\delta A_{1\pm}(\Dpm \chi_{1}\mp 
\Dpm dx \omega_{1})
\\
{*}\Dpm ({*}\mp 1)\Dpm \delta A_{1\pm}(x) \omega_{1} &=
\Round(x)^{-1}(\partial_{x}\pm\lambda_{1})(\partial_{x}\mp 
\lambda_{1})\delta A_{1\pm}\omega_{1}
}
\aeq{
\Dpm \delta A_{2\pm}(x) \omega_{2} &= \partial_{x}\delta 
A_{2\pm} dx \omega_{2}
+ \delta A_{2\pm}\lambda_{2}\chi_{2}
\\
({*}\mp 1)\Dpm \delta A_{2\pm}(x) \omega_{2} &=(\partial_{x}\mp 
\lambda_{2})\delta A_{2\pm}
(\chi_{2}\mp dx \omega_{2})\\
\Dpm ({*}\mp 1)\Dpm \delta A_{2\pm}(x) \omega_{2} &=
\partial_{x}(\partial_{x}\mp \lambda_{2})\delta A_{2\pm} dx\chi_{2}
+(\partial_{x}\mp \lambda_{2})\delta A_{2\pm}(\Dpm \chi_{2}\mp 
\Dpm dx \omega_{2})
\\
{*}\Dpm ({*}\mp 1)\Dpm \delta A_{2\pm}(x) \omega_{2} &=
\partial_{x}(\partial_{x}\mp \lambda_{2})\delta A_{2\pm} 
\Round(x)^{-1}\omega_{2}
\\
&\qquad{}+(\partial_{x}\mp \lambda_{2})\delta 
A_{2\pm}(4\Round(x)^{-1}f_{\pm}i dx (P-Q)\pm \Round(x)^{-1} \lambda_{2} \omega_{2})
\nonumber\\
{*}\Dpm ({*}\mp 1)\Dpm \delta A_{2\pm}(x) \omega_{2} &=
\Round(x)^{-1}(\partial_{x}\pm\lambda_{2})(\partial_{x}\mp 
\lambda_{2})\delta A_{2\pm}\omega_{2}
\\
&\qquad{}+ 4\Round(x)^{-1}f_{\pm}(\partial_{x}\mp \lambda_{2})\delta 
A_{2\pm} i dx (P-Q) 
\nonumber
}

We expand $\delta A_{\pm}$ in a basis of $U(2)$ invariant 
$su(2)$-valued 1-forms on $S^{4}$,
\eq
\delta A_{\pm} = \delta A_{\pm0} (x) idx (P-Q)
+ \delta f_{\pm}(x)\,\omega+\delta A_{\pm1} (x)
\,\omega_{1}+\delta A_{\pm2}(x)\, \omega_{2}
\,.
\en
From Appendix~\ref{app:laplacianformulas},
\aeq{
{*} \Dpm  ({*}\mp1) \Dpm \delta A_{\pm} &=
\Round(x)^{-1}4f_{\pm} [-2   f_{\pm} \delta A_{0\pm} 
+ (\partial_{x}\mp \lambda_{2})\delta A_{2\pm}]
i dx (P-Q)
\\
&\qquad{}
+\Round(x)^{-1}(\partial_{x}\pm\lambda)(\partial_{x}\mp \lambda)\delta f_{\pm} \omega
\nonumber\\
&\qquad{}
+\Round(x)^{-1}(\partial_{x}\pm\lambda_{1})(\partial_{x}\mp 
\lambda_{1})\delta A_{1\pm}\omega_{1}
\nonumber\\
&\qquad{}
+\Round(x)^{-1}(\partial_{x}\pm\lambda_{2})
[-2f_{\pm} \delta A_{0\pm}
+(\partial_{x}\mp \lambda_{2})\delta A_{2\pm}
]
\omega_{2}
\nonumber
}
where
\eq
\lambda = 2(1-2f_{\pm}) \qquad \lambda_{1} = 2(1+f_{\pm})
\qquad \lambda_{2} = 2(1-f_{\pm})
\,.
\en
\ateq{2}{
(\partial_{x}\pm\lambda) &=f_{\pm}^{-1}(1-f_{\pm})^{-1}\partial_{x}f_{\pm}(1-f_{\pm})
\qquad&
(\partial_{x}\mp\lambda) &= f_{\pm}(1-f_{\pm})\partial_{x}f_{\pm}^{-1}(1-f_{\pm})^{-1}
\\
(\partial_{x}\pm\lambda_{1}) 
&=f_{\pm}^{-1}(1-f_{\pm})^{2}\partial_{x}f_{\pm}(1-f_{\pm})^{-2}
\qquad&
(\partial_{x}\mp\lambda_{1}) &= f_{\pm}(1-f_{\pm})^{-2}\partial_{x}f_{\pm}^{-1}(1-f_{\pm})^{2}
\\
(\partial_{x}\pm\lambda_{2}) &=f_{\pm}^{-1}\partial_{x}f_{\pm}
&
(\partial_{x}\mp\lambda_{2}) &= f_{\pm}\partial_{x}f_{\pm}^{-1}
}
so
\aeq{
{*} \Dpm  ({*}\mp1) \Dpm \delta A_{\pm} &=
\Round(x)^{-1}4f_{\pm}^{2} [-2  \delta A_{0\pm} 
+ \partial_{x}(f_{\pm}^{-1}\delta A_{2\pm})]
i dx (P-Q)
\\
&\qquad{}
+\Round(x)^{-1}f_{\pm}^{-1}(1-f_{\pm})^{-1}\partial_{x}
f_{\pm}^{2}(1-f_{\pm})^{2}
\partial_{x}f_{\pm}^{-1}(1-f_{\pm})^{-1}\delta f_{\pm} \omega
\nonumber\\
&\qquad{}
+\Round(x)^{-1}f_{\pm}^{-1}(1-f_{\pm})^{2}\partial_{x}
f_{\pm}^{2}(1-f_{\pm})^{-4}
\partial_{x}f_{\pm}^{-1}(1-f_{\pm})^{2}
\delta A_{1\pm}\omega_{1}
\nonumber\\
&\qquad{}
+\Round(x)^{-1}f_{\pm}^{-1}\partial_{x}
\left (
f_{\pm}^{2}
[-2 \delta A_{0\pm}
+\partial_{x}(f_{\pm}^{-1}\delta A_{2\pm})
]
\right )
\omega_{2}
\nonumber
}

\section{\texorpdfstring{$SU(2){\rightarrow} SU(3){\rightarrow}S^{5}$
pulled back along $[-1,1]{\times} S^{4}{\rightarrow} 
S^{5}$}{SU(2)->SU(3)->S5 pulled back along [-1,1]xS4}}
\label{app:DetailsComputer}

\subsection{\texorpdfstring{$SU(3)/SU(2)= S^{5}$}{SU(3)/SU(2)=S5}}
$S^{5}$ is represented as the unit sphere in $\Complexes \oplus 
\Complexes^{2}$.  The unit vectors are written
\eq
w =
\begin{pmatrix}
w_{0}\\
\wb
\end{pmatrix}
\qquad
|w_{0}|^{2}+ \wb^{\dagger}\wb =   1
\en
The north pole in $S^{5}$ is
\eq
n = 
\begin{pmatrix}
1\\
\mathbf{0}
\end{pmatrix}
\en
$SU(3)$ acts by block matrices on $\Complexes \oplus 
\Complexes^{2}$.
$SU(2)$ is identified with the subgroup of $SU(3)$ leaving $n$ 
fixed, the block matrices of the form
\eq
\begin{pmatrix}
1 & \mathbf0^{\dagger} \\
\mathbf0 & g
\end{pmatrix}
\,.
\en
We will write this $SU(3)$ matrix simply as $g$.

$w\in S^{5}$ is identified with the $SU(2)$ coset 
\eq
\{G\in SU(3): \; G n = w\}
\,.
\en

\subsection{\texorpdfstring{$U(2)$ acts on $SU(3)\rightarrow 
S^{5}$}{U(2) acts on SU(3)->S5}}
$U\in U(2)$ acts as a symmetry of the bundle $SU(3)\rightarrow S^{5}$ by
\eq
G \mapsto 
\begin{pmatrix}
1 & \mathbf0^{\dagger} \\
\mathbf0 & U
\end{pmatrix}
G
\begin{pmatrix}
1 & \mathbf0^{\dagger} \\
\mathbf0 & (\det U)^{-1/2} \,\mathbf1 
\end{pmatrix}
\en
\eq
\begin{pmatrix}
w_{0}\\
\wb
\end{pmatrix}
\mapsto
\begin{pmatrix}
w_{0}\\
U \wb
\end{pmatrix}
\en
where the sign of $(\det U)^{-1/2}$ is immaterial, because $-\mathbf 1$ 
acts trivially on connections, being in the center of $SU(2)$.

\subsection{A map \texorpdfstring{$[-1,1]\times S^{4}\rightarrow 
S^{5}$}{[-1,1]xS4->S5}}

We construct a nontrivial loop of connections
by pulling back along a map $[-1,1]\times S^{4}\rightarrow S^{5}$,
\eq
(s,\vec y)\mapsto w(s,\vec y)
\en
\eq
w(s,\vec y)
=
\begin{pmatrix}
\cos \theta + i s \sin \theta \\
\zb \sin \theta \sqrt{1-s^{2}}
\end{pmatrix}\,,
\qquad
\vec y=(\cos \theta, \zb \sin \theta)
\,.
\en
This map is manifestly $U(2)$-invariant so, 
for each $s\in [-1,1]$, the pulled back connection $A(s)$ on $S^{4}$ is 
$U(2)$ invariant.
At $s=\pm 1$, the pulled back connection over $S^{4}$ is flat,
so we get a closed loop in $\mathcal{A}/\mathcal{G}$.

\subsection{Trivialize}
We find a formula for $A(s)$ by trivializing $SU(3)\rightarrow S^{5}$
over a convenient region of $S^{5}$, giving,
for each of the $SU(2)$-bundles in the loop,
a trivialization over $S^{4}\backslash \text{south pole}$.
Recall
\eq
w = 
\begin{pmatrix}
w_{0}\\
\wb
\end{pmatrix}
=
\begin{pmatrix}
w_{0}\\
\zb\sqrt{1-|w_{0}|^{2}}
\end{pmatrix}
\,,
\qquad
P =  \zb\zb^{\dagger}\,,
\quad Q=1-P
\,.
\en
Define a partial section $S^{5}\rightarrow SU(3)$
\eq
w\mapsto G(w) = 
\begin{pmatrix}
w_{0} & -\wb^{\dagger} \\
\wb & \bar w_{0} P + Q
\end{pmatrix}
\,.
\en
which is regular on $S^{5}$ except where $|w_{0}|=1$, $w_{0}\ne 1$.
For each $s$ except $s\pm 1$,
it is regular on $S^{4}$ away from the south 
pole $\theta = \pi$.

The invariant connection in $SU(2)\rightarrow SU(3)\rightarrow S^{5}$ 
takes the explicit form
\eq
d + A_{\mathit{inv}}(w) =d_{\omega} + \Delta A_{\mathit{inv}}(w)=  d + P_{V} G(w)^{-1} d G(w)
\en
where $P_{V}$ is the invariant projection on $su(2)\subset su(3)$.
We calculate,
\eq
\Delta A_{\mathit{inv}}(w) = \left [
\frac12 (w_{0}d\bar w_{0}-\bar w_{0} dw_{0})
-(1-|w_{0}|^{2})\zb^{\dagger}d\zb \right ]
\frac12 ( P -Q)
+\bar w_{0}dP P -w_{0} PdP
\,.
\en

\subsection{A formula for \texorpdfstring{$\Delta A(s)$}{Delta A(s)}}
We substitute $w_{0}= cos \theta + i s \sin \theta$
and, for each $s$, restrict the connection to $S^{4}$
\eq
\Delta  A(s)=
-\frac12 i s d\theta ( P -Q)
+(\cos \theta -is\sin \theta)\eta_{-} +(\cos \theta +is\sin \theta)\eta_{+}
+\left [1-\frac12 (1-s^{2}) \sin^{2}\theta\right ] \eta_{3}
\en
At the midpoint of the loop, $s=0$,
\eq
\Delta A(0) = 
\cos \theta \, (\eta_{+}+\eta_{-})
+\left (1-\frac12 \sin^{2}\theta\right ) \eta_{3}
\label{eq:midpoint}
\en

\subsection{Discrete symmetries of the loop}

In addition to the $U(2)$ symmetry, there are two discrete symmetries.

\subsubsection{Reflection symmetry}
The bundle $SU(2)\rightarrow SU(3)\rightarrow S^{5}$
has a discrete symmetry
\eq
G\mapsto
\begin{pmatrix}
-1 & 0 \\
0 & \mathbf1
\end{pmatrix}
G
\begin{pmatrix}
1 & 0 \\
0 & i \mathbf1
\end{pmatrix}
\en
which acts on $S^{5}$ by
\eq
w
\mapsto
Rw =
\begin{pmatrix}
-w_{0}\\
\wb
\end{pmatrix}
\en
so on $[-1,1]\times S^{4}$ by
\eq
s\mapsto -s\,,
\qquad
(y_{0},\yb)
\mapsto
(-y_{0},\yb)
\,.
\en

\subsubsection{Conjugation symmetry}

Complex conjugation acts on $SU(2)$ by
\eq
\bar g = g_{1} g g_{1}^{-1} \qquad
g_{1} = 
\begin{pmatrix}
0 & 1 \\
-1 & 0
\end{pmatrix}
\,.
\en
Define, for $G\in SU(3)$,
\eq
G_{c} = \bar G g_{1} \,.
\en
Then, for $g\in SU(2)$,
\eq
(G g)_{c} = \bar G \bar g g_{1} = \bar G g_{1} g = G_{c} g
\en
so $G\mapsto G_{c}$
is a symmetry of the bundle $SU(2)\rightarrow SU(3)\rightarrow S^{5}$.
It acts on $S^{5}$ by
\eq
w
\mapsto
\bar w
\en
so on $I\times S^{4}$ by
\eq
s\mapsto -s\,,
\qquad
(y_{0},\yb)
\mapsto
 (y_{0},\bar \yb)
\,.
\en

\subsection{Action of the discrete symmetries}

\subsubsection{Action of the reflection symmetry}
\eq
G(Rw)  = G(w)_{r} g_{r}(w)^{-1}
\en
\eq
\begin{pmatrix}
-w_{0} & -\wb^{\dagger} \\
\wb & -\bar w_{0} P + Q
\end{pmatrix}
=
\begin{pmatrix}
-w_{0} & i\wb^{\dagger} \\
\wb & i(\bar w_{0} P + Q)
\end{pmatrix}
g_{r}(w)^{-1}
\en
\eq
g_{r}(w)^{-1} =
\begin{pmatrix}
-\bar w_{0} & \wb^{\dagger} \\
-i \wb & -i(w_{0} P + Q)
\end{pmatrix}
\begin{pmatrix}
-w_{0} & -\wb^{\dagger} \\
\wb & -\bar w_{0} P + Q
\end{pmatrix}
=\begin{pmatrix}
1 & 0 \\
0 & i (P - Q)
\end{pmatrix}
\en
\eq
g_{r} = -i(P-Q)
\en
\aeq{
d_{\omega}+ \Delta A(Rw) &= g_{r}(w) (d_{\omega}+ \Delta A(w)) g_{r}^{-1}(w) \\
&= d_{\omega}+ g_{r}(w)  \Delta A(w) g_{r}^{-1}(w) 
}
since
\eq
(d +\omega) P = g d g^{-1} g P(\eb) g^{-1} = P (d+\omega)
\en
so
\eq
\Delta A(Rw)
= g_{r} \Delta A A(w) g_{r}^{-1}
\en
We have
\eq
g_{r} \eta g_{r}^{-1} = -\eta
\qquad
g_{r} \eta^{\dagger} g_{r}^{-1} = -\eta^{\dagger}
\qquad
g_{r} \eta_{3} g_{r}^{-1} = \eta_{3}
\en
so,
writing
\eq
\Delta A(s) = 
A_{0}(s,\theta) d\theta\, i(P-Q)
+ f(s,\theta) \eta
-\bar f(s,\theta) \eta^{\dagger}
+ f_{3}(s,\theta) \eta_{3}
\en
Then the reflection symmetry is
\aeq{
A_{0}(s,\theta) &= - A_{0}(-s,\pi-\theta)\\
f(s,\theta) &= -  f(-s,\pi-\theta)\\
f_{3}(s,\theta) &=  f_{3}(-s,\pi-\theta)
}

\subsubsection{Action of the conjugation symmetry}

$G(\bar w)$ and $G(w)_{c}$ belong to the same $SU(2)$ coset, so there 
is $g_{c}\in SU(2)$ such that
\eq
G(\bar w) =  G(w)_{c} g_{c}(w)^{-1}
= \overline{G(w)} g_{1}g_{c}(w)^{-1}
= G(\bar w)g_{1}g_{c}(w)^{-1}
\en
so
\eq
g_{c} = g_{1}
= 
\begin{pmatrix}
0 & 1 \\
-1 & 0
\end{pmatrix}
\,.
\en
$G\mapsto G_{c}$ is a symmetry, so
\eq
d+ A(\bar w) = g_{c} (d + A(w)) g_{c}^{-1}
\en
or
\eq
A(w) = g_{c}  A(\bar w) g_{c}^{-1}
\,.
\en
Complex conjugation acting on $su(2)$ gives
\eq
g_{c} \bar \eta_{3} g_{c}^{-1} =\eta_{3}
\qquad
g_{c} \bar \eta g_{c}^{-1} =-\eta^{\dagger}
\qquad
g_{c} \bar \eta^{\dagger} g_{c}^{-1} =-\eta
\en
In particular, $g_{c} \bar \omega g_{c}^{-1} =\omega$,
so the conjugation symmetry is
\eq
\Delta A(w) = g_{c}  \Delta A(\bar w) g_{c}^{-1}
\en
\aeq{
A_{0}(s,\theta) &= -  A_{0}(-s,\theta) \\
f(s,\theta) &=  \bar f(-s,\theta) \\
f_{3}(s,\theta) &=  f_{3}(-s,\theta)
\,.
}

\subsubsection{Summary of the actions of the discrete symmetries}

Each connection $A(s)$ along the loop has the discrete symmetry
that combines reflection and conjugation,
\aeq{
A_{0}(s,\theta) &=  A_{0}(s,\pi-\theta)\\
f(s,\theta) &= -\bar f(s,\pi-\theta)\\
f_{3}(s,\theta) &=  f_{3}(s,\pi-\theta)
\,.
}
In addition, there is a discrete symmetry that reflects the loop,
leaving the midpoint $s=0$ fixed,
\aeq{
A_{0}(s,\theta) &= -  A_{0}(-s,\theta) \\
f(s,\theta) &=  \bar f(-s,\theta)\\
f_{3}(s,\theta) &=  f_{3}(-s,\theta)
\,.
}

\subsubsection{Discrete symmetries at the midpoint 
\texorpdfstring{$s=0$}{s=0}}
The two discrete symmetries of the loop leave fixed the midpoint, 
$s=0$, so they are symmetries of the connection $ A(0)$.
The conjugation symmetry is
\eq
A_{0}(0,\theta)  =  0 \,,\qquad 
f(0,\theta) =  \bar f(0,\theta)
\,.
\en
The reflection symmetry is
\eq
f(0,\theta) =  - f(0,\pi-\theta) \,,
\qquad
f_{3}(0,\theta) =  f_{3}(0,\pi-\theta)
\,.
\en
Both can be checked in equation~\ref{eq:midpoint}.

\subsection{Nontriviality of the loop}

\subsubsection{Trivialize over \texorpdfstring{$S^{4}\backslash 
\text{north pole}$}{S4\\ north pole}}

Define a second partial section
which is nonsingular on $S^{4}$ except at the north pole,
\eq
G^{-}(w) = 
\begin{pmatrix}
w_{0} & - \wb^{\dagger} \\
\wb & \bar w_{0} P - Q
\end{pmatrix}
= G(w) 
\begin{pmatrix}
1 & 0 \\
0 &  P - Q
\end{pmatrix}
\,.
\en
The pulled back connection is
\aeq{
d_{\omega}+  \Delta A^{-}(s) &= (P-Q) (d_{\omega} + \Delta A(s)) (P-Q)
\\
&= d_{\omega} + (P-Q) \Delta A(s) (P-Q)\\
\Delta A^{-}(s) &=(P-Q) \Delta A(s) (P-Q)
}

\subsubsection{Transform \texorpdfstring{$ A(s)$ to $A_{0}=0$}{A(s) 
to A0=0} gauge}
Gauge transform $ A(s)$ to $A_{0}=0$ gauge, using a gauge transformation that 
is regular at  $\theta=0$, the north pole,
\aeq{
d_{\omega}+\Delta \tilde A(s) &= e^{-\frac12 i s \theta (P-Q)} (d_{\omega} 
+\Delta A(s)) e^{\frac12 i s \theta (P-Q)}
\\
&= d_{\omega} +\frac12 i s d\theta (P-Q)+ e^{-\frac12 i s \theta  (P-Q)} \Delta A(s) e^{\frac12 i s \theta (P-Q)}
}
so, in $A_{0}=0$ gauge, the loop is
\eq
\Delta \tilde A(s) =
(\cos \theta +is\sin \theta)e^{-is\theta} \eta
-(\cos \theta  -is\sin \theta)e^{is\theta}\eta^{\dagger}
+\left [1-\frac12 (1-s^{2}) \sin^{2}\theta\right ]\, \,\eta_{3}
\en

\subsubsection{Transform \texorpdfstring{$ A^{-}(s)$ to 
$A_{0}=0$}{A-(s) to A0=0} gauge}
Gauge transform $ A^{-}(s)$ to $A_{0}=0$ gauge, using a gauge transformation that 
is regular at $\theta=\pi$, the south pole,
\aeq{
d_{\omega} + \Delta \tilde A^{-}(s) &= e^{-\frac12  i s (\theta-\pi)(P-Q)} 
(d_{\omega}+ \Delta A^{-}(s)) e^{\frac12 i  s (\theta-\pi) (P-Q)}
\\
&=
\phi(s)
(D+\Delta \tilde A(s))\phi(s)^{-1}
}
with
\eq
\phi(s) = e^{-\frac12 i  s (\theta-\pi) (P-Q)}i^{-1} (P-Q)e^{\frac12  i s \theta (P-Q)}
= e^{\frac12 i (s-1) \pi (P-Q)}
\en

\subsubsection{Patching map at the equator is the suspension of the Hopf fibration}
Now we have a loop of bundles over $S^{4}$, made from
trivial bundles over the two hemispheres patched together by 
the loop of gauge transformations $\phi(s)$, which we can write
\eq
\phi(s)
= 
g(\zb)
\begin{pmatrix}
e^{\frac12 i\pi (s-1)} & 0\\
0 & e^{-\frac12 i\pi (s-1)}
\end{pmatrix}
g(\zb)^{-1}
\en
which is the suspension of the Hopf fibration.
So the loop of connections is nontrivial.

\end{appendices}

\addcontentsline{toc}{section}{Acknowledgments}
\section*{Acknowledgments}
I thank V. Calian of the Natural Science Institute, University of Iceland, 
for discussions of exact solutions of nonlinear diffusion equations.
I am grateful to T. Mrowka for pointing out references on solving the 
Yang-Mills equations with symmetry and on the Yang-Mills flow.
%

\addcontentsline{toc}{section}{References}
\bibliographystyle{unsrt}
\bibliography{Literature}

\begin{thebibliography}{10}

\bibitem{Singer1978}
I.~M. Singer.
\newblock {Some remarks on the Gribov ambiguity.}
\newblock {\em Commun. Math. Phys.}, 60:7--12, 1978.

\bibitem{Hopf1931}
Heinz Hopf.
\newblock {\"Uber die Abbildungen der dreidimensionalen Sph\"are auf die
  Kugelfl\"ache.}
\newblock {\em Math. Ann.}, 104:637--665, 1931.

\bibitem{Freudenthal1937}
Hans Freudenthal.
\newblock {\"Uber die Klassen der Sph\"arenabbildungen. I. Gro{\ss}e
  Dimensionen.}
\newblock {\em Compos. Math.}, 5:299--314, 1937.

\bibitem{Eckmann1942thesis}
B.~Eckmann.
\newblock {Zur Homotopietheorie Gefaserter R\"aume}.
\newblock {\em Comm. Math. Helv.}, 14:141--192, 1942.

\bibitem{Pontryagin1950}
L.~S. Pontryagin.
\newblock Homotopy classification of the mappings of an {$(n+2)$}-dimensional
  sphere on an {$n$}-dimensional one.
\newblock {\em Doklady Akad. Nauk SSSR (N.S.)}, 70:957--959, 1950.

\bibitem{Whitehead1950}
George~W. Whitehead.
\newblock The {$(n+2)^{\rm nd}$} homotopy group of the {$n$}-sphere.
\newblock {\em Ann. of Math. (2)}, 52:245--247, 1950.

\bibitem{Friedan2003}
Daniel Friedan.
\newblock {A tentative theory of large distance physics}.
\newblock {\em JHEP}, 10:063, 2003.

\bibitem{Belavin1975}
A.~A. Belavin, Alexander~M. Polyakov, A.~S. Shvarts, and Yu.~S. Tyupkin.
\newblock {Pseudoparticle solutions of the Yang-Mills equations}.
\newblock {\em Phys. Lett.}, B59:85--87, 1975.

\bibitem{Donaldson1985}
S.~K. Donaldson.
\newblock Anti self-dual {Y}ang-{M}ills connections over complex algebraic
  surfaces and stable vector bundles.
\newblock {\em Proc. London Math. Soc. (3)}, 50(1):1--26, 1985.

\bibitem{Struwe1994}
Michael Struwe.
\newblock The {Y}ang-{M}ills flow in four dimensions.
\newblock {\em Calc. Var. Partial Differential Equations}, 2(2):123--150, 1994.

\bibitem{Schlatter1996}
Andreas Schlatter.
\newblock Global existence of the {Y}ang-{M}ills flow in four dimensions.
\newblock {\em J. Reine Angew. Math.}, 479:133--148, 1996.

\bibitem{Schlatter1997}
Andreas Schlatter.
\newblock Long-time behaviour of the {Y}ang-{M}ills flow in four dimensions.
\newblock {\em Ann. Global Anal. Geom.}, 15(1):1--25, 1997.

\bibitem{Schlatter1998}
Andreas~E. Schlatter, Michael Struwe, and A.~Shadi Tahvildar-Zadeh.
\newblock Global existence of the equivariant {Y}ang-{M}ills heat flow in four
  space dimensions.
\newblock {\em Amer. J. Math.}, 120(1):117--128, 1998.

\bibitem{Sibner1989}
L.~M. Sibner, R.~J. Sibner, and K.~Uhlenbeck.
\newblock Solutions to {Y}ang-{M}ills equations that are not self-dual.
\newblock {\em Proc. Nat. Acad. Sci. U.S.A.}, 86(22):8610--8613, 1989.

\bibitem{Donaldson1986}
S.~K. Donaldson.
\newblock Connections, cohomology and the intersection forms of
  {$4$}-manifolds.
\newblock {\em J. Differential Geom.}, 24(3):275--341, 1986.

\bibitem{Taubes1983}
Clifford~Henry Taubes.
\newblock Stability in {Y}ang-{M}ills theories.
\newblock {\em Comm. Math. Phys.}, 91(2):235--263, 1983.

\bibitem{Puettmann2003}
Thomas Puettmann and A.~Rigas.
\newblock {Presentations of the first homotopy groups of the unitary groups}.
\newblock {\em Comment. Math. Helv.}, 78:648--662, 2003.

\bibitem{Urakawa1988}
Hajime Urakawa.
\newblock Equivariant theory of {Y}ang-{M}ills connections over {R}iemannian
  manifolds of cohomogeneity one.
\newblock {\em Indiana Univ. Math. J.}, 37(4):753--788, 1988.

\bibitem{Bor1990}
Gil Bor and Richard Montgomery.
\newblock {$SO(3)$ invariant Yang-Mills fields which are not self-dual.}
\newblock {Hamiltonian systems, transformation groups and spectral transform
  methods, Proc. CRM Workshop, Montr\'eal/Can. 1989, 191-198 (1990).}, 1990.

\bibitem{Sadun1990}
Lorenzo Sadun and Jan Segert.
\newblock Constructing non-self-dual {Y}ang-{M}ills connections on {$S^4$} with
  arbitrary {C}hern number.
\newblock In {\em Differential geometry: geometry in mathematical physics and
  related topics ({L}os {A}ngeles, {CA}, 1990)}, volume~54 of {\em Proc.
  Sympos. Pure Math.}, pages 529--537. Amer. Math. Soc., Providence, RI, 1993.

\bibitem{Sadun1991}
Lorenzo Sadun and Jan Segert.
\newblock Non-self-dual {Y}ang-{M}ills connections with nonzero {C}hern number.
\newblock {\em Bull. Amer. Math. Soc. (N.S.)}, 24(1):163--170, 1991.

\bibitem{Sadun1992}
Lorenzo Sadun and Jan Segert.
\newblock Non-self-dual {Y}ang-{M}ills connections with quadrupole symmetry.
\newblock {\em Comm. Math. Phys.}, 145(2):363--391, 1992.

\bibitem{Sadun92sta}
Lorenzo Sadun and Jan Segert.
\newblock Stationary points of the {Y}ang-{M}ills action.
\newblock {\em Comm. Pure Appl. Math.}, 45(4):461--484, 1992.

\bibitem{Bor1992}
Gil Bor.
\newblock Yang-{M}ills fields which are not self-dual.
\newblock {\em Comm. Math. Phys.}, 145(2):393--410, 1992.

\bibitem{Parker1992a}
Thomas~H. Parker.
\newblock Nonminimal {Y}ang-{M}ills fields and dynamics.
\newblock {\em Invent. Math.}, 107(2):397--420, 1992.

\bibitem{Parker1992}
Thomas~H. Parker.
\newblock A {M}orse theory for equivariant {Y}ang-{M}ills.
\newblock {\em Duke Math. J.}, 66(2):337--356, 1992.

\bibitem{Sadun1994}
Lorenzo Sadun.
\newblock A symmetric family of {Y}ang-{M}ills fields.
\newblock {\em Comm. Math. Phys.}, 163(2):257--291, 1994.

\bibitem{sage}
W.\thinspace{}A. Stein et~al.
\newblock {\em {S}age {M}athematics {S}oftware ({V}ersion 4.3.4)}.
\newblock The Sage Development Team, 2009.
\newblock {\tt http://www.sagemath.org}.

\bibitem{Friedan2009Pisa}
D.~Friedan.
\newblock Preliminary evidence for a stable 2-sphere in the {Y}ang-{M}ills flow
  for {$SU(3)$} gauge fields on {$S^4$}.
\newblock Talk at workshop: Geometric Flows in Mathematics and Theoretical
  Physics, Pisa, June 24, 2009.
\newblock Slides at http://www.physics.rutgers.edu/pages/friedan/.

\bibitem{Chaves1996}
Lucas~M. Chaves and A.~Rigas.
\newblock Complex reflections and polynomial generators of homotopy groups.
\newblock {\em J. Lie Theory}, 6(1):19--22, 1996.

\bibitem{Atiyah1978}
M.~F. Atiyah, N.~J. Hitchin, and I.~M. Singer.
\newblock Self-duality in four-dimensional {R}iemannian geometry.
\newblock {\em Proc. Roy. Soc. London Ser. A}, 362(1711):425--461, 1978.

\bibitem{Atiyah1971}
M.~F. Atiyah and I.~M. Singer.
\newblock The index of elliptic operators. {IV}.
\newblock {\em Ann. of Math. (2)}, 93:119--138, 1971.

\bibitem{Nucci199249}
M.~C. Nucci and P.~A. Clarkson.
\newblock The nonclassical method is more general than the direct method for
  symmetry reductions. an example of the fitzhugh-nagumo equation.
\newblock {\em Physics Letters A}, 164(1):49 -- 56, 1992.

\bibitem{VCalian2009}
V.~Calian.
\newblock Private coversations, 2009-2010.

\bibitem{Witten1982}
E.~Witten.
\newblock {An SU(2) anomaly}.
\newblock {\em Phys. Lett.}, B117:324--328, 1982.

\end{thebibliography}
%
%
%

\end{document}